%% file: 2017-ESA-B-SUBMISSION_CFLAT-EVALUATION.tex
\documentclass[a4paper,USenglish]{lipics-v2016}

\usepackage{wrapfig,graphicx,xcolor,url,hyperref,booktabs,amssymb}

\usepackage{colortbl,multirow,morefloats,bchart}
\usepackage{color,soul}

\usepackage{rotating}

\usepackage[ddmmyyyy,hhmmss]{datetime}

\usepackage{mystyle_lipics-compatible}

\title{Improved Oracles for Time-Dependent Road Networks\footnote{%
	Partially supported by
	EU FP7/2007-2013 under grant agreements no.~609026 (project MOVESMART),
	no.~621133 (project HoPE), and by DFG grant WA 654/23-1 within FOR 2083.	
	}
}

\titlerunning{Improved Oracles for Time-Dependent Road Networks} 

\author[1,4]{Spyros Kontogiannis}
\author[1,4]{Georgia Papastavrou}
\author[3,4]{Andreas Paraskevopoulos}
\author[2]{Dorothea Wagner}
\author[3,4]{Christos Zaroliagis}

\affil[1]{%
	Department of Comp. Science \& Engineering,
	University of Ioannina,
	Greece
  \\
  \hfill[\texttt{kontog@cse.uoi.gr}~,~\texttt{gioulycs@gmail.com}]}

\affil[2]{%
	Karlsruhe Institute of Technology,
	Germany
	\hfill[\texttt{dorothea.wagner@kit.edu}] \\ ~}

\affil[3]{%
	Department of Comp. Eng. \& Informatics,
	University of Patras, 
	Greece
	\\
	\hfill[\texttt{$\{$~paraskevop~,~zaro~$\}$@ceid.upatras.gr}]}

\affil[4]{%
	Computer Technology Institute and Press ``Diophantus'',
	Greece
	}

\authorrunning{Kontogiannis,~Papastavrou,~Paraskevopoulos,~Wagner,~Zaroliagis}

\Copyright{S.~Kontogiannis, G.~Papastavrou, A.~Paraskevopoulos, D.~Wagner and C.~Zaroliagis}

\subjclass{%
	F.2.2 Nonnumerical Algorithms and Problems; 
}

\keywords{%
	Time-dependent shortest paths;
	FIFO property;
	Distance oracles.
}

\EventEditors{}
\EventNoEds{1}
\EventLongTitle{}
\EventShortTitle{\scriptsize Last latexed: \today}
\EventAcronym{}
\EventYear{}
\EventDate{}
\EventLocation{}
\EventLogo{}
\SeriesVolume{}
\ArticleNo{}

\begin{document}

\maketitle

\vspace*{-0.5cm}
\begin{abstract}

A novel landmark-based oracle ($\alg{CFLAT}$) is presented, which provides earliest-arrival-time route plans in time-dependent road networks.
To our knowledge, this is the first oracle that preprocesses combinatorial structures (collections of time-stamped min-travel-time-path trees)
rather than travel-time functions.
The preprocessed data structure is exploited by a new query algorithm ($\alg{CFCA}$) which computes (and pays for it), apart from earliest-arrival-time estimations, the \emph{actual connecting path} that preserves the theoretical approximation guarantees.
To make it practical and tackle the main burden of landmark-based oracles (the large preprocessing requirements), $\alg{CFLAT}$ is extensively engineered.
A thorough experimental evaluation on two real-world benchmark instances shows that $\alg{CFLAT}$ achieves a significant improvement on preprocessing, approximation guarantees and query-times, in comparison to previous landmark-based oracles, whose query
algorithms do \emph{not} account for the path construction. It also achieves competitive query-time performance and approximation guarantees compared to state-of-art speedup heuristics for time-dependent road networks, whose query-times in most cases do \emph{not} account for path construction.

\end{abstract}

\section{Introduction}

The surge for efficient solutions (min-cost paths) in networks with temporal characteristics is a highly challenging research goal,
due to both the large-scale and the time-varying nature of the underlying arc-cost metric.
Along this line, the development of practical algorithms for providing earliest-arrival-time route plans in large-scale road networks
accompanied with a time-dependent arc-travel-time metric (known as \term{Time-Dependent Route Planning} -- TDRP),
has received a lot of attention in the last decade.
TDRP is a hard challenge, both theoretically and in practice. For certain tractable cases,
there is an analogue of Dijkstra's algorithm (called \term{Time-Dependent Dijkstra} -- $\alg{TDD}$)
to solve the problem in quasi-linear time, which is already too much for a route-planning application
supporting real-time query responses in \emph{large-scale} road networks.
Time-dependence is also by itself a quite important degree of complexity, both in space and in query-time requirements. 
These two challenges have been tackled in the past either by oracles, or by speedup heuristics.
	An \emph{oracle} is a preprocessed and succinctly stored data structure encoding min-cost path information
for carefully selected pairs of vertices. This data structure is accompanied with a query algorithm, which responds
to arbitrary queries in time \emph{provably better} than the corresponding Dijkstra-time and, if approximate
solutions are also an option, with a \emph{provable} approximation guarantee (stretch).
	Analogously, a \emph{speedup heuristic} preprocesses arc-cost metrics which are custom-tailored to road networks,
and then uses a query algorithm for responding to (exact or approximate) min-cost path queries in time that is in
practice \emph{several orders of magnitude} faster than the running time of Dijkstra's algorithm.

\remove{
In this work we present, engineer and experimentally evaluate a novel \emph{landmark-based} oracle ($\alg{CFLAT}$),
whose objective is to tackle the main burden of such oracles, the large preprocessing requirements, without
compromising either the preprocessing scalability, or the competitiveness of query-response times and approximation guarantees.
}

\vspace*{-12pt}
\subparagraph*{Modeling Instances, Problem Statement \& Related Work.}

We model road network instances by directed graphs in which every arc $a=uv$ depicts an uninterrupted portion of a road segment and is accompanied by an arc-travel-time \emph{function} $D[a]$ determining the time to traverse $a$, given the departure-time from its tail $u$. These functions are assumed to be continuous, piecewise-linear (pwl), periodic with one-day period, and are succinctly represented as sequences of consecutive \term{breakpoints}, i.e., (departure-time,arc-travel-time) pairs.
This model is typical in the literature when we seek for route plans for private cars (e.g.,
\cite{%
	2009-Delling-Wagner,%
	2010-Demiryurek-Kashani-Shahabi,%
	2011-Delling_TDSHARC,%
	2012-Nannicini-Delling-Liberti-Schultes,%
	2012-Delling-Nannicini,%
	2013-Batz-Geisberger-Sanders-Vetter,%
	2014-Foschini-Hershberger-Suri,%
	2014-Omran-Sack,%
	2014-Kontogiannis-Zaroliagis,2016-Kontogiannis-Wagner-Zaroliagis,%
	2015-Kontogiannis-Michalopoulos-Papastavrou-Paraskevopoulos-Wagner-Zaroliagis,%
	2016-Baum-Dibbelt-Pajor-Wagner,%
	2016-Kontogiannis-Michalopoulos-Papastavrou-Paraskevopoulos-Wagner-Zaroliagis}).
For an arbitrary pair $(o,d)$ of origin-destination points, there are two main algorithmic challenges:
(i)  $TDRP(o,d,t_o)$ concerns the computation of a minimum travel-time $od$-path for a given departure-time $t_o$, i.e., the \emph{evaluation} of the minimum-travel-time function $D[o,d](t_o)$ from $o$ to $d$;
(ii) $TDRP(o,d)$ concerns the construction and succinct representation of the entire function $D[o,d]$, for \emph{all} possible departure-times (e.g., for future instantaneous evaluations).
A crucial property that makes $TDRP(o,d,t_o)$ tractable is the FIFO property, 
according to which delaying the departure-time from the tail of an arc cannot possibly cause an earlier arrival at its head (i.e., the arcs behave as FIFO queues).
For FIFO-abiding instances, a time-dependent variant of Dijkstra's algorithm ($\alg{TDD}$) running in quasi-linear time is known \cite{1969-Dreyfus,1990-Orda-Rom}.
Without the FIFO property the problem can become extremely hard, depending on the adopted waiting policy at the vertices of the network \cite{1990-Orda-Rom}.
As for $TDRP(o,d)$, this is known to be hard even when the FIFO property holds \cite{2014-Foschini-Hershberger-Suri}.
Fortunately, if (good) upper-approximations $\overline{\Delta}[o,d]$ of the minimum-travel-time functions $D[o,d]$ are an option, then there exist polynomial-time and space-efficient \emph{one-to-one} \cite{2012-Dehne-Omran-Sack,2014-Foschini-Hershberger-Suri,2014-Omran-Sack}, or \emph{one-to-all}
\cite{%
	2016-Kontogiannis-Michalopoulos-Papastavrou-Paraskevopoulos-Wagner-Zaroliagis,%
	2016-Kontogiannis-Wagner-Zaroliagis,%
	2014-Kontogiannis-Zaroliagis}
approximation algorithms.
	
As a quality measure, independent of the query at hand, the \term{relative error} is typically used, i.e., the \term{maximum absolute error} (MAE) divided by the optimal travel-time; the MAE is the worst-case difference of an optimal travel-time from the proposed (path's) travel-time.

Several speedup heuristics, with remarkable success in road networks possessing scalar arc-cost metrics, have been extended to the case of TDRP. Some of them \cite{2012-Delling-Nannicini,2007-Delling-Wagner,2012-Nannicini-Delling-Liberti-Schultes} are based on (scalar) lower bounds of travel-time functions (e.g., free-flow travel-times) to orient the search for a good route.
$\alg{TDCALT}$ \cite{2012-Delling-Nannicini} yields reasonable query-response times for $TDRP(o,d,t_o)$, and
$\alg{TDSHARC}$ \cite{2011-Delling_TDSHARC} provides in reasonable time solutions to $TDRP(o,d)$, even for continental-size networks.
$\alg{TDCRP}$ \cite{2016-Baum-Dibbelt-Pajor-Wagner} is currently one of the most successful speedup heuristics, whose main feature is \emph{customizability}, i.e., almost real-time adaptation to changes in the arc-cost metric.
$\alg{TCH}$ \cite{2013-Batz-Geisberger-Sanders-Vetter} also achieves remarkable query times, both for $TDRP(o,d,t_o)$ and for $TDRP(o,d)$, even for continental-size networks. All the above mentioned heuristics only compute (estimations of) erarliest-arrival-times, excluding the overhead for constructing the corresponding connecting path. 
The only heuristics that also account the path construction in their query-times are provided in \cite{2016-Strasser}, with quite competitive performances.

In parallel to speedup heuristics, there has been a recent trend to provide oracles for TDRP, with \emph{provable} theoretical performance and approximation guarantees \cite{2016-Kontogiannis-Wagner-Zaroliagis,2014-Kontogiannis-Zaroliagis}, which have been experimentally evaluated on real-world instances \cite{%
	2015-Kontogiannis-Michalopoulos-Papastavrou-Paraskevopoulos-Wagner-Zaroliagis,%
	2016-Kontogiannis-Michalopoulos-Papastavrou-Paraskevopoulos-Wagner-Zaroliagis}.
The most successful one, $\alg{FLAT}$ \cite{2016-Kontogiannis-Michalopoulos-Papastavrou-Paraskevopoulos-Wagner-Zaroliagis,2016-Kontogiannis-Wagner-Zaroliagis}, demonstrated in practice noticeable query times and relative errors, much better than the theoretical guarantees, thus being competitive to the aforementioned speedup heuristics, justifying further research on providing even better oracles for TDRP, for the additional reason that oracles also ensure scalability.

\vspace*{-15pt}
\subparagraph*{Contributions and Outline.}
We present, engineer and experimentally evaluate $\alg{CFLAT}$ (Section~\ref{section:CFLAT}), a novel \emph{landmark-based} oracle for TDRP whose objective is to tackle the main burden of such oracles, the large preprocessing requirements, without compromising either the preprocessing scalability, the competitiveness of query-response times, or the approximation guarantees.
To our knowledge, $\alg{CFLAT}$ is the first oracle for time-dependent networks that preprocesses only time-evolving \emph{combinatorial structures}: it maintains a carefully selected collection of time-stamped min-cost-path trees 
which can assure good approximation guarantees while minimizing the required space.
Computing (and storing) less during preprocessing, unavoidably leads to more demanding work per query in real-time.
Nevertheless, our novel query algorithm ($\alg{CFCA}$) manages to achieve better query times and significantly improved practical performance compared to previous oracles, despite the fact that it actually computes a connecting path, and not just an estimation of a good upper bound on the minimum travel-time for the query at hand, as is done by almost all other oracles and speedup techniques for TDRP.
Our specific contributions are threefold:
\textbf{(i)}
We propose $\alg{CTRAP}$ (Section~\ref{section:CTRAP}), a novel \emph{approximation method} which stores only min-cost-path trees for carefully selected landmark vertices and sampled departure-times.
Apart from the obvious economy of space due to omitting certain attributes (travel-time values), the novelty of this approach is that it exploits the fact that there are significantly fewer changes in the combinatorial structure, than in the functional description of the optimal solution (earliest arrival-times at a destination).
Moreover, we avoid multiple copies of the same preprocessed information, by organizing the destinations from a landmark into groups of (roughly) equidistant vertices, for which  the common departure-times sequence is stored only once.
We then proceed with the \emph{landmark selection policies} (Section \ref{section:experimental-evaluation}) considered by $\alg{CFLAT}$. Apart from the most successful ones in \cite{2016-Kontogiannis-Michalopoulos-Papastavrou-Paraskevopoulos-Wagner-Zaroliagis}, we also consider new policies based on the betweeness-centrality measure. Due to the significant reduction in space requirements, we are in a position to select much larger landmark sets, which allows us to showcase the full scalability of $\alg{CFLAT}$ in trading smoothly preprocessing requirements with query response times and approximation guarantees.
\textbf{(ii)} 
We propose $\alg{CFCA(N)}$ (Section~\ref{section:CFCA}), a novel \emph{query algorithm} that exploits the preprocessed information of $\alg{CFLAT}$:
For a query $(o,d,t_o)$, it starts by growing a $\alg{TDD}$ ball from $o$ at time $t_o$, until the $N$ closest landmarks are settled.
It then marks a small subset of relevant arcs, using the $N$ settled landmarks as ``attractors'' that orient the discovery of certain paths from $d$ back to $o$. This is reminiscent of the $\alg{ARCFLAGS}$ algorithm for static metrics \cite{2009-Hilger-Koehler-Moehring-Schilling}, but the choice of the relevant arcs is done ``on the fly'', since this information is also time-dependent.
In the final step, it continues growing the initial $\alg{TDD}$ ball, but only  within the subgraph of marked arcs, until the destination $d$ is settled within this subgraph. $\alg{CFCA(N)}$ achieves the same theoretical approximation
guarantee with the query algorithm $\alg{FCA(N)}$ of $\alg{FLAT}$, but in practice it is much better than $\alg{FCA(N)}$.
\textbf{(iii)} 
We conduct a thorough \emph{experimental evaluation} of $\alg{CFLAT}$ (Section~\ref{section:experimental-evaluation}), on two well established real-world instances, the urban area of Berlin and the national road network of Germany. Our findings are perceptible.
For Berlin, the preprocessing requirements are less than $3.306$sec and $2.521$MB ($0.69$MB compressed) per landmark.
Thus, if space is our primary concern, we can preprocess $250$ random landmarks in less than $14$min, consuming $0.7$GB ($0.17$GB compressed) space, whereas the query performance (average query time and relative error) varies from
$0.565$msec and $2.418\%$ (for $N=1$), to
$3.330$msec and $0.136\%$ (for $N=6$).
With $16$K landmarks the query performance varies from
$0.076$msec and $0.192\%$ (for $N=1$),
to
$0.226$msec and $0.022\%$  (for $N=6$).
As for Germany, the preprocessing requirements are $29.322$sec and $26.8$MB ($8.07$MB compressed) per landmark. 
\remove{
For $3$K landmarks, the query performance varies
from $0.733$msec and $0.911$\%  (for $N=1$),
to $4.787$msec and $0.029$\% (for $N=6$).
}
For $4K$ landmarks, we achieve a query performance varying from
$0.683$msec and $0.831$\% (for $N=1$),
to
$4.104$msec and $0.031$\% (for $N=6$).

\section{The $\alg{CFLAT}$ Oracle}
\label{section:CFLAT}

A landmark-based oracle selects a set $L\subseteq V$ of \term{landmarks} and preprocesses travel-time information (\term{summaries}) between them and all (or some) reachable destinations. A query algorithm exploits these summaries for responding to earliest-arrival-time queries $(o,d,t_o)$, from an origin $o$ and departure-time $t_o$ to a destination $d$, in time that is \emph{provably} efficient (e.g., sublinear in the size of the instance). The oracle is also accompanied with a \emph{theoretically proved} approximation guarantee (a.k.a. stretch) for the quality of the recommended routes.

In Section~\ref{section:CFLAT-description} we present our novel oracle, $\alg{CFLAT}$. Before doing that, we recap in Section~\ref{section:FLAT-recap} $\alg{FLAT}$, an oracle upon which $\alg{CFLAT}$ builds and achieves remarkable improvements.

\subsection{Recap of $\alg{FLAT}$}
\label{section:FLAT-recap}

\begin{wrapfigure}{r}{0.5\textwidth}
\vspace{-40pt}
\centerline{\includegraphics[width=0.5\textwidth]{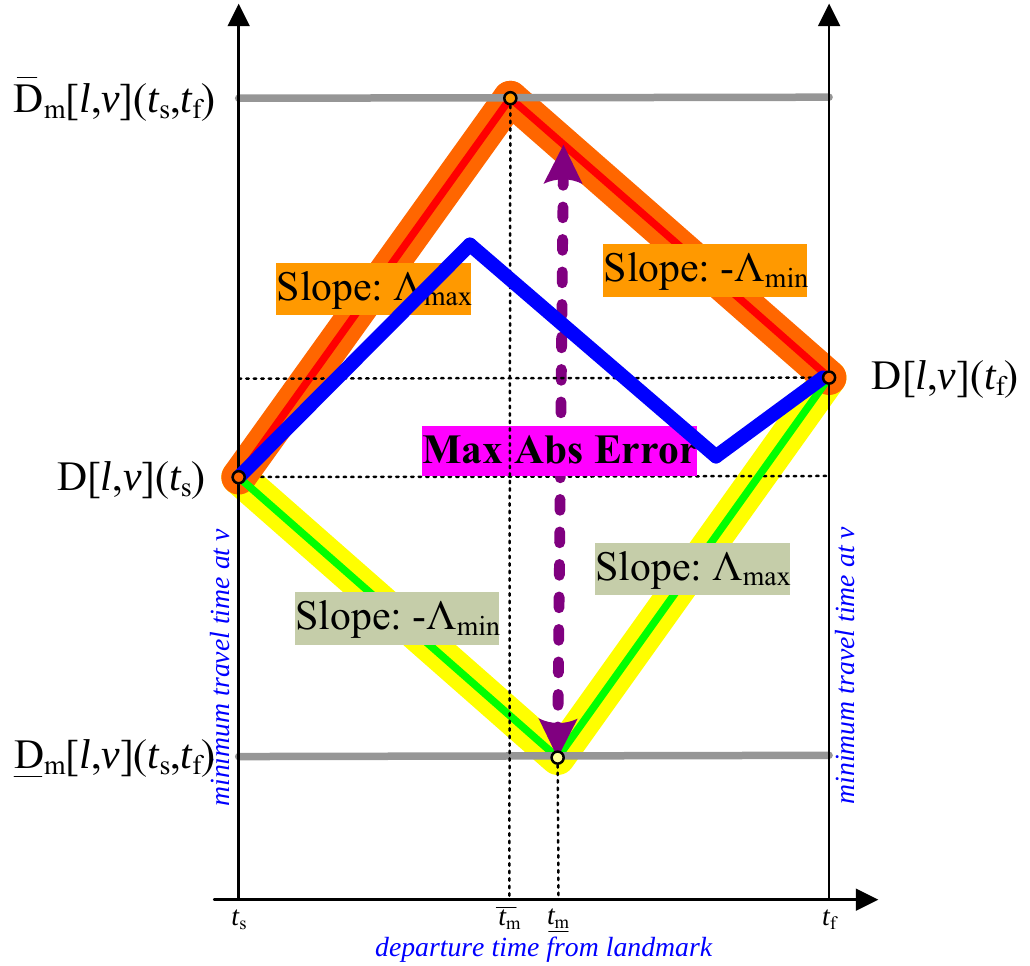}}
\caption{\label{fig:trapezoidal-approximation}
	Upper-approximation $\overline{\delta}_k[\ell,v]$ (thick-orange) and lower-approximation $\underline{\delta}_k[\ell,v]$  (thick-green) of $D[\ell,v]$ (blue), within $[t_s, t_f)$.
}
\end{wrapfigure}

$\alg{FLAT}$ is, to date, the most successful oracle for TDRP in road networks, and was originally presented and analyzed in
\cite{2016-Kontogiannis-Wagner-Zaroliagis}.
A variant of $\alg{FLAT}$ was implemented and experimentally evaluated in
\cite{2016-Kontogiannis-Michalopoulos-Papastavrou-Paraskevopoulos-Wagner-Zaroliagis}.
In this work, we consider (and refer to as $\alg{FLAT}$) to that variant.
Its main building block is the $\alg{TRAP}$ approximation method: Given a landmark $\ell$, the period $[0,T)$ is split into intervals of an (arbitrarily chosen) length $3,200$sec. The endpoints of these intervals are used as sampled departure-times. The corresponding min-cost-path trees rooted at $\ell$ are computed, producing travel-time values for all reachable destinations $v$.
For each interval $[t_s,t_f)$, an upper-approximating function $\overline{\delta}$ is considered, which is the lower-envelope of a line of max slope ($\La_{\max}$) passing via $\langle t_s,D[\ell,v](t_s)\rangle$ and a line of min slope ($-\La_{\min}$) passing via $\langle t_f,D[\ell,v](t_f)\rangle$ (cf. Figure~\ref{fig:trapezoidal-approximation}).
Observe that $\overline{\delta}$ considers an \term{intermediate breakpoint} $\langle\overline{t}_m,\overline{D}_m\rangle$, the intersection of the two lines, which is \emph{not} the outcome of an actual sampling. This intermediate breakpoint is only stored when $v$ becomes deactivated (i.e., within this interval there is no need for further sample points, see next paragraph).
A similar lower-approximating function $\underline{\delta}$ is considered, which is the upper-envelope of a min-slope line passing via $\langle t_s,D[\ell,v](t_s)\rangle$ and a max-slope line passing via $\langle t_f,D[\ell,v](t_f)\rangle$.

A closed-form expression of the worst-case error (\term{maximum absolute error} -- MAE) is used to determine whether $\overline{\delta}$ is a sufficient upper-approximation of $D[\ell,v]$ within $[t_s,t_f)$, given a required approximation guarantee $\varepsilon > 0$.
If this is the case, $v$ becomes \term{deactivated} for this subinterval, meaning that no more sampled trees will be of interest for $v$ within it.
$\alg{TRAP}$ continues by choosing finer sampling intervals, first of length $1,600$sec, then $800$sec, $400$sec, etc., computing min-cost-path trees only for the new departure-time samples in each round, until eventually there is no active destination for any of subintervals of the currently chosen length.
The concatenation of all the upper-approximations for the smallest active subintervals of $v$ is considered by $\alg{TRAP}$ as the required $(1+\varepsilon)$-upper-approximation $\overline{\Delta}[\ell,v]$ (called a \term{travel-time summary}) of $D[\ell,v]$ within $[0,T)$. $\overline{\Delta}[\ell,v]$ is stored as a sequence of pairs of breakpoints, i.e., (departure-time,travel-time) pairs, in increasing order w.r.t. departure-times.
During the preprocessing, $\alg{FLAT}$ calls $\alg{TRAP}$ to produce travel-time summaries, from
a carefully selected set of landmark vertices towards all reachable destinations.

Upon a query $(o,d,t_o)$ $\alg{FLAT}$ calls $\alg{FCA(N)}$\footnote{%
	In \cite{2016-Kontogiannis-Michalopoulos-Papastavrou-Paraskevopoulos-Wagner-Zaroliagis} it was called $\alg{FCA}^+$, with a fixed number $N=6$ of landmarks to settle.
}, 
a query algorithm which grows a $\alg{TDD}$ ball from $o$ with departure-time $t_o$, until either $d$ or the first $N$ landmarks are settled.
It then returns either the exact route (when $d$ is settled), or the best-of-$N$ (w.r.t. the \emph{theoretical guarantees}) $od$-path passing via one of the $N$ settled landmarks and being completed (from $\ell$ to $d$) by exploiting the preprocessed summaries for $d$.
Since $\alg{FCA(N)}$ does not need all summaries to be concurrently available in memory, the preprocessed data blocks representing travel-time summaries of $\alg{FLAT}$ were compressed, and only summaries of the landmarks required per query were decompressed on the fly. The $\alg{zlib}$ library was used for this purpose, leading to a reduction of $10\%$ in the required space.
More details on $\alg{FLAT}$ are provided in \cite{2016-Kontogiannis-Michalopoulos-Papastavrou-Paraskevopoulos-Wagner-Zaroliagis,2016-Kontogiannis-Wagner-Zaroliagis}.

\vspace*{-10pt}
\subsection{Description of $\alg{CFLAT}$}
\label{section:CFLAT-description}

We now present $\alg{CFLAT}$, which can be considered as the combinatorial analogue of $\alg{FLAT}$. At a high level, $\alg{CFLAT}$ works as follows. In a preprocessing phase, it constructs and compactly stores min-cost-path trees at carefully sampled departure-times, rooted at each landmark $\ell \in L$. A query $(o,d,t_o)$ is answered by first growing a $\alg{TDD}$ ball from $o$ at time $t_o$, until either $d$ or a small number of landmarks are settled. In the latter case, starting from $d$, a suitably small subgraph is constructed (consisting of certain paths going from $d$ back to $o$, using the settled landmarks as ``attractors''), until a settled vertex of the initial $\alg{TDD}$ ball is reached. Then, a continuation of growing the initial $\alg{TDD}$ ball on the resulted small subgraph returns an $od$ path that turns out to approximate very well the optimal $od$ path.

\vspace*{-10pt}
\subsubsection{The Approximation Method $\alg{CTRAP}$ and $\alg{CFLAT}$ Preprocessing}
\label{section:CTRAP}

$\alg{CTRAP}$ computes and stores only min-cost-path trees at carefully sampled departure-times, rather than actual breakpoints of the corresponding minimum-travel-time functions. The algorithm's pseudocode is provided in the appendix (cf. Section~\ref{section:CTRAP-pseudocode}).
We present here only a sketch of the main steps
as well as the key new insights, compared to $\alg{TRAP}$. $\alg{CFLAT}$ preprocessing consists simply in calling $\alg{CTRAP}(\ell,\varepsilon)$ for each landmark $\ell \in L$.
\\[2pt]
\begin{footnotesize}
\begin{tabular}{|p{0.5cm}p{12.75cm}|}
\hline
\rowcolor{gray!25}
\multicolumn{2}{|l|}{\textbf{procedure} $\alg{CTRAP}(\ell,\varepsilon)$}
\\ \hline \hline
\rowcolor{gray!5}
\multicolumn{2}{|p{0.975\textwidth}|}{\textbf{STEP 1:} Keep sampling finer departure-times from $[0,T)$, as in $\alg{TRAP}$, until all destinations achieve relative error less than $\varepsilon$ and become inactive.}
\\
\rowcolor{gray!5}
\textbf{1.1:}		& Store (pruned at inactive nodes) min-cost-path trees from $\ell$, for all departure-times.
\\
\rowcolor{gray!5}
\textbf{1.2:} 	& Omit intermediate breakpoints.
\\[5pt]
\rowcolor{gray!5}
\multicolumn{2}{|l|}{\textbf{STEP 2:} Merge consecutive breakpoints with identical predecessors.}
\\[5pt]
\rowcolor{gray!5}
\multicolumn{2}{|l|}{\textbf{STEP 3:} Avoid multiple copies of common departure-time sequences.}
\\ \hline
\end{tabular}
\end{footnotesize}
\\[2pt]
When executed from a landmark $\ell$, $\alg{CTRAP}$ works as follows:
Step 1 resembles $\alg{TRAP}$, the only difference being that $\alg{CTRAP}$ keeps only the immediate predecessors (parents) per active destination $v$ in the sampled min-cost-path trees. In particular, a pair of sequences is created, $PRED[\ell,v]$ for predecessors and $DEP[\ell,v]$ for the corresponding sampled departure-times, per landmark-destination pair $(\ell,v)\in L\times V$.
Step 2 cleans up each pair of sequences, by merging consecutive breakpoints for which the predecessor is the same.
Step 3 organizes the destinations from a landmark $\ell$ into groups with the same departure-times sequence, so that multiple copies of the same sequence are avoided.
In the rest of this section, we describe in more detail the key new insights and algorithmic steps of $\alg{CTRAP}$, compared to $\alg{TRAP}$ \cite{2016-Kontogiannis-Michalopoulos-Papastavrou-Paraskevopoulos-Wagner-Zaroliagis,2016-Kontogiannis-Wagner-Zaroliagis}.

\vspace*{-15pt}
\subparagraph*{Store min-cost-path trees.}

For each leg of $\overline{\Delta}[\ell,v]$, we store pairs $\langle t_{\ell}, PRED[\ell,v](t_{\ell}) \rangle$ of departure-times $t_{\ell}$ from $\ell$ and the predecessor of $v$ in the corresponding min-cost-path tree rooted at $(\ell,t_{\ell})$, omitting the actual min-travel-time values $D[\ell,v](t_{\ell})$. This modification makes the oracle aware only of the min-cost-path-tree structures created during the repeated sampling procedure.	
Additionally, rather than storing repeatedly the IDs of predecessors, which would be space consuming in networks with millions of vertices, we only store the position of the corresponding arc in the list of incoming arcs to a vertex $v$. Since the maximum in-degree in the road instances we have at our disposal is at most $7$, we only need to consume $1$ byte per storage for a predecessor.
We could even consume $3$ bits per predecessor, which could then be packed into only two bytes containing also the corresponding departure-time value (by an appropriate discretization of the departure-time values).
We prefer \emph{not} to combine predecessors with departure-times in the same bit string, because we shall exploit later the extensive repetition of identical sequences of departure-times, which nevertheless would be lost for strings also containing the predecessors.
It was observed in both benchmark instances that about one half of all possible destinations per landmark $\ell$ appear to have a \emph{unique} predecessor throughout the entire period of departure-times, $[0,T)$.
For them we store their unique predecessor only once. For the remaining destinations though, even with only two possible predecessors, we have to store the entire sequence of predecessor-changes.

\vspace*{-15pt}
\subparagraph*{Omit intermediate breakpoints.}
\label{section:omit-intermediate-breakpoints}

$\alg{TRAP}$  computes, and explicitly stores, intermediate breakpoints $(\overline{t}_m , \overline{D}_m)$ between consecutive sampled breakpoints of $D[\ell,v]$, as the intersection points of the two legs involved in the definition of $\overline{\delta}[\ell,v](t)$ (cf. Figure~\ref{fig:trapezoidal-approximation}), for each pair $(\ell,v)$ and those intervals where the MAE is sufficiently small and $v$ becomes deactivated.
In $\alg{CTRAP}$ we choose \emph{not} to keep these intermediate breakpoints and restrict the preprocessed information only to the actual samples. We let the query algorithm deal with the missing information, whenever needed.
This way we avoid storing approximately $10$M (for Berlin) and $100$M (for Germany) of intermediate breakpoints per landmark.

\vspace*{-15pt}
\subparagraph*{Merge sequences of breakpoints with identical predecessors.}

$\alg{CTRAP}$'s next algorithmic intervention is based on the observation that the vast majority of all destinations appear to have on average $2$ alternating predecessors throughout the entire period $[0,T)$.
To save space, we choose to merge \emph{consecutive} sampled breakpoints for $v$ of the form
$\langle t_{\ell}, x = PRED[\ell,v](t_{\ell})\rangle$ and
$\langle t'_{\ell}, x = PRED[\ell,v](t'_{\ell})\rangle$, i.e.,
possessing the same predecessor.
This leads to a reduction in the number of breakpoints to store, but also has a negative influence on the similarities of the departure-times sequences, and thus on the repetitions that we could avoid (see next heuristic).
However, there is still positive gain by applying both this heuristic and that for avoiding multiple copies of departure-times sequences.

\vspace*{-15pt}
\subparagraph*{Avoid multiple copies of common departure-time sequences.}
\label{section:avoid-multiple-copies-of-DEP-sequence}

$\alg{CTRAP}$'s next key insight is based on the fact that it is a repeated-sampling method which probes (at common departure-times for all destinations) min-cost-path trees from a landmark $\ell$, starting from a coarse-grained sampling towards more fine-grained samples of the entire period $[0,T)$, until the MAE guarantee is satisfied for all reachable destinations from $\ell$.
A destination $v$ may not care for all these departure-times, because the value of MAE may be satisfied at an early stage for it. This indeed depends on the actual minimum travel-time $\min\{ D[\ell,v](t_s),D[\ell,v](t_f) \}$ at the endpoints of each given subinterval $[t_s,t_f)$.
For each landmark-destination pair $(\ell,v)$, we store the sequences $DEP[\ell,v]$ of necessary departure-times and $PRED[\ell,v]$ of the corresponding predecessors.
The crucial observation is that destinations which are (roughly) at the same distance from $\ell$ are anticipated to have the same sequence of sampled departure-times, possibly differing only in their sequences of predecessors.
It is clearly a waste of space to store two identical sequences $DEP[\ell,v] = DEP[\ell,u]$ more than once, even if the corresponding sequences of predecessors differ. Thus, we store each departure-times sequence as soon as it first appears for some destination $v$, and consider $v$ as the \term{representative} of all other destinations $u$ for which $DEP[\ell,u] = DEP[\ell,v]$. For each non-representative destination $u$, we store $PRED[\ell,u]$ and the corresponding representative $v$.
	%
Our next challenge is to efficiently compare departure-times sequences. To avoid a potential blow-up of the preprocessing time, we do not compare them point-by-point.
Instead, we assign to every sampled departure-time $t_{\ell}$ two $\iuar$\footnote{$\iuar$ = independently and uniformly at random, without repetitions.} chosen floating-point numbers $w_1(t_{\ell}), w_2(t_{\ell})$ from the interval $[1.0, 100.0]$. Each destination $u$ adds the two values $w_1(t_{\ell})\cdot t_{\ell}$ and $w_2(t_{\ell})\cdot t_{\ell}$ to its own hash keys, i.e., $H_1[u] = H_1[u] + w_1(t_{\ell})\cdot t_{\ell}$ and $H_2[u] = H_2[u] + w_2(t_{\ell})\cdot t_{\ell}$, \emph{only when} $t_{\ell}$ is indeed a necessary sample for $u$. Otherwise, the hash keys of $u$ remain intact.
 At the end of the sampling process, we sort lexicographically the hash pairs of all destinations, in order to discover families of common departure-times sequences. We deduce that two destinations possess the same sequence when both their hash pairs match, in which case we verify this allegation by comparing them point by point.
We observed that, for both benchmark instances, $80$\% of all destinations with at least two predecessors can be represented w.r.t departure-times by the remaining $20$\% of (representative) destinations.

\vspace*{-15pt}
\subparagraph*{Indexing preprocessed information.}

For retrieving efficiently the summaries from a landmark $\ell$ to each destination $v$, we maintain a vector of pointers per landmark, one pointer per destination, providing the address for the starting location of the summary for $v$. The pointers are in ascending order of vertex ID. The lookup time is $\Order{1}$ and the required space for this indexing scheme is $\Order{n\cdot |L|}$ additional bytes, where $L$ is the chosen landmark set.

\vspace*{-15pt}
\subparagraph*{Speeding up preprocessing time.}

Handling only min-cost-path trees also has a collateral effect of speeding up the required preprocessing time.
The reason for this is that we do not compute explicitly, each and every time that we sample travel-time values from $\ell$,
the exact shapes of the corresponding minimum-travel-time functions per destination.
The travel-time summaries provided by $\alg{FLAT}$ were created based on this explicit computation of all the earliest-arrival \emph{functions} per destination $v$, from each landmark $\ell$.
In contrast, the min-cost-path summaries of $\alg{CFLAT}$ are created without having to compute earliest-arrival functions. This leads to a reduction in the preprocessing time of more than $60\%$.

\vspace*{-12pt}
\subsubsection{The Query Algorithm $\alg{CFCA(N)}$}
\label{section:CFCA}
\vspace*{-6pt}

$\alg{CFCA(N)}$ is based on $\alg{FCA(N)}$ \cite{2016-Kontogiannis-Michalopoulos-Papastavrou-Paraskevopoulos-Wagner-Zaroliagis}, but is fundamentally different from it in the sense that it exploits min-cost-path trees, and also considers the $od$-path construction as part of it, which was not the case for $\alg{FCA(N)}$, and indeed for most of the query algorithms in the literature.
$N$ indicates the number of landmarks to be settled by $\alg{CFCA(N)}$ around the origin $o$.
The pseudocode of the algorithm is presented in the next paragraph. $\alg{CFCA(N)}$ works as follows. In case that the destination $d$ is already settled in Step 1, the resulting (exact) $od$-path can be computed by backtracking towards the origin, following the pointers to all predecessors.
Otherwise, we proceed as follows.
For each settled landmark $\ell$, we have an optimal $o\ell$-path guaranteeing arrival-time $t_\ell = t_o + D[o,\ell](t_o)$ at $\ell$.
Since we do not have at our disposal travel-time values from $\ell$ towards $d$, or any other vertex, we are not able to compare $\ell v$-paths based on their (approximate) lengths.
On the other hand, for the given departure-times $t_{\ell}$ and any vertex $v$, we can tell the predecessor(s) of $v$ in the (at most two per landmark) most relevant min-cost-path trees, the ones at the consecutive sampled departure-times $t^{-}_{\ell}$ and $t^{+}_{\ell}$ of each $DEP[\ell,v]$ for which it holds that $t_{\ell}\in [t^{-}_{\ell}, t^{+}_{\ell})$.
\\[2pt]
\begin{footnotesize}
\begin{tabular}{|p{0.5cm}p{12.75cm}|}
\hline
\rowcolor{gray!25}
\multicolumn{2}{|l|}{\textbf{procedure} $\alg{CFCA(N)}$}
\\ \hline \hline
\rowcolor{gray!5}
\multicolumn{2}{|p{0.975\textwidth}|}{\textbf{STEP 1:} A $\alg{TDD}$ ball is grown from $(o,t_o)$, until $N$ landmarks are settled.}
\\
\rowcolor{gray!5}
\textbf{1.1:}
& \IF $d$ is already settled \THEN \RETURN optimal solution.
\\
\rowcolor{gray!5}
\textbf{1.2:}
& For each settled landmark $\ell$, $t_{\ell} = t_o + D[o,\ell](t_o)$.
\\[5pt]
\rowcolor{gray!5}
\multicolumn{2}{|p{0.975\textwidth}|}{\textbf{STEP 2:} An appropriate subgraph is recursively created from $d$.}
\\
\rowcolor{gray!5}
\textbf{2.1:}
& $Q = \{~ d ~\}$ \hfill\COMMENT{$Q$ is a FIFO queue}
\\
\rowcolor{gray!5}
\textbf{2.2:}
&
\WHILE $\neg Q.Empty()$ \DO:
\\
\rowcolor{gray!5}
\textbf{2.3:}
& \TAB\IF $v = Q.Pop()$ is not explored from STEP 1's $\alg{TDD}$ ball \THEN:
\\
\rowcolor{gray!5}
\textbf{2.4:}
&\TAB\TAB \FOR each settled landmark $\ell$ of STEP 1 \DO:
\\
\rowcolor{gray!5}
\textbf{2.5:}
& \TAB\TAB\TAB
Mark the arcs $\langle PRED[\ell,v](t^{-}_{\ell}),v\rangle$ and $\langle PRED[\ell,v](t^{+}_{\ell}),v\rangle$ leading to $v$, where $[t^{-}_{\ell},t^{+}_{\ell})$ is the unique interval in $DEP[\ell,v]$ containing $t_{\ell}$.
\\
\rowcolor{gray!5}
\textbf{2.6:}
& \TAB\TAB\TAB
$Q.Push(PRED[\ell,v](t^{-}_{\ell}))$;
$Q.Push(PRED[\ell,v](t^{+}_{\ell}))$
\\
\rowcolor{gray!5}
\textbf{2.7:}
& \TAB\TAB\END\FOR
\\
\rowcolor{gray!5}
\textbf{2.8:}
& \END\WHILE
\\[5pt]
\rowcolor{gray!5}
\multicolumn{2}{|p{0.975\textwidth}|}{\textbf{STEP 3:} \RETURN optimal od-path in the induced subgraph by ($\alg{TDD}$ ball of) STEP 1 and STEP 2.}
\\ \hline
\end{tabular}
\end{footnotesize}
\\[2pt]
$\alg{CFCA(N)}$ marks (per settled landmark $\ell$) the connecting arcs from these most relevant predecessor(s) $PRED[\ell,v](^{-}_{\ell})$ and $PRED[\ell,v](^{+}_{\ell})$, towards $v$.
All these discovered predecessors w.r.t. the $N$ settled landmarks are inserted (if not already there) in a FIFO queue, which was initialized with $d$, so that, upon their extraction from the queue, they can provide in turn their own  predecessors, etc.
The recursive search for predecessors stops as soon as a vertex $x$ in the explored area of the initial $\alg{TDD}$ ball of Step 1 is reached. $\alg{CFCA}$ marks then also the arcs of the corresponding short (not necessarily the shortest though, since $x$ is explored but not necessarily settled) $ox$-path. This way we are guaranteed that in the subgraph of marked arcs there is already an $od$-path which has been oriented by $(\ell,t_{\ell})$ and passes via $x$. Step 2 of $\alg{CFCA(N)}$ terminates when the FIFO queue becomes empty, i.e., we no longer have to process intermediate vertices which are unexplored by Step 1.
The actual path construction takes place in Step 3, which considers the subgraph induced by the marked arcs and continues growing the $\alg{TDD}$ ball from $(o,t_o)$ within this subgraph.
This path construction indeed leads to significantly smaller relative errors, since the resulting $od$-path is not only the best \emph{prediction} among a given set of $N$ paths induced by the $N$ settled landmarks (as in $\alg{FLAT}$), but actually the \emph{optimal} $od$-path within the induced sugbgraph.

The worst-case approximation guarantee of $\alg{CFCA}(1)$ is $(1+\eps+\psi)$ (identical to that of $\alg{FCA}$ \cite{2016-Kontogiannis-Wagner-Zaroliagis}), where $\eps$ is $\alg{CTRAP}$'s approximation guarantee and $\psi$ is a constant depending on $\eps$ and the travel-time metric (but not on the size) of the network. Note that we could theoretically improve the stretch of $\alg{CFCA(N)}$ to $(1+\s)$, for any constant $\s > \eps$,
and get a PTAS, by using in Step 1 the $\alg{RQA}$ algorithm \cite{2016-Kontogiannis-Wagner-Zaroliagis}. We choose \emph{not} to do so, because our previous experimental evaluation with $\alg{FLAT}$
\cite{2016-Kontogiannis-Michalopoulos-Papastavrou-Paraskevopoulos-Wagner-Zaroliagis} has shown that $\alg{FCA(N)}$ in practice dominates $\alg{RQA}$.

\section{Experimental Evaluation}
\label{section:experimental-evaluation}

\vspace*{-5pt}
\subparagraph*{Experimental Setup and Goal.}

Our algorithms were implemented in C++ (GNU GCC version 5.4.0) and Ubuntu Linux (16.04 LTS).
All the experiments were conducted on a $6$-core Intel(R) Xeon(R) CPU E5-2643v3 3.40GHz machine,
with $128$GB of RAM. We used $12$ threads for the parallelization of the preprocessing phase.
$\alg{CFCA}$ was always executed on a single thread. For the sake of comparison, we used the same set of $50,000$ queries, $\iuar$
chosen from $V\times V\times [0,T)$ in each instance, for all possible landmark sets.
The PGL library \cite{2013-PGL} was used for graph representation and operations.
Two benchmark instances were used, the first concerning the city of Berlin, and the second the national  road network of Germany. More details on the availability of code and data are provided in Appendix~\ref{section:benchmark-instances-details}.

The main goal of our experimental evaluation was to investigate the scalability of $\alg{CFLAT}$: how smoothly does
it trade higher preprocessing requirements for better approximation guarantees and query-times. To demonstrate this,
we aim at showcasing the performance of $\alg{CFCA(N)}$ for several types and sizes of landmark sets.
We also choose to increase the typical size of the used landmark sets in our comparison of different
landmark selection policies.

\vspace*{-15pt}
\subparagraph*{Landmark Selection Policies.}

Although the preprocessing requirements are proportional to $|L|$ (number of landmarks),
they are essentially invariant of the landmark selection policy.
However, as previous experimental evaluation indicated \cite{2016-Kontogiannis-Michalopoulos-Papastavrou-Paraskevopoulos-Wagner-Zaroliagis},
the performance of the query algorithms has a strong dependence on the type of the landmarks.
A key observation was that the \emph{sparsity} of landmarks (not being too close to each other) as well as their \emph{importance}, are crucial parameters.
Therefore, in this work we insist in almost all cases (except for the \textsc{random} landmark sets which are used as baseline) on selecting the landmarks sparsely throughout the network.
As for their importance, when such information is available, we also consider the selection of landmarks at junctions of an important road segment (as in \cite{2016-Kontogiannis-Michalopoulos-Papastavrou-Paraskevopoulos-Wagner-Zaroliagis}). Finally, we consider a new measure of vertex significance,
the (approximate) betweeness-centrality measure.
In particular, we consider the following landmark selection policies:

%
\noindent$\diamond$~\underline{\textsc{random}} ($R$): $\iuar$ choice of landmarks.

%
\noindent$\diamond$~\underline{\textsc{sparse-random}} ($SR$): Incremental $\iuar$ choice of landmarks, where each chosen landmark excludes a free-flow neighborhood of vertices around it from future landmark selections.

%
\noindent$\diamond$~\underline{\textsc{important-random}} ($IR$): A variant of $R$ which moves each random landmark to its nearest important vertex within a free-flow neighborhood of size $100$. This policy is only applicable for the instance of Berlin which provides road-segment importance information.

%
\noindent$\diamond$~\underline{\textsc{sparse-kahip}} ($SK$): We use the $\alg{KaFFPa}$ algorithm of the \textsc{kahip} partitioning software (v1.00) \cite{2014-KAHIP}, setting the parameters so that there are many more boundary vertices than the required number of landmarks. The landmarks are incrementally and $\iuar$ chosen among the boundary vertices. Each landmark excludes a free-flow neighborhood from future selections.

%
\noindent$\diamond$~\underline{\textsc{kahip-cells}} ($KC$).
Starting with a \textsc{kahip} partition, one landmark per cell is incrementally and $\iuar$ chosen, excluding a free-flow neighborhood from future selections.

%
\noindent$\diamond$~\underline{\textsc{betweeness-centrality}} ($BC$):
Vertices are ordered in non-increasing \emph{approximate betweeness-centrality} (ABC) values~\cite{2007-Bader-Kintali-Madduri-Mihail}. Landmarks are selected incrementally according to ABC values, excluding a free-flow neighborhood from future selections.

%
\noindent$\diamond$~\underline{\textsc{kahip-betweeness}} ($KB$): For a \textsc{kahip} partition, incrementally choose as landmark the vertex with the highest ABC value in a cell, excluding a neighborhood from future selections.


We finally consider the following systematic naming of the landmark sets. Each set is encoded as $XY$, where $X\in\{ R, SR, IR, SK, KC, BC, KB \}$ determines the type of landmark set, and $Y\in\{ 250, 500, 1K, 2K, 3K, 4K, 8K, 16K, 32K \}$ determines its size.

\vspace*{-15pt}
\subparagraph*{Evaluation of $\alg{CFLAT}$ @ Berlin.}

For Berlin we have considered all types of landmarks. For each of them, we have used as  baseline the size $Y=4K$.
$\{ R, SR, IR, SK \}$ were considered also in \cite{2016-Kontogiannis-Michalopoulos-Papastavrou-Paraskevopoulos-Wagner-Zaroliagis} (but for smaller sizes), whereas $\{KC, BC, KB \}$ are new types.
Especially for $R$ we tried all possible values for $Y$, in order to showcase the scalability of $\alg{CFLAT}$ and its smooth trade-off of preprocessing requirements, query-times and stretch factors.
Concerning \emph{vertex-importance} (only available in Berlin), we considered as important those vertices which are incident to roads of category at most $3$.
As for \emph{sparsity}, we set the sizes of the excluded free-flow ball per selected landmark to $150$ vertices for $SR$, $100$ for $IR$, $50$ for $SK$, $20$ for $KC$, $150$ for $BC$, and $20$ for $KB$.
For \textsc{kahip} based landmark sets ($SK$, $KC$ and $KB$) we used the following parameters: The number of cells to partition the graph was set to $4,000$, having $13,256$ boundary vertices in total. For $SK$ we chose randomly $4,000$ boundary vertices as landmarks. For $KC$ and $KB$ we chose one landmark per cell.

\begin{figure}[htb!]
\centerline{%
	\includegraphics[height=4cm]{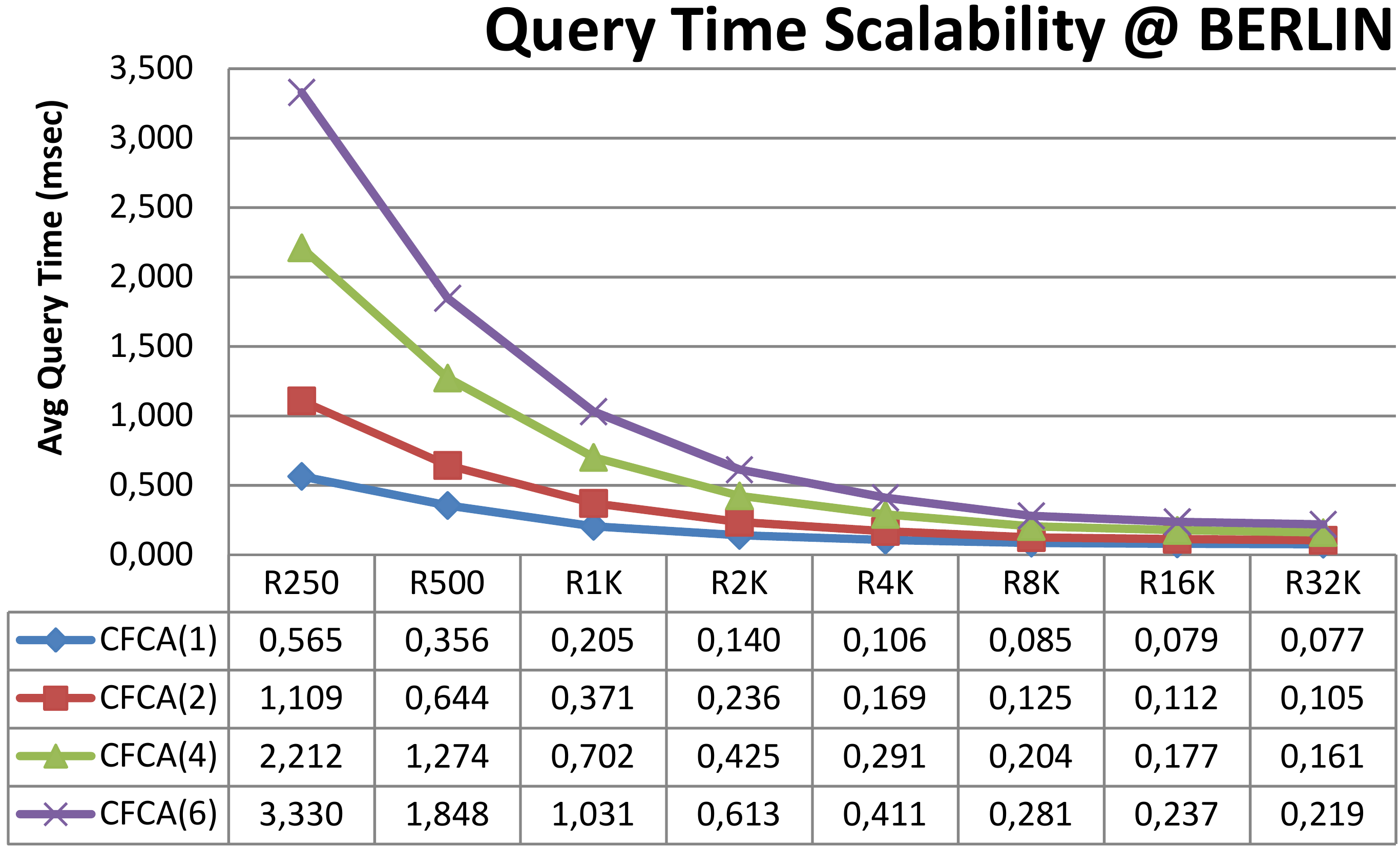}
	\includegraphics[height=4cm]{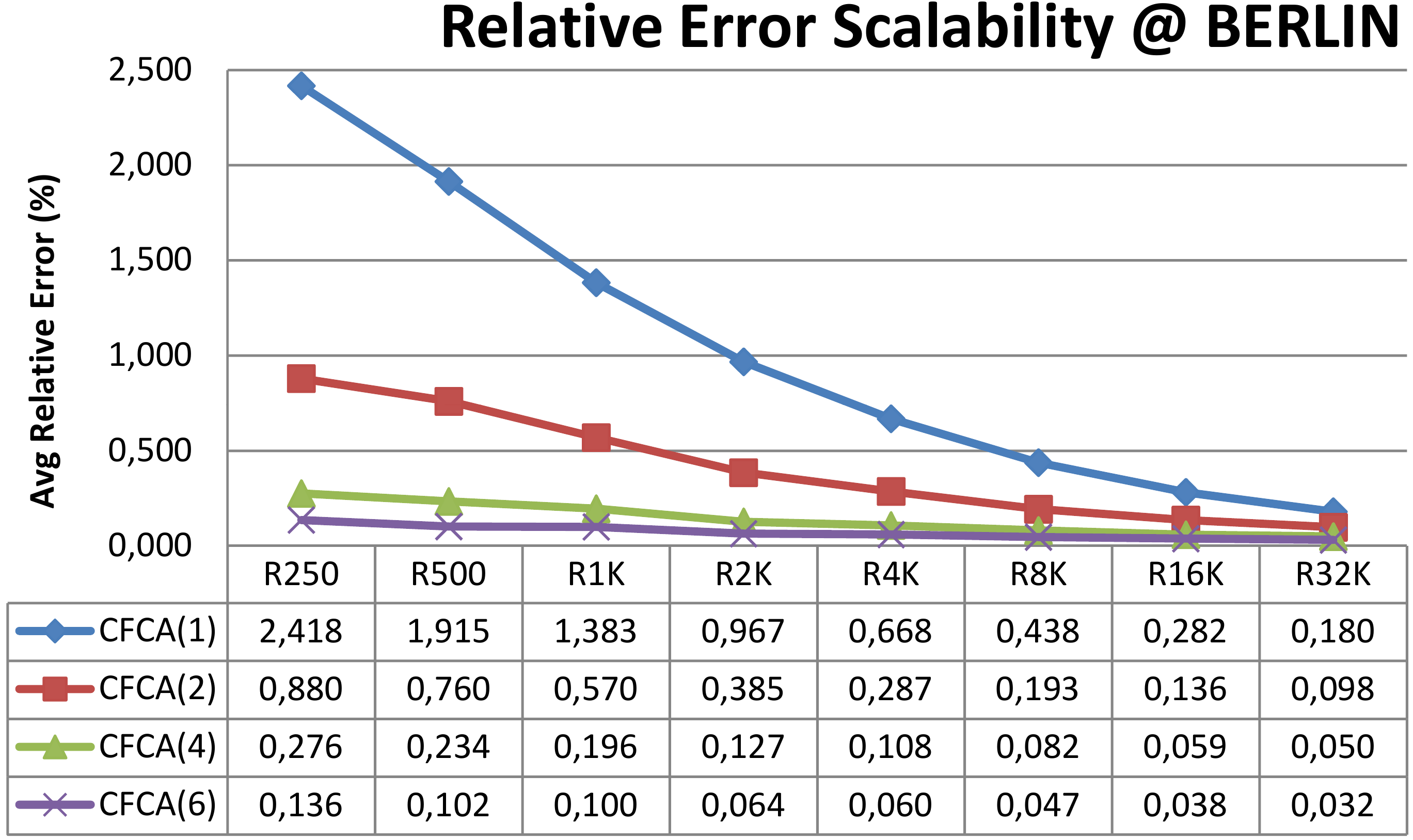}
}
\caption{\label{figure:BERLIN_CFLAT-SCALABILITY-RANDOM}
	Performance of $\alg{CFCA(N)}$ in Berlin,
	for random landmarks and $50,000$ \emph{random queries}.	
}
\end{figure}
We first conducted an experiment to test the scalability of $\alg{CFCA}$'s performance as a function of $N$ and the number of landmarks, always for R-type landmarks.
As is evident from Figure~\ref{figure:BERLIN_CFLAT-SCALABILITY-RANDOM}, the average errors decrease linearly and the query-times decrease quadratically, as we double the number of landmarks.
Additionally, notable ``quick-and-dirty'' answers are possible with only $250$ landmarks, which require total
space $0.7$GiB ($0.17$GiB after compression), cf. Figure~\ref{figure:BERLIN+GERMANY_PREPROCESSING-RANDOM}.
In particular, the query performance (average query time and relative error) varies from
$0.565$msec and $2.418$\% ($N=1$), to
$3.330$msec and $0.136$\% ($N=6$).
If query time is the main goal, then for BC8K+R8K, the query performance of $\alg{CFCA}$ varies from
$0.076$msec	and $0.19$\% ($N=1$), to
$0.226$msec	and $0.022$\% ($N=6$).
Since the average query-time for $\alg{TDD}$ is $107.466$msec\footnote{
$\alg{TDD}$ is executed here on the original instance, even before the vertex contraction. In \cite{2016-Kontogiannis-Michalopoulos-Papastavrou-Paraskevopoulos-Wagner-Zaroliagis} it was executed on the contracted graph, hence the slightly smaller execution times of $\alg{TDD}$ in that work. Nevertheless, we believe that this is the appropriate measurement to make for $\alg{TDD}$, for sake of comparison with other works, and also since the contraction of degree-2 vertices is  part of the preprocessing phase.
}, the achieved speedup is more than $1,414$.

	%
\begin{figure}[htb!]
\centerline{%
	\includegraphics[height=4.35cm]{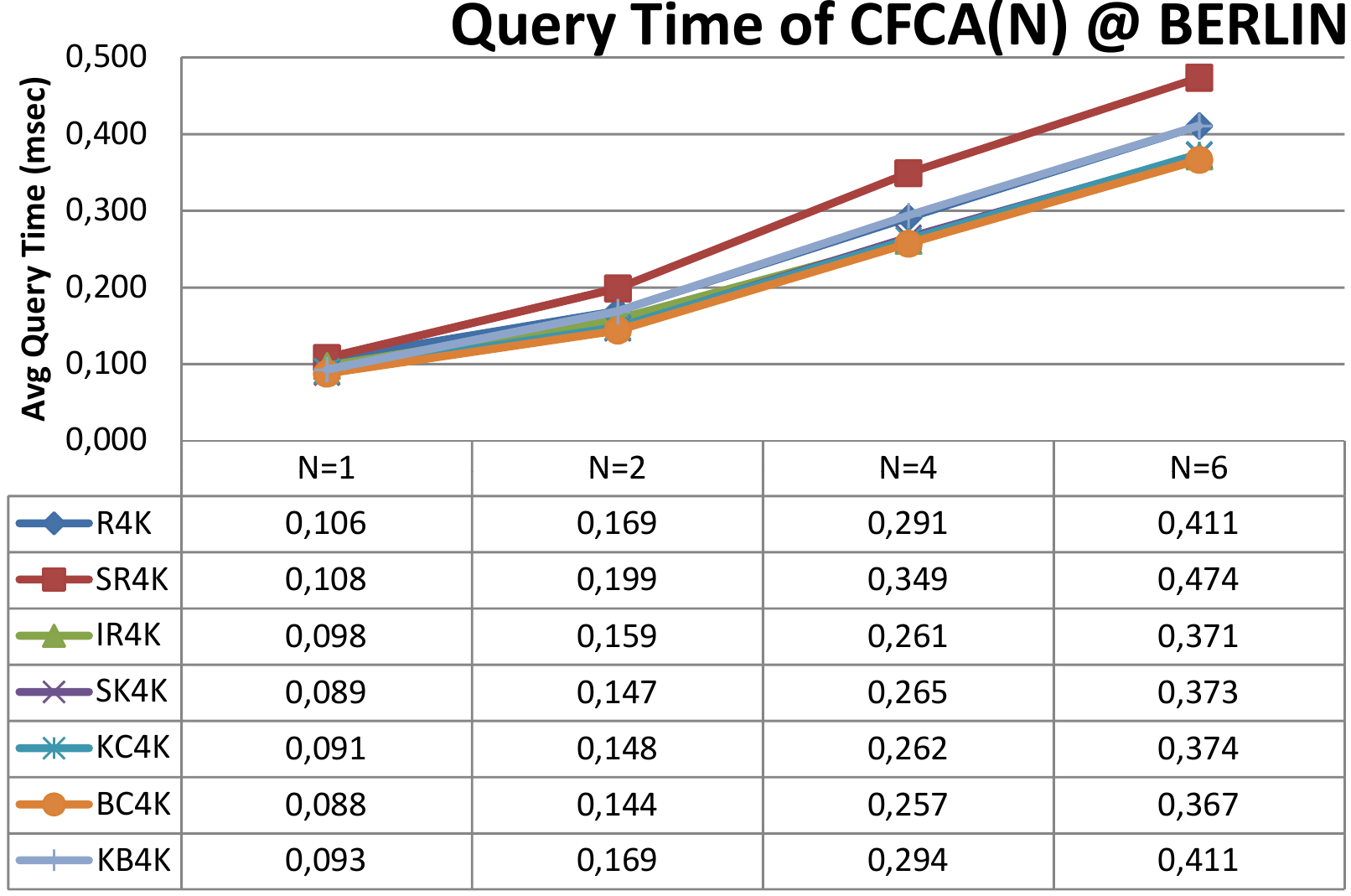}
	\includegraphics[height=4.35cm]{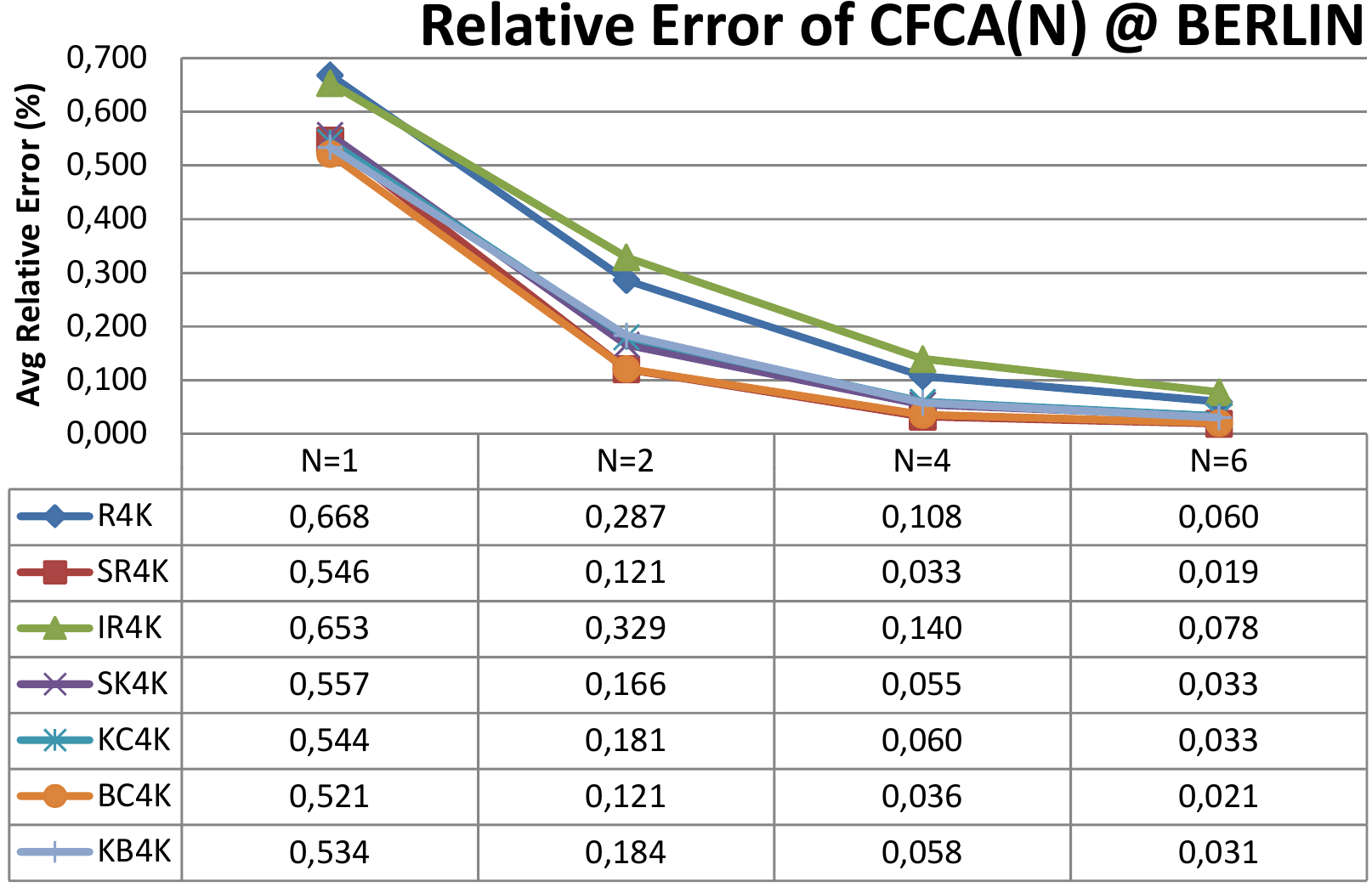}
}
\caption{\label{figure:Berlin-CFLAT-TIME_ERROR-4K-LANDMARKS}
	Performance of $\alg{CFCA(N)}$ in Berlin, for $4K$ landmarks and $50,000$ \emph{random queries}.}
\end{figure}
Our next experiment compares landmark types of size $4$K each (cf. Figure~\ref{figure:Berlin-CFLAT-TIME_ERROR-4K-LANDMARKS}). Concerning query-times, the best curve is that of BC4K. As for relative errors, SR4K and BC4K are clear winners. Further experiments are reported in Section~\ref{section:cfca-app}.
In comparison with $\alg{FLAT}$, the query-performance of $\alg{CFCA}(1)$ for BC4K is comparable ($0.088$msec and $0.521$\%) to that of $\alg{FCA}(1)$ ($0.081$msec and $0.771$\%) in \cite{2016-Kontogiannis-Michalopoulos-Papastavrou-Paraskevopoulos-Wagner-Zaroliagis}.
We also tested hybrid landmark sets. Interestingly, we achieved our best query performance with the hybrid set BC8K+R8K, which varies
from $0.076$msec and $0.192$\% (for $N=1$),
to $0.226$msec and $0.022$\% (for $N=6$).
It is also observed that, as we mix BC-landmarks with R-landmarks, the more BC landmarks we get the better for the relative error, whereas query-time is favored by more R-landmarks (cf. Figure~\ref{figure:BERLIN_CFLAT-16K-R+BC-MIX}).

\vspace*{-15pt}
\subparagraph*{Evaluation of $\alg{CFLAT}$ @ Germany.}

\begin{figure}[htb!]
\centerline{%
	\includegraphics[height=3.95cm]{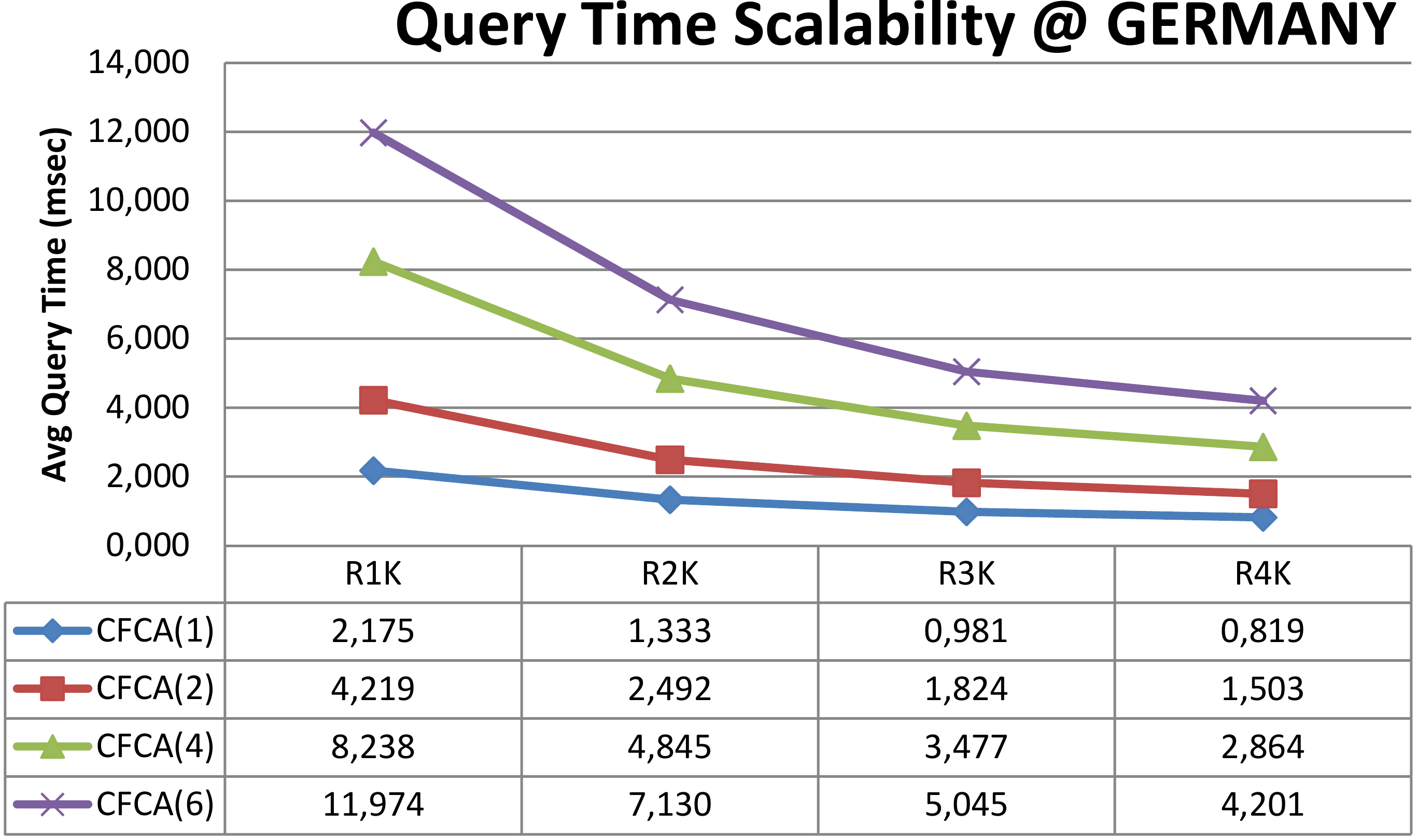}
	\includegraphics[height=3.95cm]{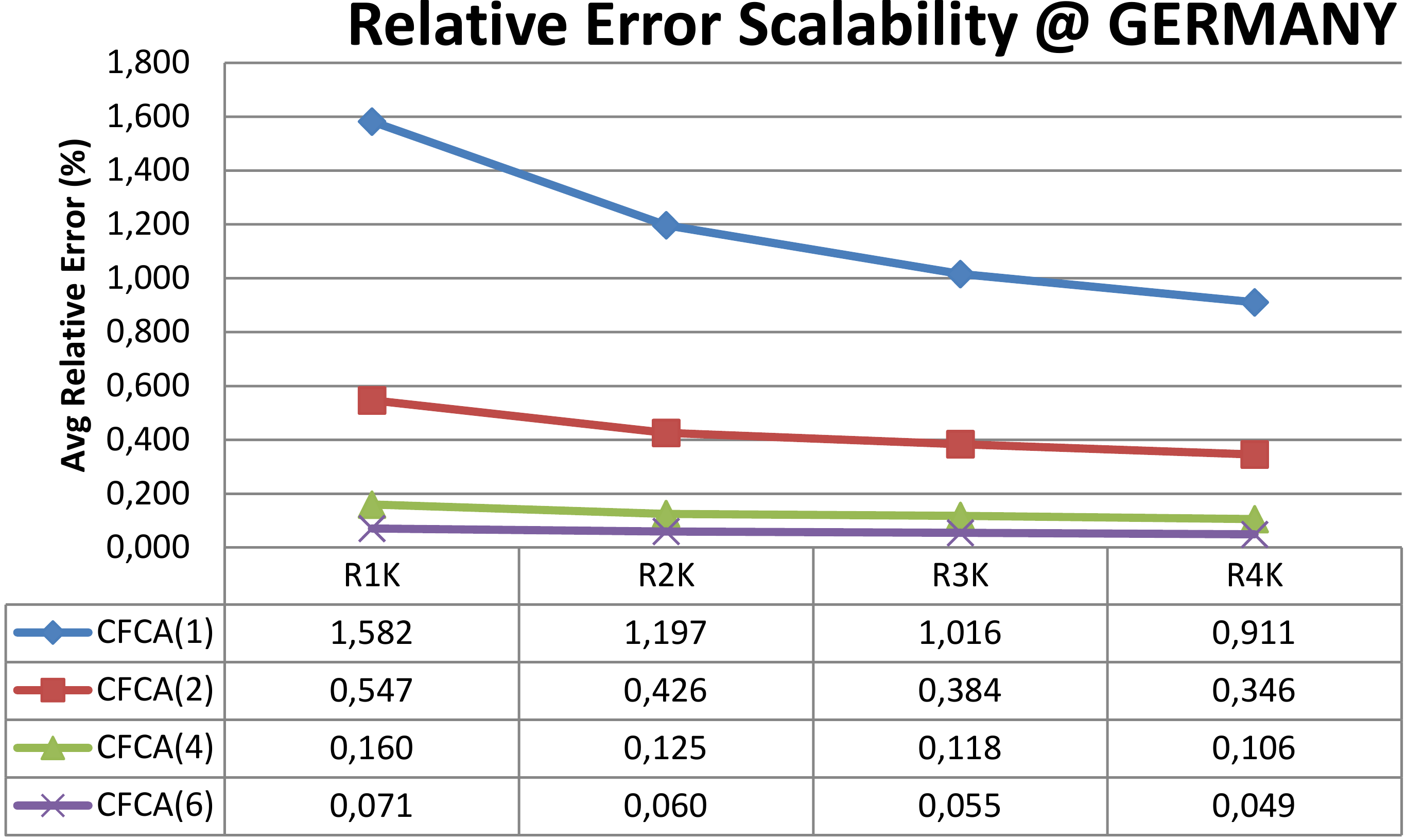}
}
\caption{\label{figure:GERMANY_CFLAT-SCALABILITY-RANDOM}
	Performance of
	$\alg{CFCA(N)}$ in Germany,
	for random landmarks and $50,000$ \emph{random queries}.	
}
\end{figure}

We considered R-landmark sets of sizes from $1$K to $4$K. The rest of the landmark sets were of size $3$K, with excluded neighborhood size $1,200$ vertices for SR3K, $350$ for SK3K, and $1,000$ for BC3K.
We started again with a demonstration of the scalability of $\alg{CFCA}$ on R-landmark sets, as a function of the number of landmarks (cf. Figure~\ref{figure:GERMANY_CFLAT-SCALABILITY-RANDOM}). The relative errors decrease linearly  and the running times decrease quadratically, as we increase the number of landmarks.
Remarkable relative errors of $0.071$\% are achieved for $\alg{CFCA}(6)$ even with $1$K landmarks which require $26.8$GiB ($8.1$GiB compressed) space, with query-time $11.974$msec. Moreover, a ``quick-and-dirty'' answer of error at most $1.582$\% is returned in only $2.175$msec.
The best query-times and relative errors are achieved for R4K, where $\alg{CFCA}(1)$ achieves $0.819$msec and $0.911$\%, and $\alg{CFCA}(6)$ has $4.201$msec	and $0.049$\%.

\begin{figure}[tb!]
\centerline{%
	\includegraphics[height=4.35cm]{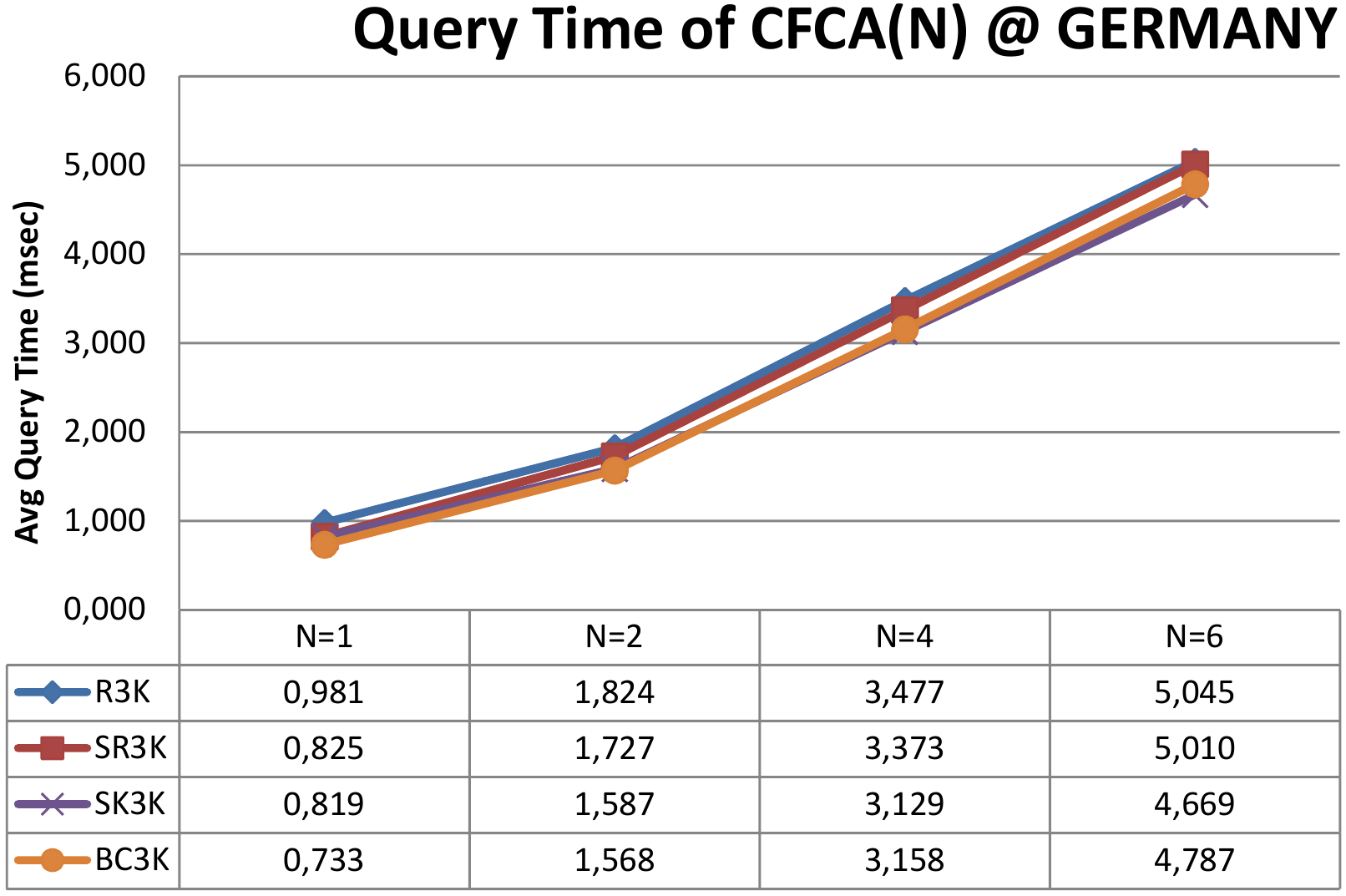}
	\includegraphics[height=4.35cm]{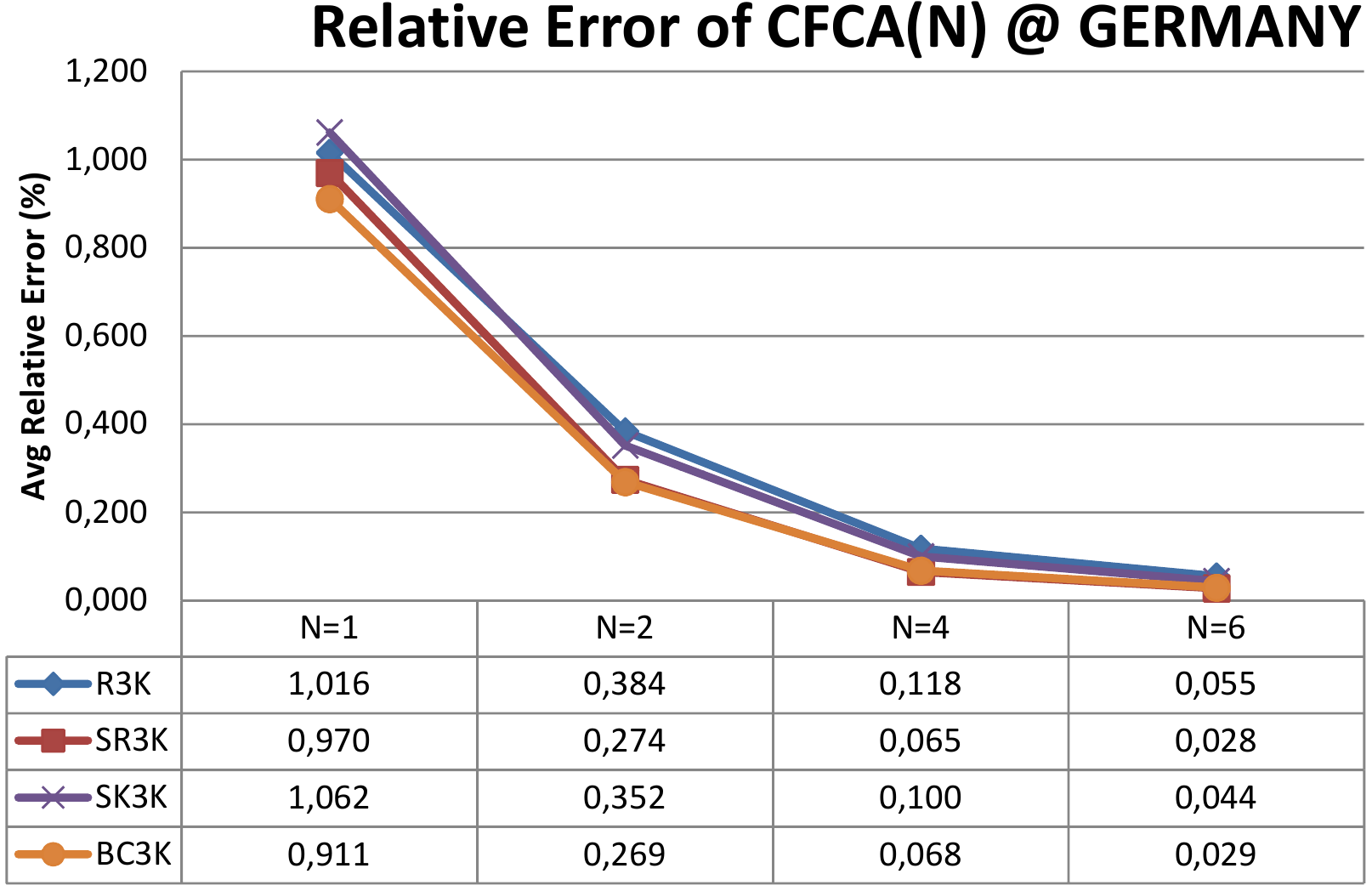}
}
\caption{\label{figure:Germany-CFLAT-TIME_ERROR-3K-LANDMARKS}
	Performance of $\alg{CFCA(N)}$ in Germany,
	for $3$K landmarks and $50,000$ \emph{random queries.}}
\end{figure}

We proceeded next with a comparison of various landmark types of size $3,000$ each (cf. Figure~\ref{figure:Germany-CFLAT-TIME_ERROR-3K-LANDMARKS}).
For Germany we have a clear winner, BC3K, w.r.t. both query-times and relative errors and $N\leq 2$. For $N\in\{4,6\}$, SK3K is the fastest and SR3K is the most accurate landmark policy.
Since the average time of $\alg{TDD}$ is $1,421.12$msec, the best speedup for $3$K landmarks is $1,938$, and the corresponding error is $0.911$\%.
Once more, the best query performance is achieved by a hybrid landmark set. In particular, for BC3K+R1K $\alg{CFCA}$'s performance varies from
$0.683$msec and $0.831$\% (for $N=1$),
to $4.104$msec and $0.031$\% (for $N=6)$, see Figure~\ref{figure:GERMANY_CFLAT-4K-R+BC-MIX}. Further experiments are reported in Section~\ref{section:cfca-app}.

\vspace*{-15pt}
\subparagraph*{Comparison with State-Of-Art.}
\label{section:sota-comparison}

Table~\ref{table:sota-comparison} presents a comparison with the most competitive speedup heuristics and oracles for TDRP. Details are provided in Section~\ref{section:discussion-on-sota}.	 We compare the performances of the following algorithms, on the instances of Berlin and Germany: 
\textbf{(1)} $\alg{TDCRP}$,  tested on a 16-core Intel Xeon E5-2670 clocked at 2.6 GHz, with 64GB of DDR3-1600 RAM, 20 MB of L3 and 256 KB of L2 cache. The reported numbers are from \cite{2016-Baum-Dibbelt-Pajor-Wagner};
\textbf{(2)} $\alg{FreeFlow}$, $\alg{TD\mbox{-}S}$ and $\alg{TD\mbox{-}S\mbox{+}A}$, tested on a 16-core Intel Xeon E5-1630 v3 clocked at 3.70GHz with 128GB of 2133GHz DDR4 RAM. The reported numbers are from \cite{2016-Strasser};
	%
\begin{table}[htb!]
\begin{scriptsize}
\vspace*{-6pt}
\hspace*{-22pt}
\begin{tabular}{|p{5pt}|c|p{1.5cm}|c|c|c|c|c|c|c|}
\hline
\rowcolor{grey!40} 
& \multicolumn{2}{c|}{\textbf{Algorithm}} & \multicolumn{3}{c|}{\textbf{Preprocessing Performance}} & \multicolumn{4}{c|}{\textbf{Query Performance}}
\\ \hline
\rowcolor{grey!40} 
& Name [ref.] & Parameters & Time 												& Work 		& Space 		& Path 							& Time & \multicolumn{2}{c|}{error (\%)}
\\
\rowcolor{grey!40} 
& &  & {\tiny h:m (\#cores)}	&  {\tiny h:m}	&  {\tiny B/node} 	&  {\tiny N/Y} 	&  {\tiny msec} &  {\tiny avg} &  {\tiny max} 
\\ \toprule\bottomrule 
\multirow{19}{*}{\rotatebox{90}{\kern-0.5cm GERMANY}}
& $\alg{TDD}$~\thiswork
& \TAB -- 				
& -- 
& --
& -- 	
& \YES & 1,421 & 0 & 0
\\ \cline{2-10}
& \multirow{4}{*}{$\alg{inex.TCH}$ \cite{2013-Batz-Geisberger-Sanders-Vetter} } 
& (0.1) 
& \multirow{4}{*}{06:18 (8)} 	
& \multirow{4}{*}{50:24} 
& 286 	
& \multirow{4}{*}{\NO} 
& 0.70 & 0.02 & 0.10
\\ \cline{3-3}\cline{6-6}\cline{8-10}
& 
& (1.0) 
& %
& %
& 214 	
& %
& 0.69 & 0.27 & 1.01
\\ \cline{3-3}\cline{6-6}\cline{8-10}
& 
& (2.5) 	
& %
& %
& 172 	
& %
& 0.72 & 0.79 & 2.44
\\ \cline{3-3}\cline{6-6}\cline{8-10}
& 
& (10.0)	
& %
& %
& 113 	
& %
& 1.06 & 3.84 & 9.75
\\ \cline{2-10}
& $\alg{KaTCH}$~\thiswork
& \TAB -- 				
& 34:29 (6) 
& 206:56	
&  9.029 	
& \NO & 26.9 & 25.6 & 1245.06
\\ \cline{2-10}
& $\alg{TDCRP}$ \cite{2016-Baum-Dibbelt-Pajor-Wagner}
& (1.0) 				
& 00:13 (16) 
& 03:28	
&  77 	
& \NO & 1.17 & 0.68 & 3.60

\\ \cline{2-10}
& $\alg{FreeFlow}$ \cite{2016-Strasser}
& \multirow{3}{*}{\TAB --} 				
& \multirow{3}{*}{00:07 (16)} 
& \multirow{3}{*}{01:57}	
& \multirow{3}{*}{n/r} 	
& \multirow{3}{*}{\YES} 
& 0.24 & 0.031 & 2.516

\\ \cline{2-2}\cline{8-10}
& $\alg{TD\mbox{-}S}$ \cite{2016-Strasser}
& 				
&  
& 
& 
& 
& 0.6 & 0.000746 & 0.989

\\ \cline{2-2}\cline{8-10}
& $\alg{TD\mbox{-}S\mbox{+}A}$ \cite{2016-Strasser}
&  				
&  
& 	
&  	
& 
& 6.36 & 0.000312 & 0.227

\\ \cline{2-10}
& $\alg{DijFreeFlow}$~\thiswork
& \TAB --
& --
& --
& --
& \YES 
& 736.24 & 0.352 & 17.569

\\ \cline{2-10}
& \multirow{4}{*}{$\alg{FLAT}$ \cite{2016-Kontogiannis-Michalopoulos-Papastavrou-Paraskevopoulos-Wagner-Zaroliagis}}
& SR2K, N=1 					
& \multirow{2}{*}{42:42 (6)} 
& \multirow{2}{*}{256:12}	
& \multirow{4}{*}{106,075} 
& \multirow{4}{*}{\NO} 
& 1.275 & 1.444 & \multirow{4}{*}{n/r}
\\ \cline{3-3}\cline{8-9}
& 
& SR2K, N=6
& %
& %
& %
& %
& 9.952 & 0.662 & 
\\ \cline{3-5}\cline{8-9}
& %
& SK2K, N=1 					
& \multirow{2}{*}{44:06 (6)} 
& \multirow{2}{*}{264:36}	
& %
& %
& 1.269 & 1.534 & 
\\ \cline{3-3}\cline{8-9}
& 
& SK2K, N=6
& %
& %
& %
& %
& 9.689 & 0.676 &

\\ \cline{2-10}
& \multirow{4}{*}{\vspace*{-15pt} $\alg{CFLAT}$~\thiswork} 
& BC4K, N=1			
& \multirow{2}{*}{32:36	(6)} 
& \multirow{2}{*}{195:29} 
& \multirow{2}{*}{30,769} 
& \multirow{4}{*}{\vspace*{-15pt} \YES}
& 0.693	& 0.858 & \multirow{4}{*}{\vspace*{-15pt} 19.154} 
\\ \cline{3-3}\cline{8-9}
& 
& BC4K, N=6		
& %
& %
& %
& %
& 3.841	& 0.049 & 
\\ \cline{3-6}\cline{8-9}
& 
& BC3K+R1K, N=1			
& \multirow{2}{*}{32:36	(6)} 
& \multirow{2}{*}{195:29} 
& \multirow{2}{*}{30,769} 
& %
& \color{black}{\bf 0.683} & 0.831 &
\\ \cline{3-3}\cline{8-9}
& 
& BC3K+R1K, N=6		
& %
& %
& %
& %
& 4.104 & \color{black}{\bf 0.031} &
\\ \toprule\bottomrule

\multirow{17}{*}{\rotatebox{90}{\kern-1.25cm BERLIN}}
& $\alg{TDD}$~\thiswork
& \TAB -- 				
& -- 
& --
& -- 	
& \YES & 107.5 & 0 & 0
\\ \cline{2-10}
& $\alg{KaTCH}$~\thiswork
& \TAB -- 				
& < 00:01 (6) 
& < 00:04	
&  0.593 	
& \NO & 0.3 & 0.41 & 47.74

\\ \cline{2-10}
& $\alg{TDCRP}$ \cite{2016-Baum-Dibbelt-Pajor-Wagner}
& (1.0)
& 00:02 (16)
& 00:28
& 67
& \NO & 0.28 & 1.47 & 2.69

\\ \cline{2-10}
& $\alg{FreeFlow}$ \cite{2016-Strasser}
& \multirow{3}{*}{\TAB --} 				
& \multirow{3}{*}{< 00:01 (16)} 
& \multirow{3}{*}{00:07}	
& \multirow{3}{*}{n/r} 	
& \multirow{3}{*}{\YES} 
& 0.09 & 0.0165 & 1.343
\\ \cline{2-2}\cline{8-10}
& $\alg{TD\mbox{-}S}$ \cite{2016-Strasser}
& 
&  
& 	
& 	
& 
& 0.23 & 0.00022 & 0.254
\\ \cline{2-2}\cline{8-10}
& $\alg{TD\mbox{-}S\mbox{+}A}$ \cite{2016-Strasser}
& 
&  
& 	
&	
&
& 3.01 & 0.000086 & 0.158

\\ \cline{2-10}
& $\alg{DijFreeFlow}$~\thiswork
& \TAB --
& --
& --
& --
& \YES & 54.608 & 0.367 & 20.42

\\ \cline{2-10}
& \multirow{4}{*}{$\alg{FLAT}$ \cite{2016-Kontogiannis-Michalopoulos-Papastavrou-Paraskevopoulos-Wagner-Zaroliagis}} 
& SR2K, N=1 					
& \multirow{2}{*}{05:12 (6)} 
& \multirow{2}{*}{31:12}	
& \multirow{2}{*}{48,389} 
& \multirow{4}{*}{\NO} 
& 0.081 & 0.771 & \multirow{4}{*}{n/r}
\\ \cline{3-3}\cline{8-9}
& %
& SR2K, N=6
& %
& %
& %
& %
& 0.586 & 0.317 &
\\ \cline{3-6}\cline{8-9}
& %
& SK2K, N=1 					
& \multirow{2}{*}{05:42 (6)} 
& \multirow{2}{*}{33:12}	
& \multirow{2}{*}{52,826} 
& %
& 0.083 & 0.781 &
\\ \cline{3-3}\cline{8-9}
& %
& SK2K, N=6
& %
& %
& %
& %
& 0.616 & 0.227 &

\\ \cline{2-10}
& 
\multirow{6}{*}{\vspace*{-45pt} $\alg{CFLAT}$~\thiswork}
& BC4K, N=1 
& \multirow{2}{*}{03:44	(6)} 
& \multirow{2}{*}{22:23} 
& \multirow{2}{*}{6,353} 
& \multirow{6}{*}{\vspace*{-45pt} \YES}
& 0.088	& 0.521 & \multirow{2}{*}{16.167}
\\ \cline{3-3}\cline{8-9}
&  
& BC4K, N=6		
&  
&  
&  
& 
& 0.367	& 0.021 & 
\\ \cline{3-6}\cline{8-10}
& %
& BC16K, N=1			
& \multirow{4}{*}{\vspace*{-30pt} 14:42	(6)} 
& \multirow{4}{*}{\vspace*{-30pt} 88:12} 
& \multirow{4}{*}{\vspace*{-30pt} 27,226} 
& %
& 0.078 & 0.227 
& \multirow{4}{*}{\vspace*{-30pt} 10.063}  
\\ \cline{3-3}\cline{8-9}
& %
& BC16K, N=6			
& %
& %
& %
& %
& 0.250	& 0.019 & 
\\ \cline{3-3}\cline{8-9}
& %
& BC8K+R8K, N=1			
&  
& 
& 
& 
& \color{black}{\bf 0.076} & 0.192 
&   
\\ \cline{3-3}\cline{8-9}
& %
& BC8K+R8K, N=6			
& %
& %
& %
& %
& 0.226	& \color{black}{\bf 0.022} & 
\\ \hline
\end{tabular}
\end{scriptsize}
\caption{\label{table:sota-comparison} Comparison with State-Of-The-Art.}
\vspace*{-18pt}
\end{table}
	%
\textbf{(3)} $\alg{inex.TCH}$, tested on an 8-Core Intel i7, clocked at 2.67 GHz, with 64 GB DDR4 RAM. The reported numbers are from~\cite{2016-Baum-Dibbelt-Pajor-Wagner};
\textbf{(4)} an open-source version of $\alg{TCH}$ ($\alg{KaTCH}$\footnote{\url{https://github.com/GVeitBatz/KaTCH}, with checksum 70b18ad0791a687c554fbfe9039edf79bc3a8ff3.}), tested (with compilation parameters -O3 and -DNDEBUG, and its default values) on our machine; 
\textbf{(5)} our own implementation of the $\alg{FreeFlow}$ heuristic (called $\alg{DijFreeFlow}$), tested on our machine (it is a static-Dijkstra execution on the Free Flow instance, with no exploitation of any speedup heuristic, and then computation of the time-dependent travel-time along the chosen path); and
\textbf{(6)} $\alg{FLAT}$ and $\alg{CFLAT}$, which were tested on our machine. The reported numbers for $\alg{FLAT}$ are from \cite{2016-Kontogiannis-Michalopoulos-Papastavrou-Paraskevopoulos-Wagner-Zaroliagis}. 
All the reported times are \emph{unscaled} (i.e., as they have been reported) and include both metric-independent and metric-dependent preprocessing of the instances. 
\emph{Work} is measured as the product of the running time with the number of cores. 
The ``path'' column indicates whether the explicit construction of a connecting path is accounted for in the reported query times. $\circ$ is a NO-answer, $\bullet$ means  YES. 
``n/r'' means that a particular value has not been reported.
The algorithms $\alg{TDD}$, $\alg{KaTCH}$, $\alg{DijFreeFlow}$ and $\alg{CFLAT}$, marked in Table 1 with \thiswork, were evaluated in the present work, on exactly the same benchmark instances and for the same sets of $50$K $\iuar$ chosen queries.


\section*{Acknowledgements}

The authors wish to thank G. Veit Batz, Julian Dibbelt and Ben Strasser for valuable and fruitful discussions.

\bibliographystyle{plain}


\input{2017-ESA-B-SUBMISSION_CFLAT-EVALUATION-refs}
\clearpage
\appendix

\input{appendix}

\end{document}

%% file: appendix.tex
\section{Preprocessing the Instances}
\label{section:preprocessing-instance}

We recap at this point some  heuristic improvements which are inherited from $\alg{FLAT}$ towards simplifying the road instance and thus saving space.

\subparagraph*{Contraction of the road network.}
\label{section:network-contraction}

The preprocessing space and time can be reduced if we only focus on a subgraph of the underlying graph representing the road network. Towards this direction, we have chosen to ``contract'' all the vertices which do not depict junctions of road segments (e.g., intermediate stops along a road segment).
We consider these vertices as \term{inactive} (only for the preprocessing phase), and we do not consider them during the subsequent preprocessing of travel-time related information, since they do not provide actual alternatives along a route using them, unless they are indeed endpoints of the query at hand. It is emphasized though, that the queries are conducted in the original graph, not just the contracted subgraph, meaning that we can query also for contracted origin-destination pairs and the returned paths do not contain shortcuts but actual road segments.

In more detail, in the \term{instance-contraction phase} we seek for maximal w.r.t. the number of arcs (possibly bidirectional) paths which have no ``vertical'' intersections, i.e., all the intermediate vertices connect only with their neighboring vertices along the path. Each such path is substituted with a \term{shortcut} (arc) connecting its endpoints, which is equipped with an arc-travel-time function equal to the corresponding exact path-travel-time function.
In fact, multiple paths with no intermediate intersections may connect the same active endpoints. In that case, a single shortcut represents more than one contracted paths, i.e. the arc-travel-time function of the shortcut is computed by applying the minimization operator on the path-travel-time functions corresponding to each of the contracted paths.
If there exists an original arc connecting two active endpoints, which are to be connected with a shortcut, we choose not to insert an additional shortcut, but to update accordingly the arc-travel-time of the already existing arc which now plays the role of a shortcut as well. The original arcs involved in the contracted paths are also considered as \emph{inactive}. All contracted vertices are ignored during the landmark-preprocessing and therefore the number of reachable destinations from a landmark is smaller. At the query phase, the contracted paths can be easily recovered, by exploiting the appropriate information kept on all shortcuts and the corresponding contracted vertices.

\subparagraph*{Almost constant legs.}

The original $\alg{TRAP}$ approximation method \cite{2016-Kontogiannis-Michalopoulos-Papastavrou-Paraskevopoulos-Wagner-Zaroliagis} introduced at least one intermediate breakpoint per interval that does not yet meet the required approximation guarantee. This is certainly unnecessary for small intervals in which the actual shortest-travel-time functions are constant.
To avoid the blow-up of the required preprocessing space, we heuristically make a ``guess'' that we have to deal with a constant shortest-travel-time function $D[\ell,v]$ within a given interval $[t_s,t_f=t_s+\tau)$ with sufficiently small length $\tau$, whenever the following holds: $D[\ell,v](t_s) = D[\ell,v](t_f) = D[\ell,v]\left(\frac{t_s+t_f}{2}\right)$.
This is justified by the fact that $D[\ell,v]$ is a continuous pwl function and it is unlikely that three different departure-times within a small interval would give the same value, unless the function is indeed constant.
Of course, one could easily construct artificial examples for which this criterion is violated, e.g. by providing a properly chosen periodic function with period $\tau/2$. On the other hand, one can easily tackle this by considering a \emph{randomly perturbed} sampling period $\tau + \delta$, for some arbitrarily small but positive random variable $\delta$. Since we engineer oracles for real-world road-networks, having three colinear points which do not belong to a leg of the sampled travel-time function is quite unlikely, therefore we choose not to randomly perturb the sampling period.

\subparagraph*{Fixed range.}

For a one-day time period, departure-times and arrival-times have a bounded value range. The same also holds for travel times which are at most one-day for any query within a country area such as Germany. Therefore, when the considered precision of the traffic data is within seconds, we handle time-values as integers in the range $\{0 ~,~ 1 ~,~\ldots ~,~ 86,399\}$, for milliseconds as integers in $\{0~,~1~,~\ldots ~,~86,399,999\}$, etc.

Any (real) time value within a single-day period, represented as a floating-point number $t_f$, can thus be converted to an integer $t_i$ with fewer bytes and a given unit of measure. For a unit measure (or scale factor) $s$, the resulting integer is $t_i = \ceil{\frac{t_f}{s}}$, requiring $\ceil{\frac{\log_2(t_f / s)}{8}}$ bytes for its storage. The division $\frac{t_f}{s}$ has quotient $\pi$ and remainder $\upsilon$ s.t., $t_f = s \cdot \pi + \upsilon$, and $t_i = \ceil{ \frac{s \cdot \pi + \upsilon}{s}} = \pi + \ceil{\frac{\upsilon}{s}} \in \{\pi, \pi+1\}$, since $0\leq \upsilon \leq s-1$. Therefore, by storing $t_i$ we actually consider the upper-approximating time $t'_f = s\cdot t_i$ of $t_f$, which causes an absolute error of at most $s$ (i.e., one unit of measure): $t'_f - t_f < s\cdot(\pi+1) - s\cdot\pi = s$.	
In our experiments, for storing the time values involved in the approximate shortest-travel-time functions, we have considered a $1.32$\emph{sec} resolution, corresponding to the appropriate scale factor $s = 1.318359375$ (when originally counting time in seconds), that requires $2$ bytes per time-value.

\section{The $\alg{CTRAP}$ approximation algorithm (pseudocode)}
\label{section:CTRAP-pseudocode}
We now present a more detailed description of $\alg{CTRAP}$. We start with the data types used in by the algorithm.
For a given landmark vertex $\ell$, a destination vertex $v$, and a subinterval $[t_s,t_f)\subseteq [0,T)$, the flag $ACTIVE[\ell,v](t_s,t_f)$ declares whether the upper-approximation $\overline{\delta}[\ell,v]$ considered by $\alg{CTRAP}$ (cf. Figure~\ref{fig:trapezoidal-approximation}) is satisfactory, given the required approximation guarantee that we consider.
The variable $\tau$ determines the current step of the sampled departure-times from $\ell$.
$PRED[\ell,v]$ and $DEP[\ell,v]$ are the sequences of predecessors and (corresponding) departure-times from $\ell$, w.r.t. the destination vertex $v$. We assure that $DEP[\ell,v]$ is always ordered in increasing departure-time values. This is done by assuming the operation $DEP[\ell,v].SortedInsertion(x)$ which places $x$ in the right position, which is then returned by the procedure. As for $DEP[\ell,v]$, we consider the insertion of a new element $u$ at an arbitrary position $pos$, $DEP[\ell,v](u,pos)$. It is mentioned at this point that these operations have been implemented in a rather straightforward manner (essentially performing linear scans on the queues), leaving for the future the consideration of more sophisticated implementations.

The boolean function $MAE[\ell,v](t_s,t_f)$ determines whether the maximum-absolute-error test is satisfied for $v$, in the interval $[t_s,t_f)$. In particular, since we already have sampled all the travel-times at $t_s$ and $t_f$, for a given approximation guarantee $\varepsilon>0$ we perform the following test, which is a sufficient condition for $\overline{\delta}[\ell,v]$ being a $(1+\varepsilon)$-upper-approximation of $D[\ell,v]$ within $[t_s,t_f)$:

\medskip
\noindent
\begin{footnotesize}
\begin{tabular}{|rp{12.7cm}|}
\hline
\rowcolor{gray!25}
\multicolumn{2}{|l|}{\textbf{procedure} $\alg{MAE}[\ell,v](t_s,t_f,\varepsilon)$}
\\ \hline \hline
\rowcolor{gray!5}
\textbf{1:}		
&		\IF $\min\{D[\ell,v](t_s), D[\ell,v](t_f) \}
			\geq \left(1 +\frac{1}{\varepsilon}\right)\La_{\max}$ \THEN \RETURN(TRUE)
\\
\rowcolor{gray!5}
\textbf{2:}
&		\ELSE	\RETURN(FALSE)
\\ \hline
\end{tabular}
\end{footnotesize}

The pseudocode of $\alg{CTRAP}$ is the following:

\medskip\noindent
\begin{footnotesize}
\begin{tabular}{|rp{12.7cm}|}
\hline
\rowcolor{gray!25}
\multicolumn{2}{|p{13.575cm}|}{\textbf{procedure} $\alg{CTRAP}(\ell,\varepsilon)$}
\\ \hline \hline
\rowcolor{gray!5}
\textbf{1:}
& 	\FOR $v\in V$ \DO $\{~ ACTIVE[\ell,v](0,T) = TRUE ~\}; \TAB \tau_{old} = T;~ \tau = 3200$
		\hfill	\COMMENT{initialization}
\\
\rowcolor{gray!5}
\textbf{2:}
& 	\WHILE $\exists v\in V,\exists k\in [0,T):~ ACTIVE[\ell,v](k\tau_{old},(k+1)\tau_{old}) == TRUE$ \DO
\\
\rowcolor{gray!5}
\textbf{3:}
& 	\TAB Sample min-cost-path trees rooted at $\ell$, only for \emph{new} departure-times $k\tau\in [0,T)$
\newline
		\COMMENT{$(w_1(k\tau),w_2(k\tau))$ is the pair of random seeds for $t_{\ell} = k\tau$.}
\newline
		 \COMMENT{$PRED[\ell,v](k\tau)$ indicates $v$'s parent in the tree routed at $(\ell,k\tau)$.}
\\
\rowcolor{gray!5}
\textbf{4:}
& 	\TAB \FOR $v\in V \AND k: k\tau\in [0,T)$ \DO
						\hfill\COMMENT{looking for still active destinations...}
\\
\rowcolor{gray!5}
\textbf{5:}
& \TAB\TAB
								\IF $ACTIVE[\ell,v](k\tau_{old},(k+1)\tau_{old}) == TRUE$ \THEN
\\
\rowcolor{gray!5}
\textbf{6:}
& \TAB\TAB\TAB $HASH[v] = HASH[v] + (w_1(k\tau),w_2(k\tau))\cdot k\tau$
		\hfill\COMMENT{Update hash keys...}				
\\
\rowcolor{gray!5}
\textbf{7:}
& 	\TAB\TAB\TAB\IF $DEP[\ell,v].NotInSequence(kt)$ \THEN
\\
\rowcolor{gray!5}
\textbf{8:}
&		\TAB\TAB\TAB\TAB $position = DEP[\ell,v].SortedInsertion(k\tau)$;
\\
\rowcolor{gray!5}
\textbf{9:}
&		\TAB\TAB\TAB\TAB $PRED[\ell,v].Insertion(parent[\ell,v](k\tau),position)$
\\
\rowcolor{gray!5}
\textbf{10:}
& 	\TAB\TAB\TAB\END\IF
\\
\rowcolor{gray!5}
\textbf{11:}
&		\TAB\TAB\TAB%
								\IF$MAE[\ell,v](k\tau,(k+1)\tau, \varepsilon) == TRUE$
								\THEN $\{~ ACTIVE[\ell,v](k\tau,(k+1)\tau) = FALSE ~\}$
\\
\rowcolor{gray!5}
\textbf{12:}
&		\TAB\TAB\END\IF
\\
\rowcolor{gray!5}
\textbf{13:}
&		\TAB
								\END\FOR
\\
\rowcolor{gray!5}
\textbf{14:}
&		\TAB$\tau = \tau / 2;~ \tau_{old} = 2\tau$								
\\
\rowcolor{gray!5}
\textbf{15:}
&		\END\WHILE
\\
\rowcolor{gray!5}
\textbf{16:}
&		\FOR $v\in V$ \DO
\\
\rowcolor{gray!5}
\textbf{17:}
&		\TAB\REPEAT \hfill\COMMENT{merge intervals with the same predecessor...}
\\
\rowcolor{gray!5}
\textbf{18:}
&		\TAB\TAB\FOR \emph{consecutive} records $(PRED[\ell,v](t),t)$ and $(PRED[\ell,v](t'),t')$ such that $PRED[\ell,v](t) == PRED[\ell,v](t')$ \DO
\\
\rowcolor{gray!5}
\textbf{19:}
&		\TAB\TAB\TAB $PRED[\ell,v].Delete(PRED[\ell,v](t'))$
\\
\rowcolor{gray!5}
\textbf{20:}
&		\TAB\TAB\TAB $DEP[\ell,v].Delete(t')$
\\
\rowcolor{gray!5}
\textbf{21:}
&		\TAB\TAB\END\FOR
\\
\rowcolor{gray!5}
\textbf{22:}
& \TAB\UNTIL $PRED[\ell,v]$ does not have identical consecutive records.
\\
\rowcolor{gray!5}
\textbf{23:}
& \END\FOR
\\
\rowcolor{gray!5}
\textbf{24:}
&		Lexicographically sort in $DEST[\ell]$ the destinations $v$ according to their hash key pairs.
\\
\rowcolor{gray!5}
\textbf{25:}
&		\FOR $v \in DEST[\ell]$ (in the previous lex-order) \DO
		\hfill\COMMENT{Avoid multiple copies of dep-time sequences...}
\\
\rowcolor{gray!5}
\textbf{26:}
&		\TAB \IF $HASH[v] == HASH[DEST[\ell].Previous[v]]$
\\ \rowcolor{gray!5}
&		\TAB	\THEN $\{~ representative[v] = DEST.Previous[v];~ DEP[\ell,v].Destroy()~\}$
\\
\rowcolor{gray!5}
\textbf{27:}
& 	\END\FOR
\\ \hline
\end{tabular}
\end{footnotesize}

\section{Benchmark Instances and Preprocessing}
\label{section:benchmark-instances-details}

Our implementations and data sets constitute part of a broader route planning
application service developed within the frame of EU-funded projects, which has been 
piloted in the cities of Berlin, Vitoria and Athens, as well as in the national road 
network of Germany. Due to complicated IPR issues, we cannot make our source code and 
benchmark data publicly available.

We proceed in this section with a detailed presentation of the benchmark instances of Berlin and Germany, on which we have conducted the experimental evaluation of $\alg{CFLAT}$.

\subparagraph*{Berlin Instance.}

The instance of Berlin (kindly provided by TomTom in the frame of common R\&D projects) consists of $473,253$ nodes and $1,126,468$ arcs.
The instance-preprocessing heuristic \ref{section:network-contraction} created $183,468$ shortcuts. Whenever more than one contracted paths shared the same endpoints, we added only one shortcut representing all these contracted paths. There were $914$ such cases in the Berlin instance.
The contracted paths that could be represented by an original arc in the graph, are $11,398$ in total.
In overall, the contraction of Berlin led to a graph of $292,356$ active vertices and $752,362$ active arcs.

\subparagraph*{Germany Instance.}

The instance of Germany (kindly provided by PTV AG in the frame of common R\&D projects) consists of $4,692,091$ nodes and $11,183,060$ arcs. After the instance-preprocessing phase we got an instance with $3,431,213$ active vertices and $11,554,840$ active arcs. The total number of the added shortcuts was $4,595,148$. We avoided the insertion of additional shortcuts in $106,464$ cases, where $6,816$ of them correspond to ``parallel'' shortcuts and the $99,648$ correspond to the existence of actual arcs connecting the endpoints of contracted paths.

\subparagraph*{ Statistics for Berlin and Germany Instances.}

Table~\ref{table:PREPROCESSING-INFO-CFLAT-Berlin} reports some significant preprocessing statistics for the two instances. In particular, the measurements are the following:
\emph{(i)} the average number of vertices per landmark whose predecessor remains constant on the min-cost-path tree throughout the whole time period,
\emph{(ii)} the remaining vertices with pwl behaviour w.r.t. their predecessor,
\emph{(iii)} the average number of unique departure-time sequences stored, instead keeping one sequence per destination with pwl predecessor, and
\emph{(iv)} the average number of intermediate points of $\alg{TRAP}$ per landmark, which we now avoid to store.

\begin{table}[htb!]
\centering
\begin{tabular}{r|c|c|c|c|}
\cline{2-5}
	& \shortstack{\\Vertices with\\Unique Pred}
	& \shortstack{\\Vertices with\\pwl Pred}
	& \shortstack{\\Unique Departure\\Time Sequences}
	& \shortstack{\\Intermediate\\Points of $\alg{TRAP}$}
\\
\cline{2-5}
R4K & $272,286$ & $20,070$ & $5,963$ & $10,663,125$
\\
\cline{2-5}
SR4K & $272,287$ & $20,069$ & $5,831$ & $10,688,275$
\\
\cline{2-5}
IR4K & $272,284$ & $20,072$ & $5,781$ & $10,672,869$
\\
\cline{2-5}
SK4K & $272,282$ & $20,074$ & $6,011$ & $10,934,712$
\\
\cline{2-5}
KC4K & $272,287$ & $20,069$ & $5,857$ & $10,758,955$
\\
\cline{2-5}
BC4K & $272,293$ & $20,063$ & $5,858$ & $10,728,776$
\\
\cline{2-5}
KB4K & $272,300$ & $20,056$ & $5,432$ & $10,643,285$
\\
\cline{2-5}
\cline{2-5}
\end{tabular}
\caption{\label{table:PREPROCESSING-INFO-CFLAT-Berlin}
	Preprocessing statistics for $\alg{CFLAT}$ Oracle for Berlin.}
\end{table}

Table~\ref{table:PREPROCESSING-INFO-CFLAT-Germany} provides the preprocessing statistics related to Germany, in the same format as in the case of Berlin.

\begin{table}[htb!]
\centering
\begin{tabular}{r|c|c|c|c|}
\cline{2-5}
	& \shortstack{\\Vertices with\\Unique Pred}
	& \shortstack{\\Vertices with\\pwl Pred}
	& \shortstack{\\Unique Departure\\Time Sequences}
	& \shortstack{\\Intermediate\\Points of $\alg{TRAP}$}
\\
\cline{2-5}
R3K & $3,201,577$ & $229,636$ & $38,102$ & $112,137,488$
\\
\cline{2-5}
SR3K & $3,201,642$ & $229,571$ & $37,212$ & $112,081,032$
\\
\cline{2-5}
SK3K & $3,201,503$ & $229,710$ & $38,068$ & $113,536,811$
\\
\cline{2-5}
BC3K & $3,201,637$ & $229,576$ & $37,207$ & $112,067,442$
\\
\cline{2-5}
\end{tabular}
\caption{\label{table:PREPROCESSING-INFO-CFLAT-Germany}
	Preprocessing statistics for $\alg{CFLAT}$ in Germany.}
\end{table}

\begin{figure}[htb!]
\centerline{%
	\includegraphics[height=4cm]{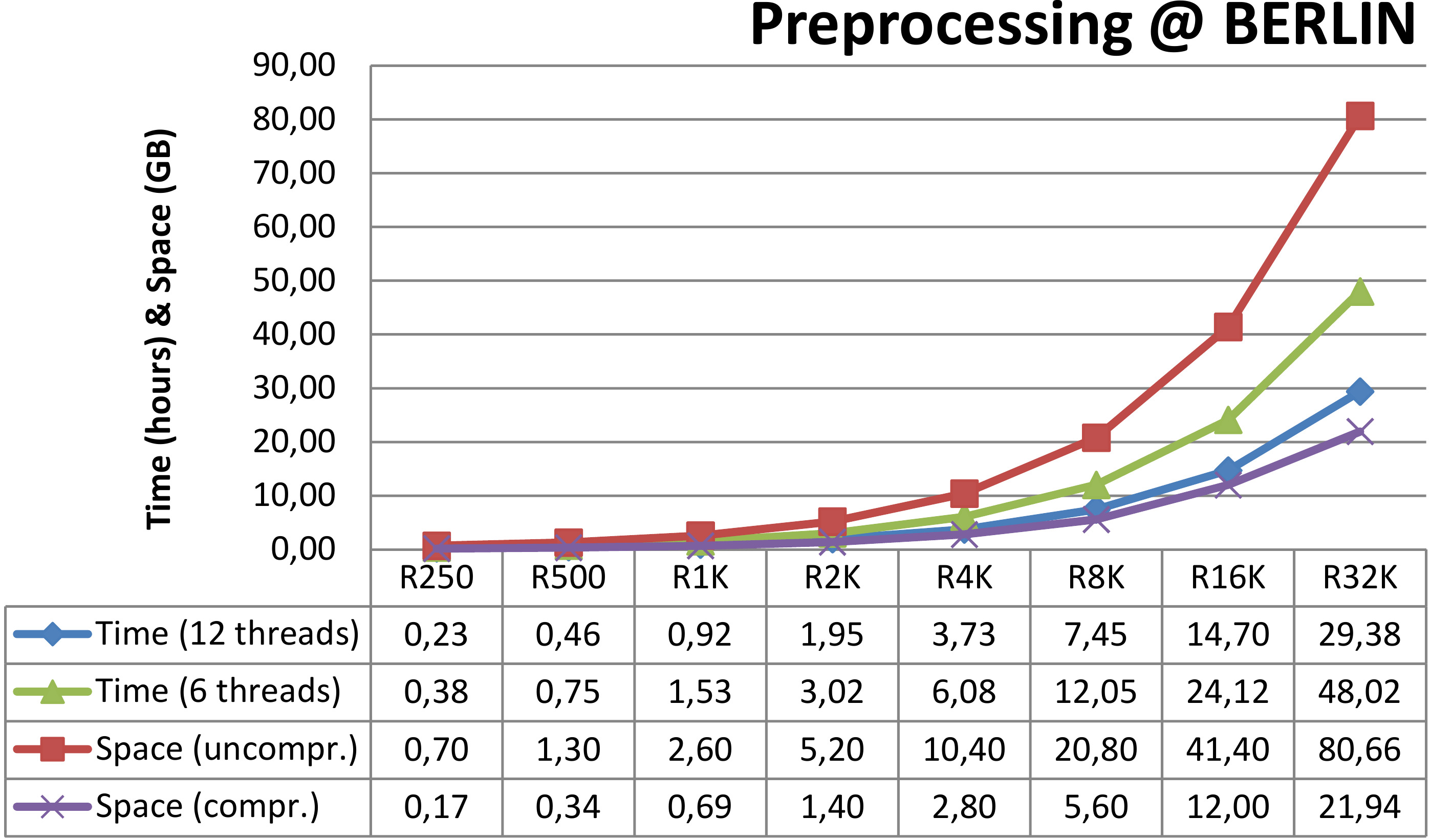}
	\includegraphics[height=4cm]{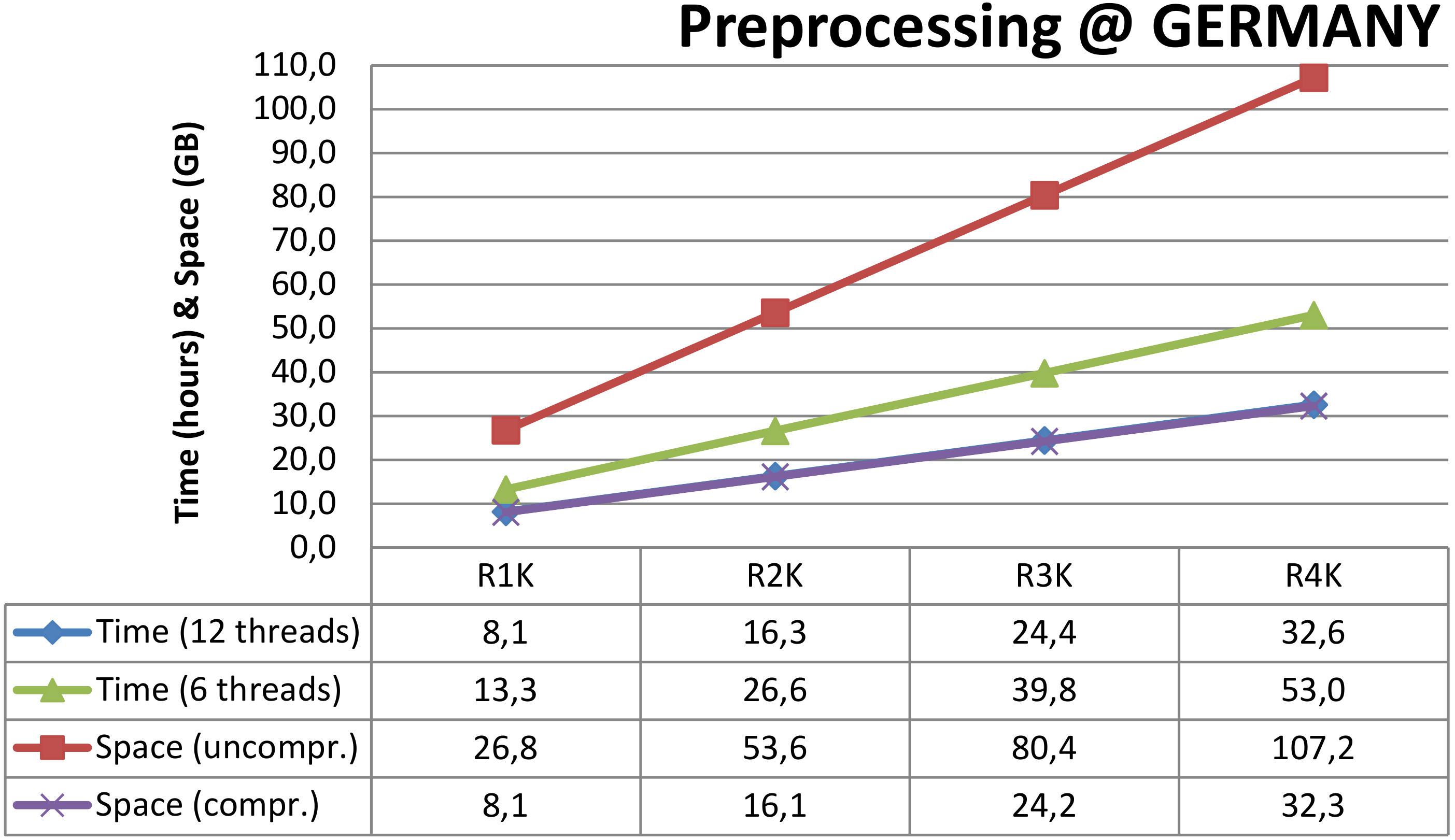}
}
\caption{\label{figure:BERLIN+GERMANY_PREPROCESSING-RANDOM}
	Preprocessing requirements for Berlin and Germany.	
}
\end{figure}

\vspace*{-12pt}
\subparagraph*{Preprocessing Requirements @ Berlin.}

We present in this section the preprocessing requirements for the construction of the summaries for $\alg{CFLAT}$, for various sizes of random (R) landmark sets (cf. Figure~\ref{figure:BERLIN+GERMANY_PREPROCESSING-RANDOM}). The requirements for other landmark types are analogous. For this preprocessing, we have used $12$ parallel threads on our $6$-core machine.

It is worth mentioning that $\alg{FLAT}$ \cite{2016-Kontogiannis-Michalopoulos-Papastavrou-Paraskevopoulos-Wagner-Zaroliagis} required \emph{uncompressed} preprocessing space $43$GB, or equivalently, \emph{compressed} size of $14$MB per landmark, and $33$h to preprocess R2K.
On the contrary, with $\alg{CFLAT}$ R32K is preprocessed in $29.38$h consuming $80.67$GB ($21.94$GB compressed) space. As for R2K, it is preprocessed in $117$min consuming only $5.2$GB ($1.4$GB compressed) space. Finally, R250 is preprocessed in $14$min, consuming only $0.7$GB ($0.17$GB compressed) space. In general, $\alg{CFLAT}$ has an average preprocessing requirement of $3.306$sec and $2.521$MB per landmark.

\vspace*{-12pt}
\subparagraph*{Preprocessing Requirements @ Germany.}

The preprocessing requirements for the constructing the summaries of $\alg{CFLAT}$ in Germany, for various sizes of R-landmark sets, are shown in Figure~\ref{figure:BERLIN+GERMANY_PREPROCESSING-RANDOM}. In general, there is a requirement for $29.33$sec and $26.8$MB per landmark, which is totally justifiable compared to Berlin, due to the larger size of the instance (by an order of magnitude).
A significant improvement over the preprocessing requirements of $\alg{FLAT}$ is again achieved. E.g., for R2K $\alg{FLAT}$ requires (uncompressed) space $100.7$GB which are constructed in $44.6$h, whereas $\alg{CFLAT}$ creates the analogous preprocessed data in $16.3$h requiring $53.6$GB ($16.1$GB compressed) space. This indeed made it possible to consider landmark sets of size up to $4,000$ in the present work.

\section{Detailed Auditing of $\alg{CFCA(N)}$'s Performance}
\label{section:cfca-app}

\begin{figure}[htb!]
\centerline{%
	\includegraphics[height=4.4cm]{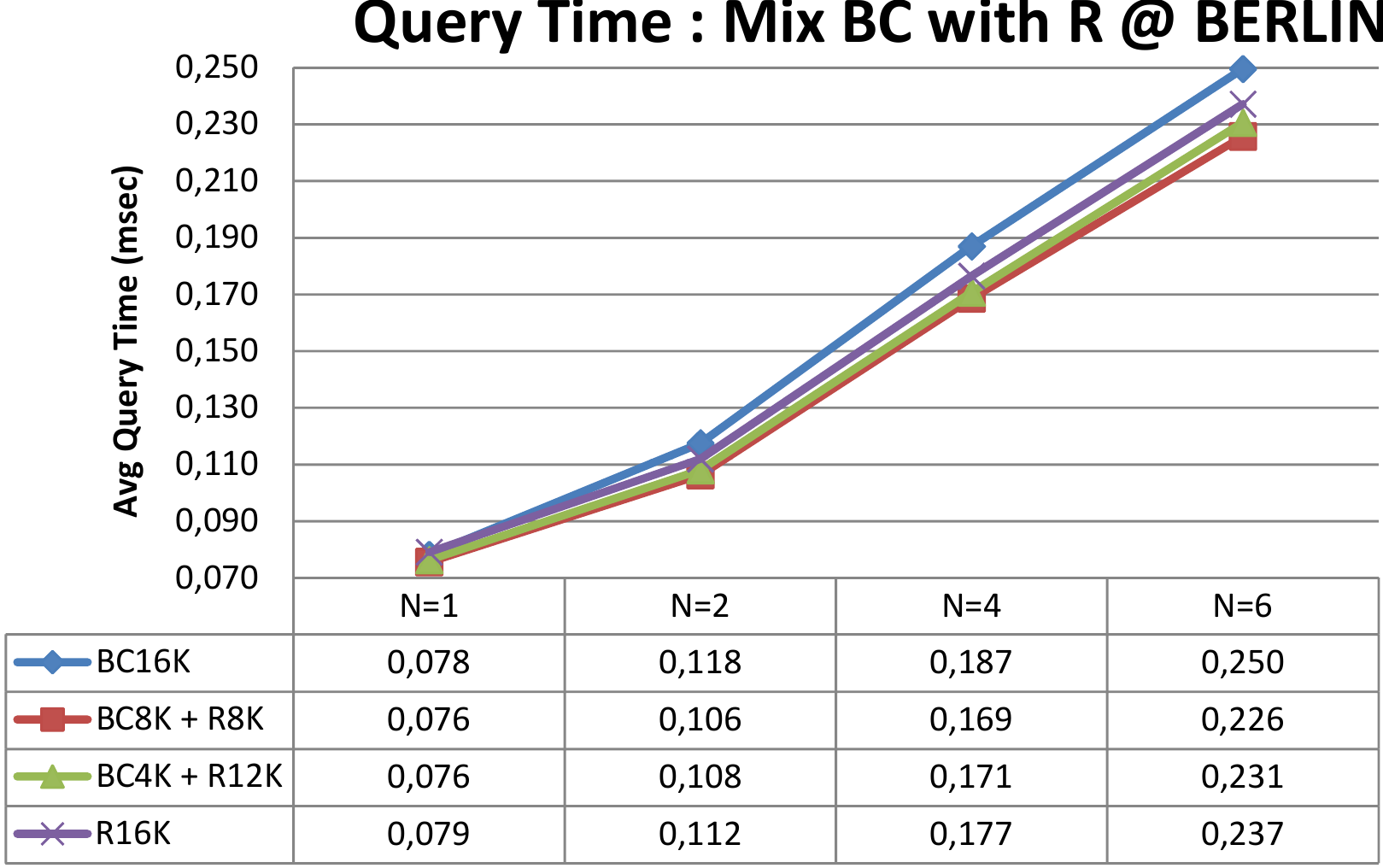}
	\includegraphics[height=4.4cm]{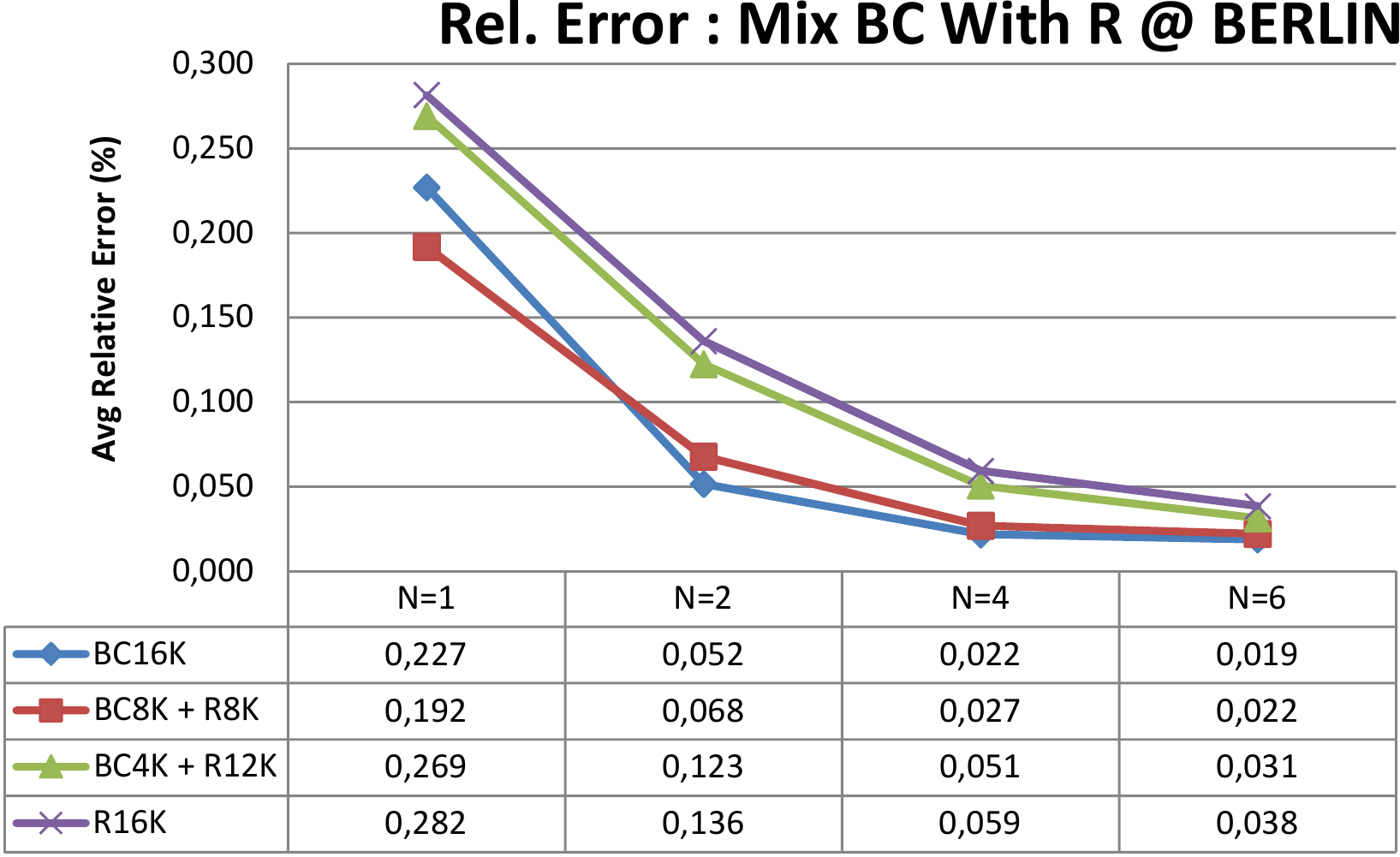}
}%
\caption{\label{figure:BERLIN_CFLAT-16K-R+BC-MIX}
	Performance of $\alg{CFCA}$ for mixtures (BC- and R-landmark types) of $16$K landmarks in Berlin, 
	and a query set of $50,000$ \emph{random queries}.
}
\end{figure}

We provide in this section more detailed experiments for the performance f $\alg{CFCA}$. We start with mixtures of BC- and R-landmark sets. As Figure~\ref{figure:BERLIN_CFLAT-16K-R+BC-MIX} shows, BC-landmarks improve mainly the relative error, whereas R-landmarks improve the query-time in Berlin. Interestingly, the best query-time is achieved by the hybrid landmark sets BC8K+R8K and BC4K+R12K, with the former having much better relative error.

\begin{figure}[htb!]
\centerline{%
	\includegraphics[height=4.4cm]{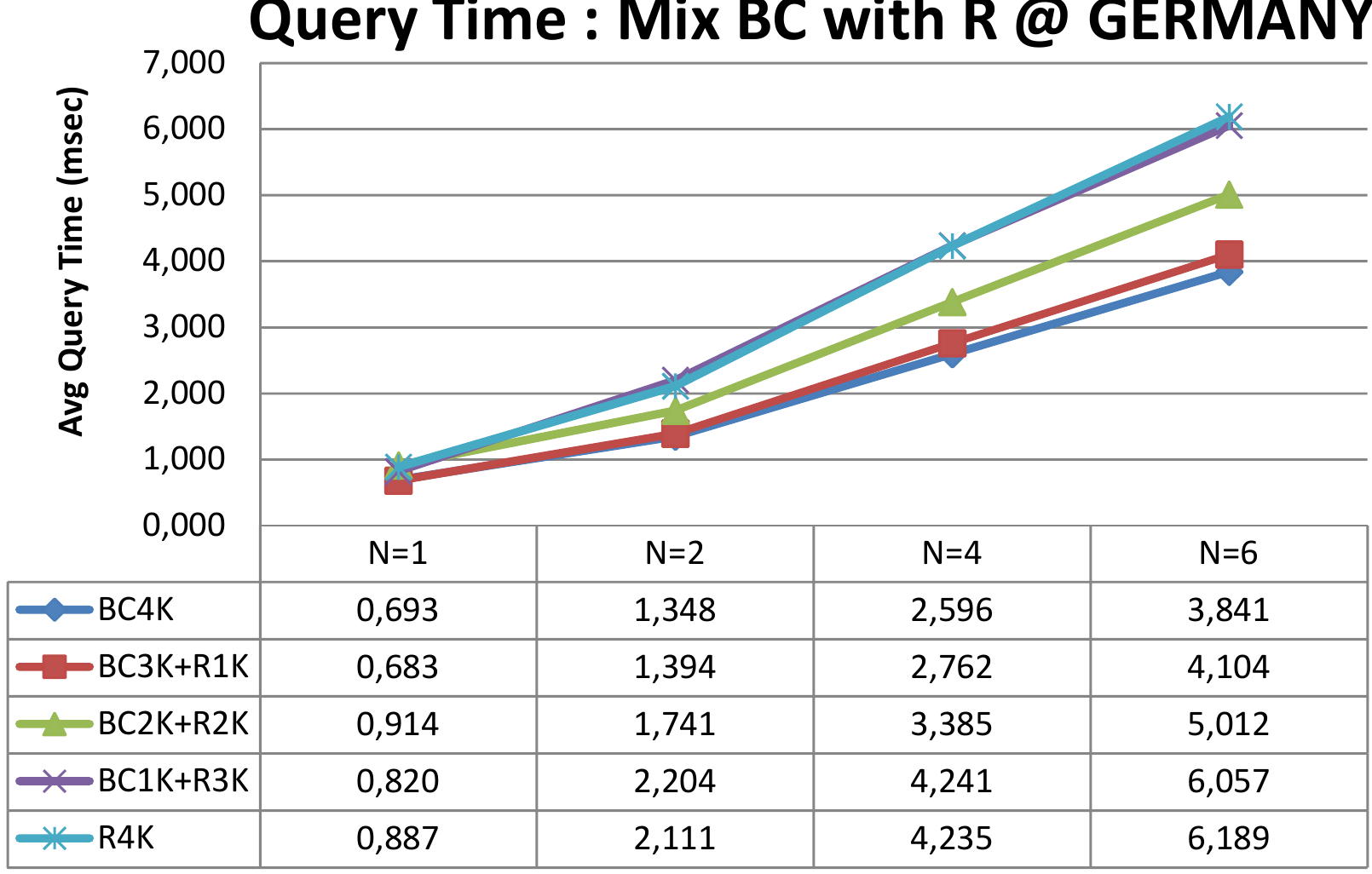}
	\includegraphics[height=4.4cm]{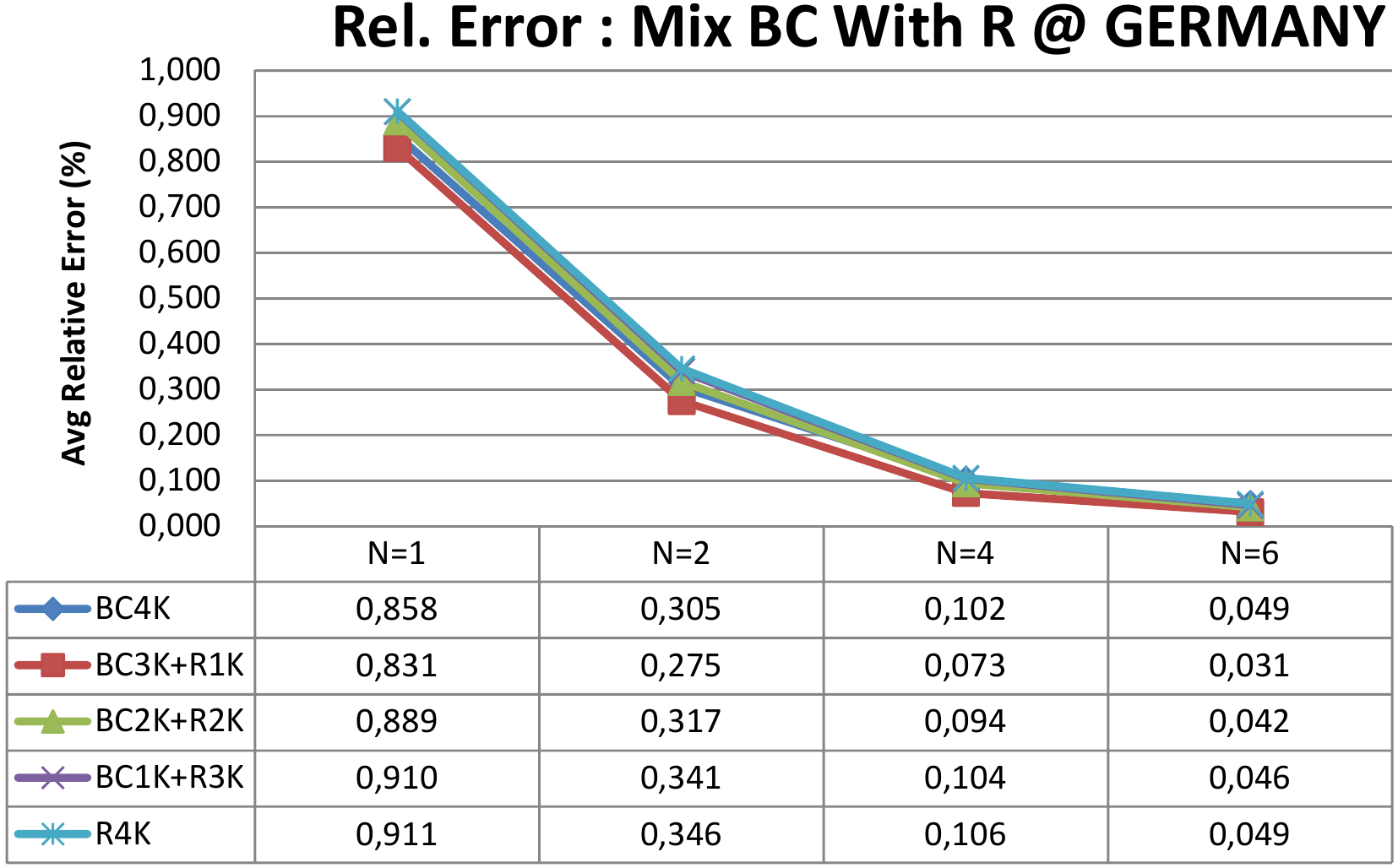}
}%
\caption{\label{figure:GERMANY_CFLAT-4K-R+BC-MIX}
	Performance of $\alg{CFCA}$ for mixtures (BC- and R-landmark types) of $4$K landmarks in Germany, 
	and a query set of $50,000$ \emph{random queries}.
}
\end{figure}

Analogous observations hold also for Germany, as it is shown in Figure~\ref{figure:GERMANY_CFLAT-4K-R+BC-MIX}. Once more the best query-time ($0.683$msec) of $\alg{CFCA}$ is achieved for BC3K+R1K.

We next audit the amount of computational effort (both in terms of Dijkstra rank, and of absolute running times) of $\alg{CFCA}$ among its major steps.

\begin{figure}[htb!]
\centerline{%
	\begin{tabular}{c}
	\includegraphics[height=4.25cm]{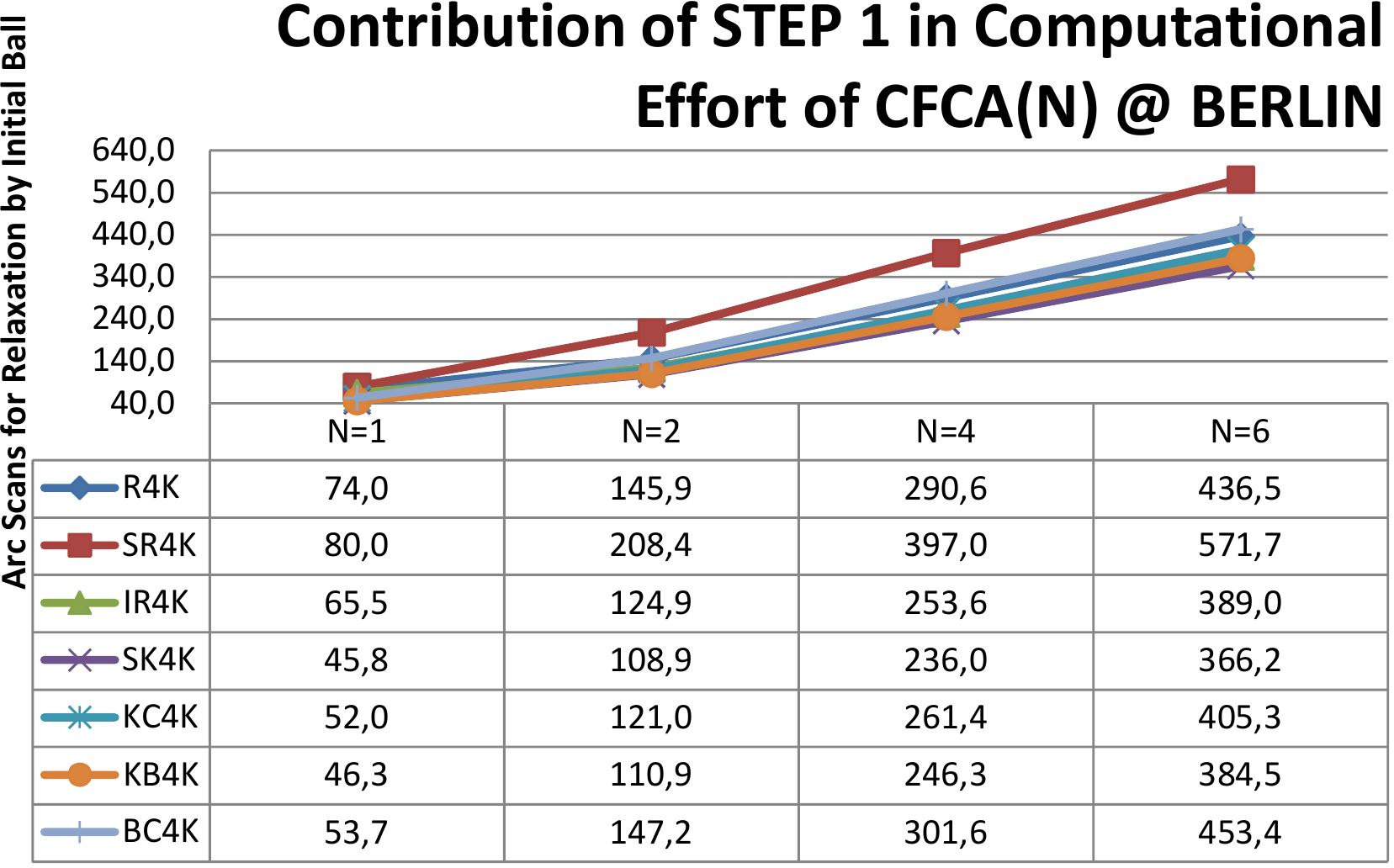}
	\includegraphics[height=4.25cm]{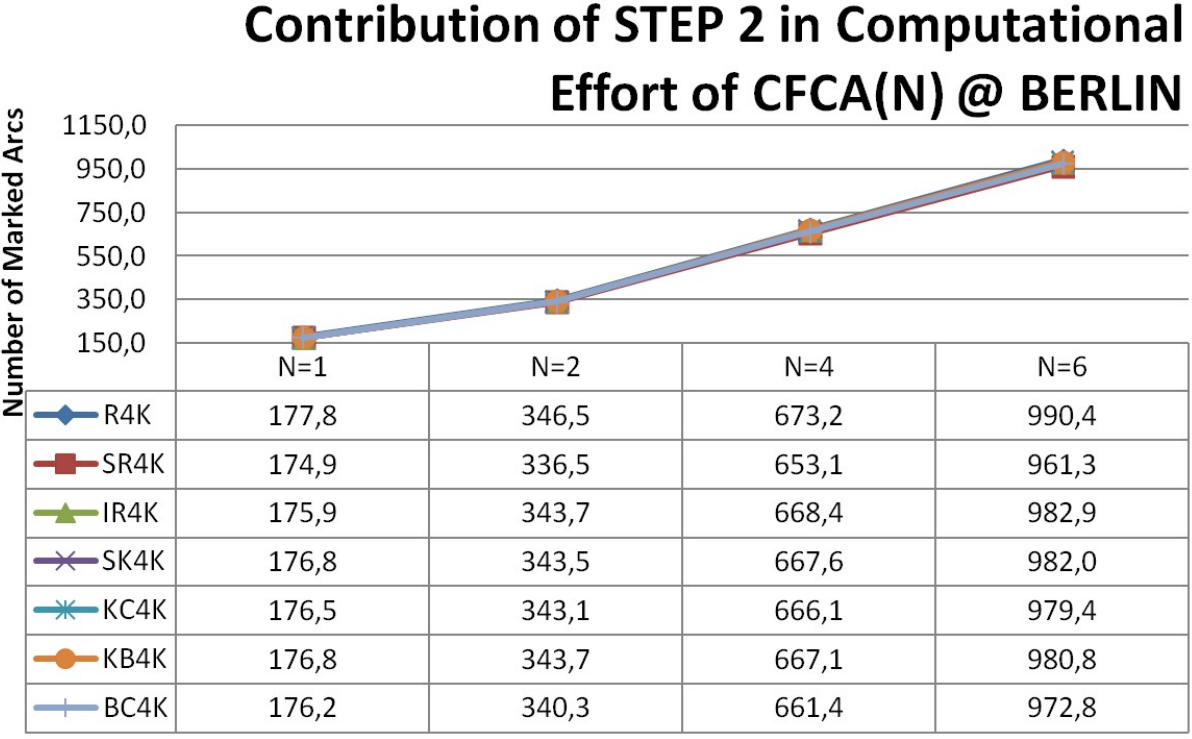}
	\\
	\includegraphics[height=4.25cm]{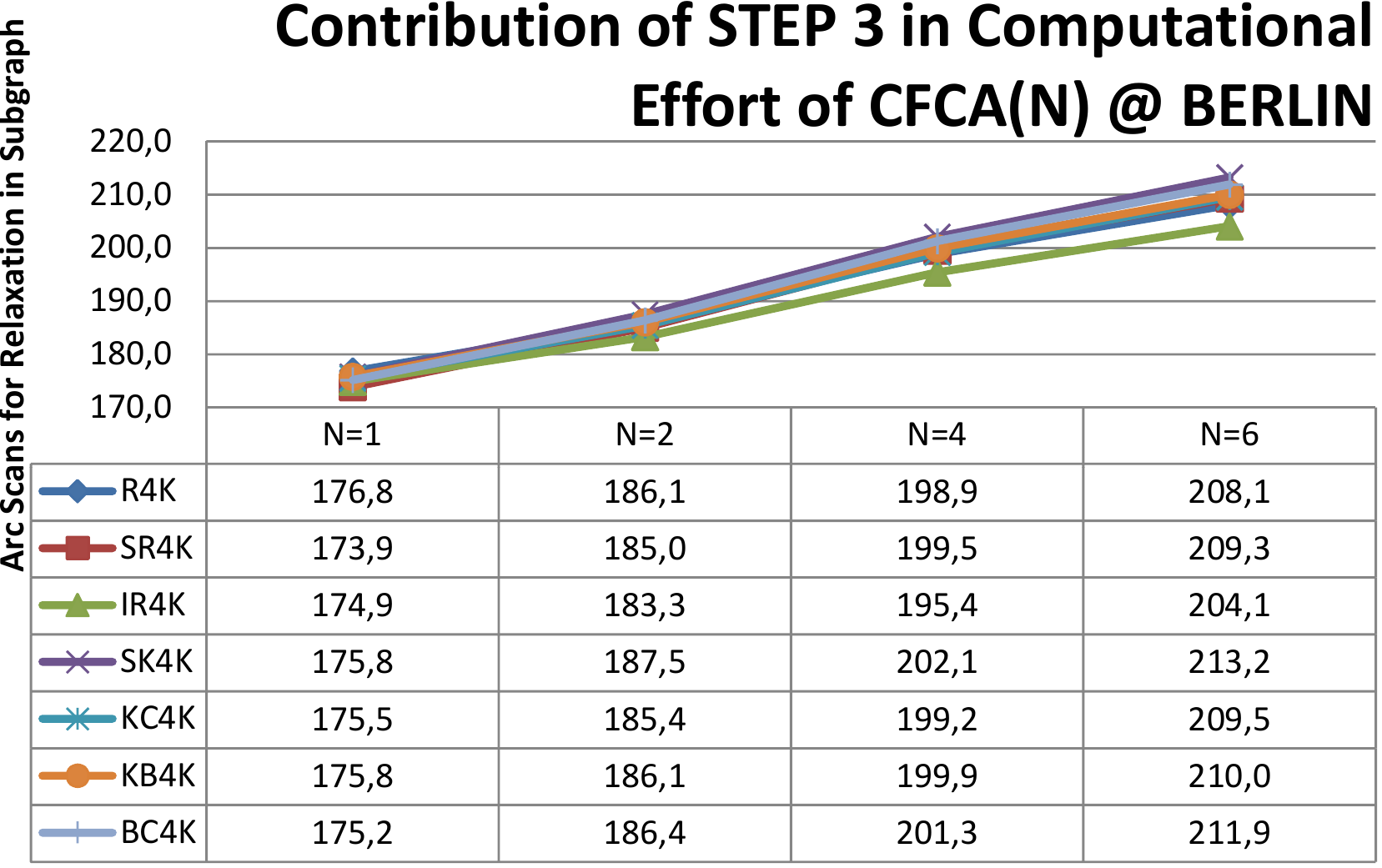}
	\end{tabular}
}%
\caption{\label{figure:BERLIN_CFLAT-PER-STEP-RANK-PERFORMANCE}
	Comparison of number of ``touched vertices'' per step of $\alg{CFCA(N)}$, at $1.32$sec resolution, for a query set of $50,000$ \emph{random queries} in Berlin.
}
\end{figure}

\begin{figure}[htb!]
\centerline{%
	\begin{tabular}{c}
	\includegraphics[height=4.25cm]{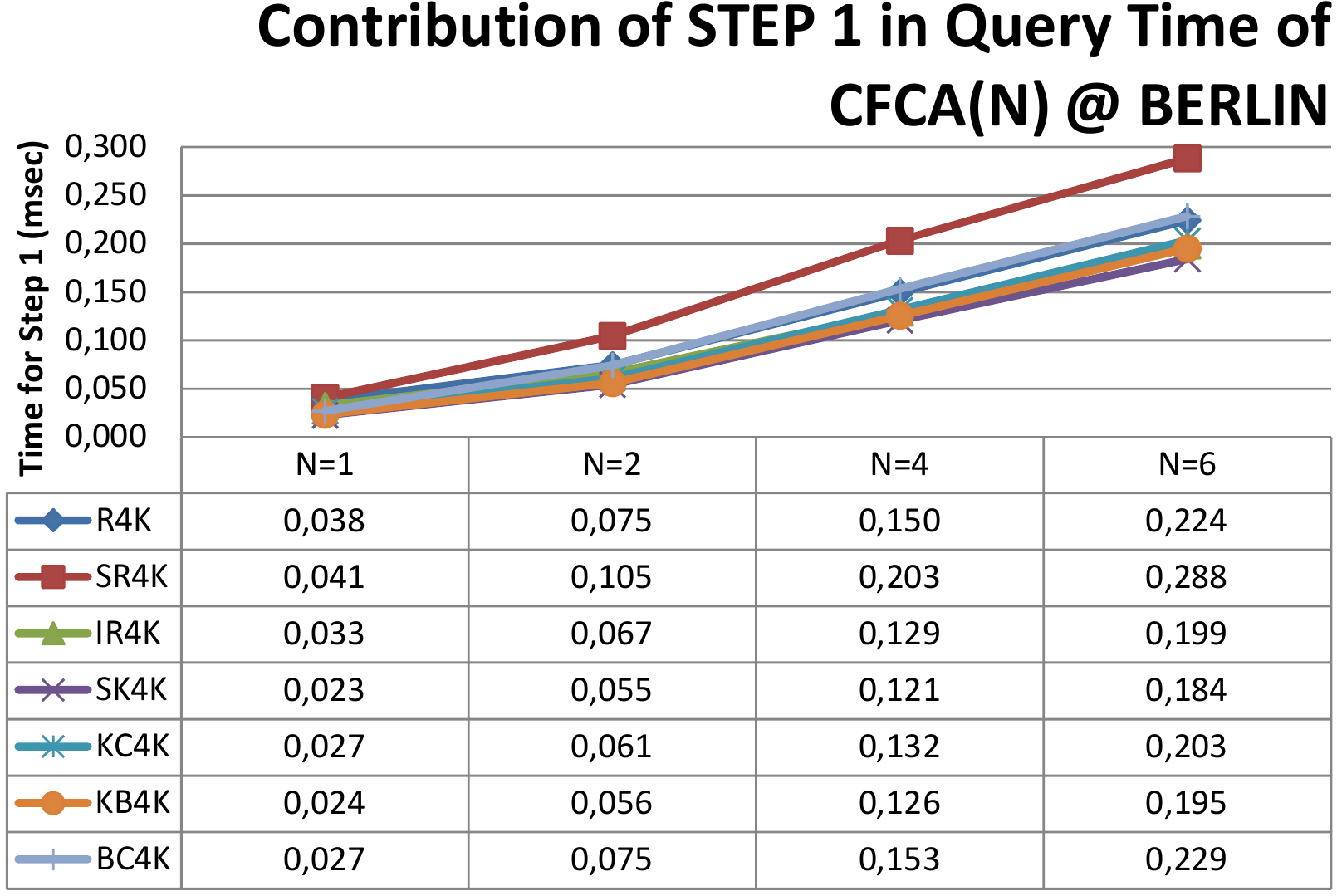}
	\includegraphics[height=4.25cm]{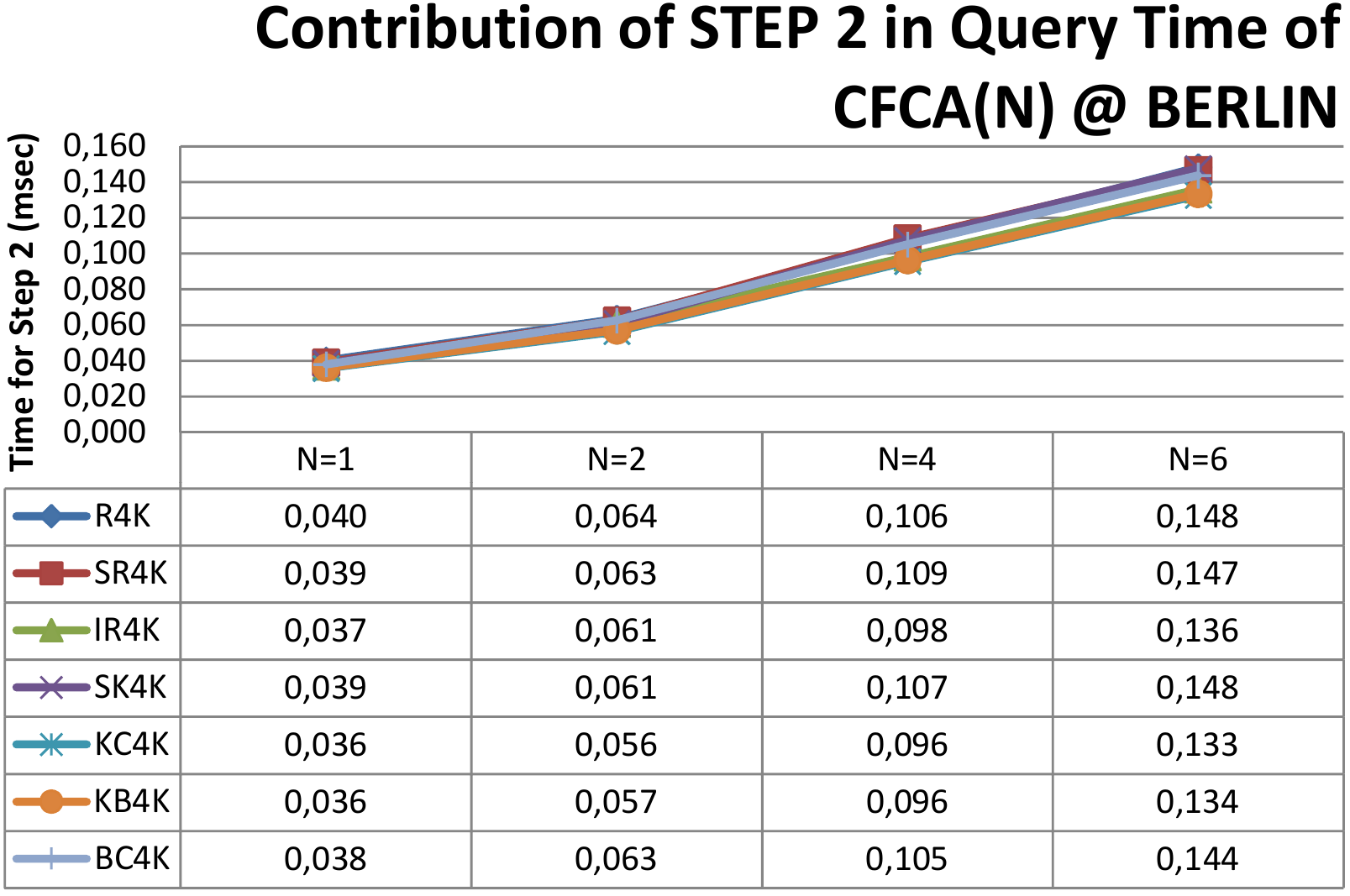}
	\\
	\includegraphics[height=4.25cm]{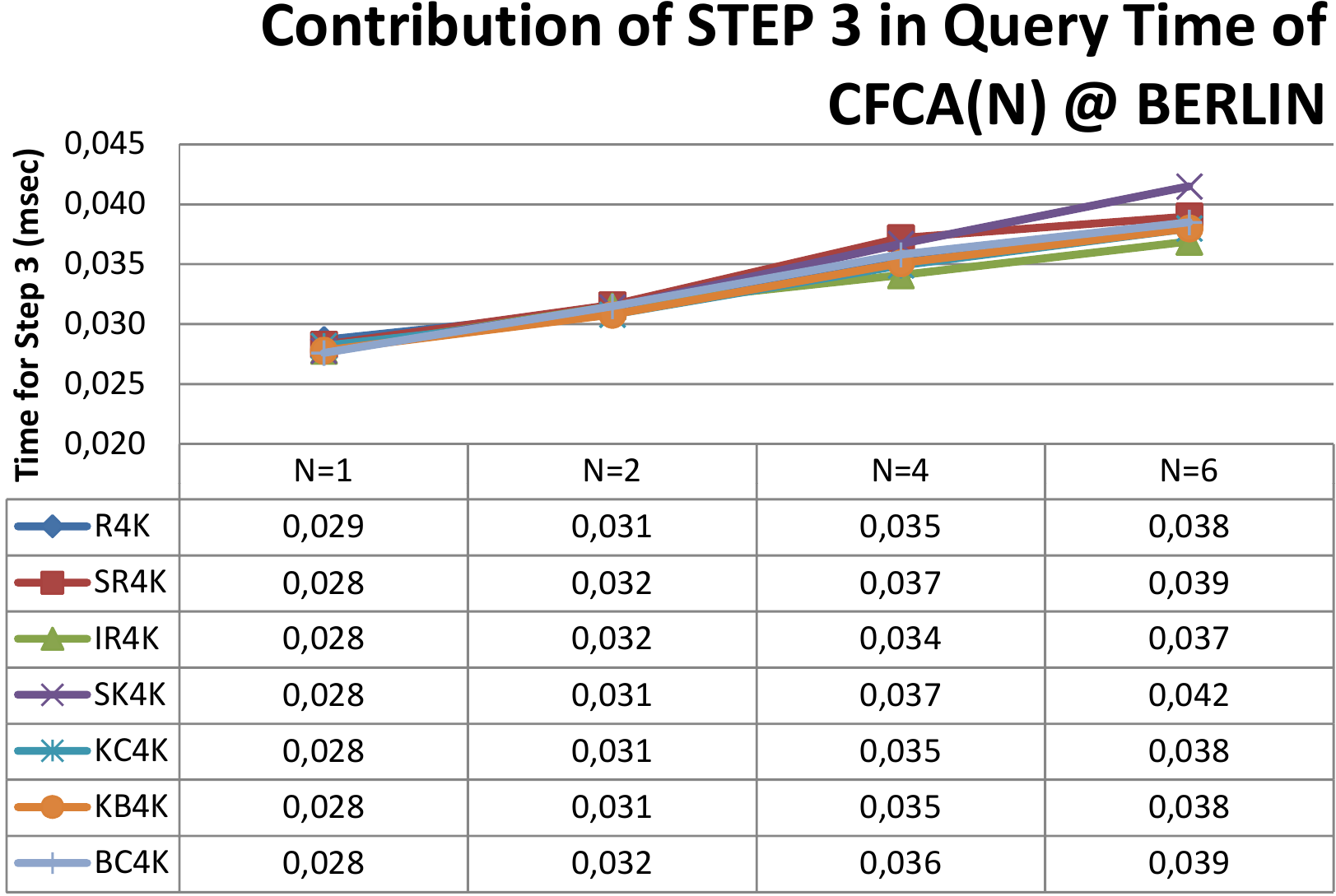}
	\end{tabular}
}
\caption{\label{figure:BERLIN_CFLAT-PER-STEP-TIME-PERFORMANCE}
	Comparison of absolute running times per step of $\alg{CFCA(N)}$, at $1.32$sec resolution, for a query set of $50,000$ \emph{random queries} in Berlin.
}
\end{figure}

Figures~\ref{figure:BERLIN_CFLAT-PER-STEP-RANK-PERFORMANCE} and ~\ref{figure:BERLIN_CFLAT-PER-STEP-TIME-PERFORMANCE} give these measurements of $\alg{CFCA(N)}$ in Berlin. I.e.,
the number of arcs checked for relaxation by the initial $\alg{TDD}$-ball from $(o,t_o)$ in Step 1,
the number of marked arcs connecting predecessors to intermediate vertices in Step 2, and
the number of arcs checked for relaxation during the extension of the $\alg{TDD}$-ball within the marked subgraph,
in order to provide the resulting $od$-path.

It is clear from Figures~\ref{figure:BERLIN_CFLAT-PER-STEP-RANK-PERFORMANCE} and \ref{figure:BERLIN_CFLAT-PER-STEP-TIME-PERFORMANCE} that only Step 1 depends on the type of landmarks that we consider. Observe also that Step 3 is essentially independent of the value of $N$, whereas the other two steps depend linearly on it.
It is worth noting that, for R4K, while the contribution of Step 1 to the overall effort of $\alg{CFCA}$, as $N$ increases, varies from $17.3$\% to $26.7$\% w.r.t.	the number of touched vertices, w.r.t. absolute times it is much more significant, varying from $35.6$\% up to $54.6$\%. This is exactly why we get a significant reduction in the query time when increasing the number of landmarks from $4$K to $8$K, but the (still significant) gain decreases as we go from $8$K to $16$K and almost vanishes when we go from $16$K to $32$K landmarks (cf.~Figure~\ref{figure:BERLIN_CFLAT-SCALABILITY-RANDOM}). At least with respect to query-times, it seems that $16$K is actually the ultimate size at which we should stop. On the other hand, the relative error keeps improving almost linearly with the number of landmarks.

Recall that the measurement does not only concern the estimation of an upper-bound on the earliest-arrival-time at (or equivalently, the shortest travel-time towards) the destination, but also the explicit construction of the corresponding $od$-path that guarantees this bound. Observe also that in absolute running times the speed-up is almost double, because the computationally most demanding step 2 only concerns accesses to the preprocessed data and there is no need for handling priority queues. Moreover, step 3 only concerns a very limited subgraph, containing only a few hundreds of arcs in overall.

\begin{figure}[htb!]
\centerline{%
	\begin{tabular}{c}
	\includegraphics[height=4.25cm]{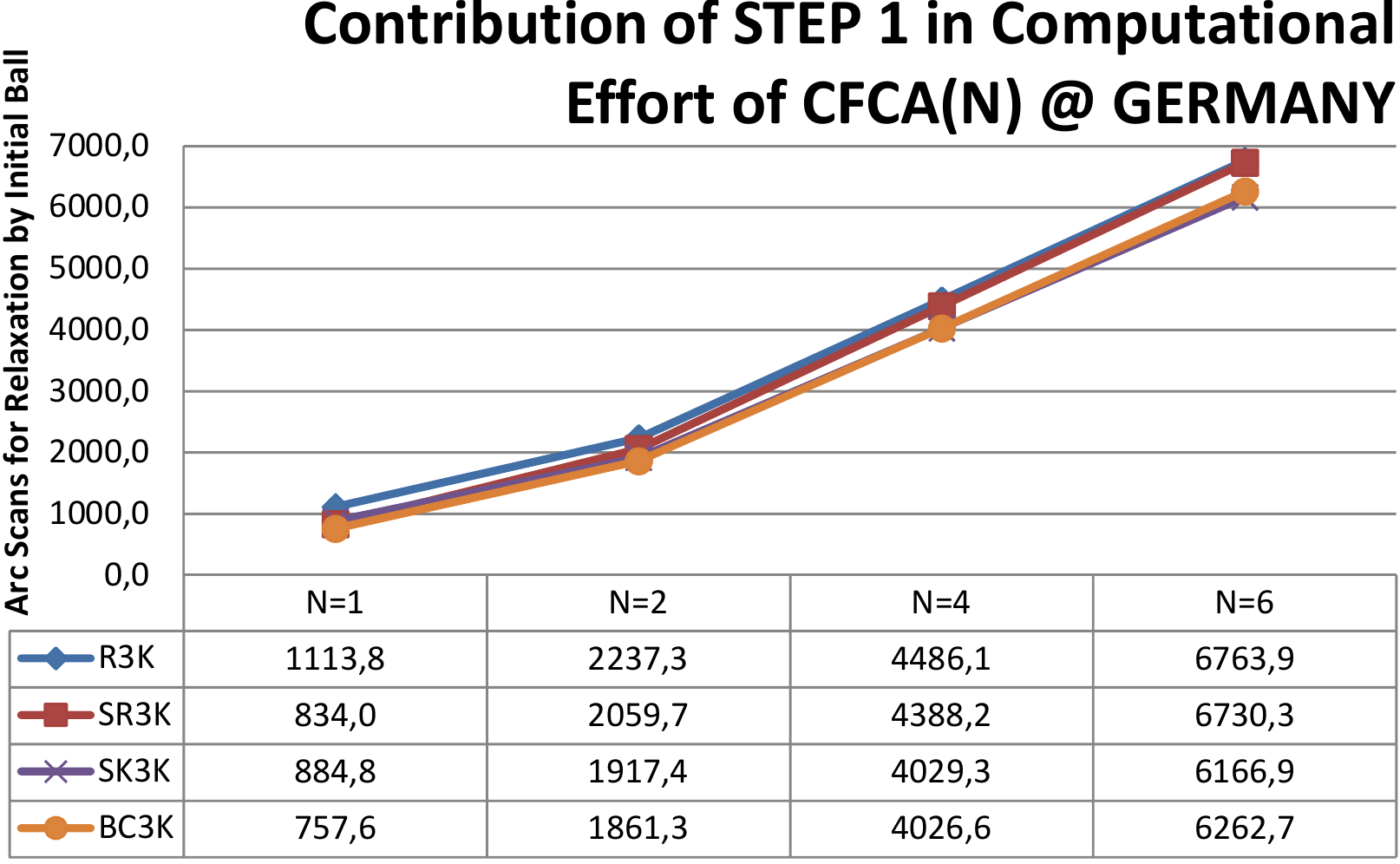}
	\includegraphics[height=4.25cm]{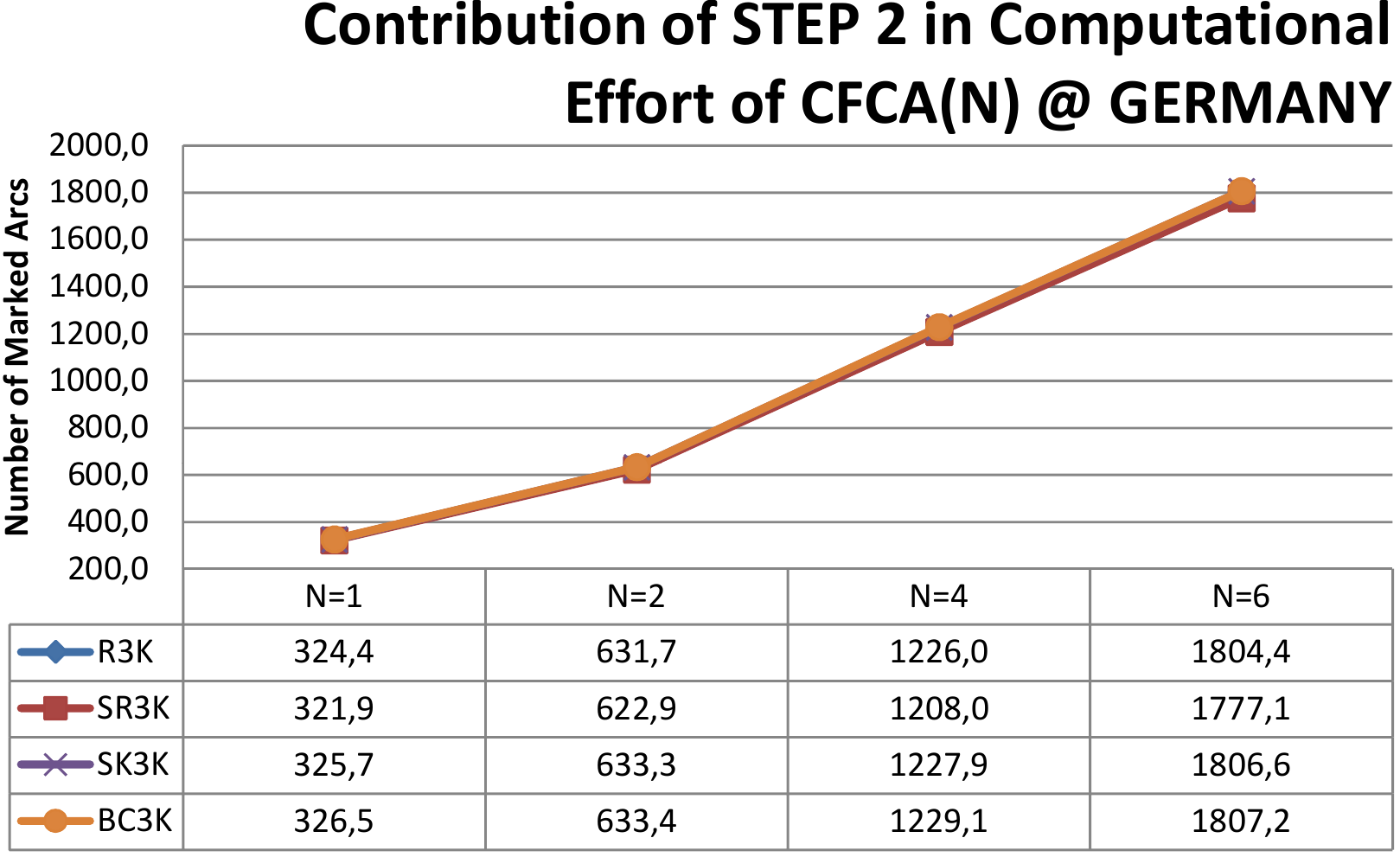}
	\\
	\includegraphics[height=4.25cm]{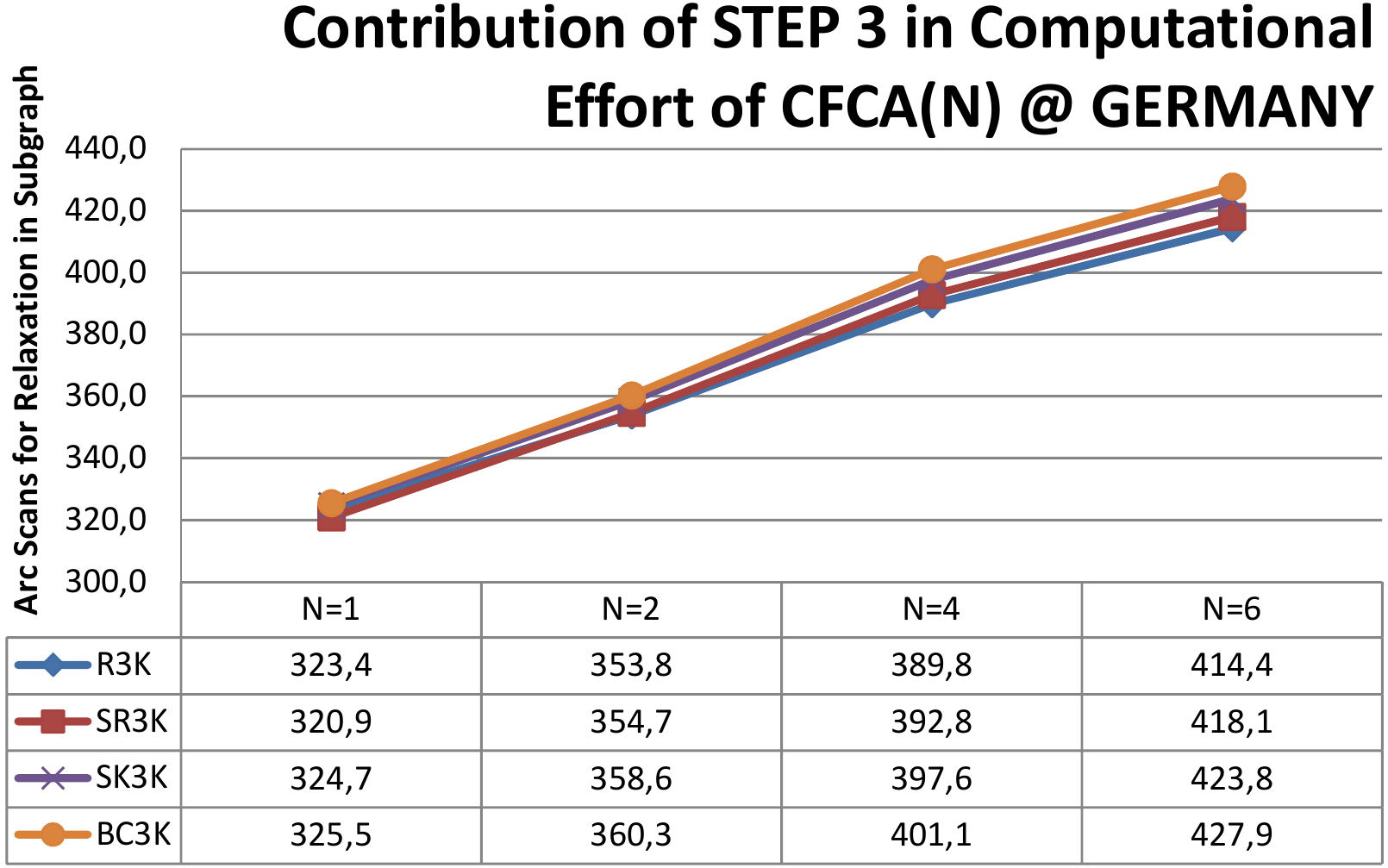}
	\end{tabular}
}%
\caption{\label{figure:GERMANY_CFLAT-PER-STEP-RANK-PERFORMANCE}
	Comparison of contributions in number of touched vertices, per step of $\alg{CFCA(N)}$, at $1.32$sec resolution, for a query set of $50,000$ \emph{random queries} in Germany.
}
\end{figure}

Figure~\ref{figure:GERMANY_CFLAT-PER-STEP-RANK-PERFORMANCE} demonstrates the analogous measurements for Germany. Again we observe the remarkable stability (and independence of the landmark set) for steps 2 and 3, as well as the linear dependence of steps 1 and 2, and the independence of step 3 on the value of $N$.

\begin{figure}[htb!]
\centerline{%
	\begin{tabular}{c}
	\includegraphics[height=4.25cm]{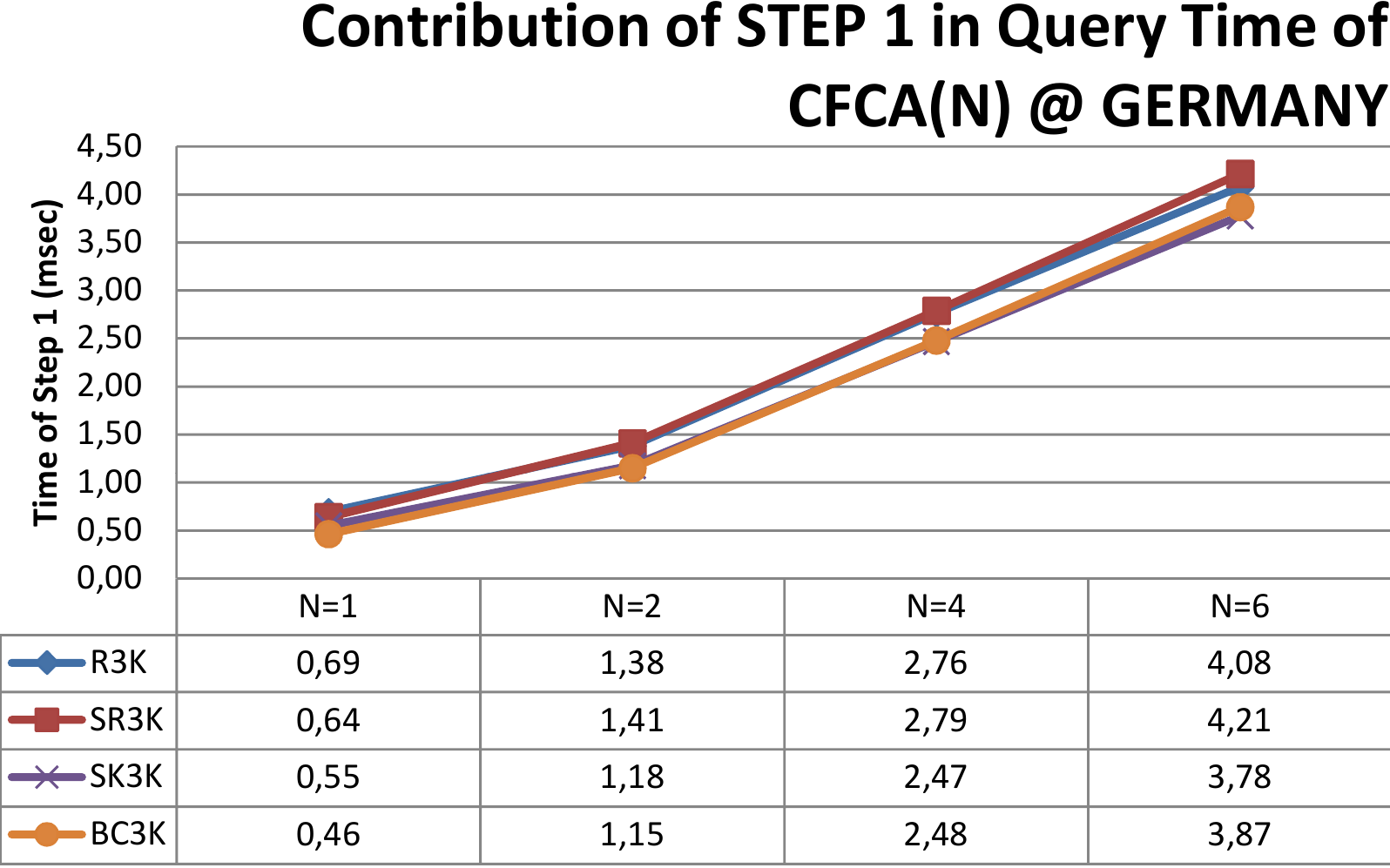}
	\includegraphics[height=4.25cm]{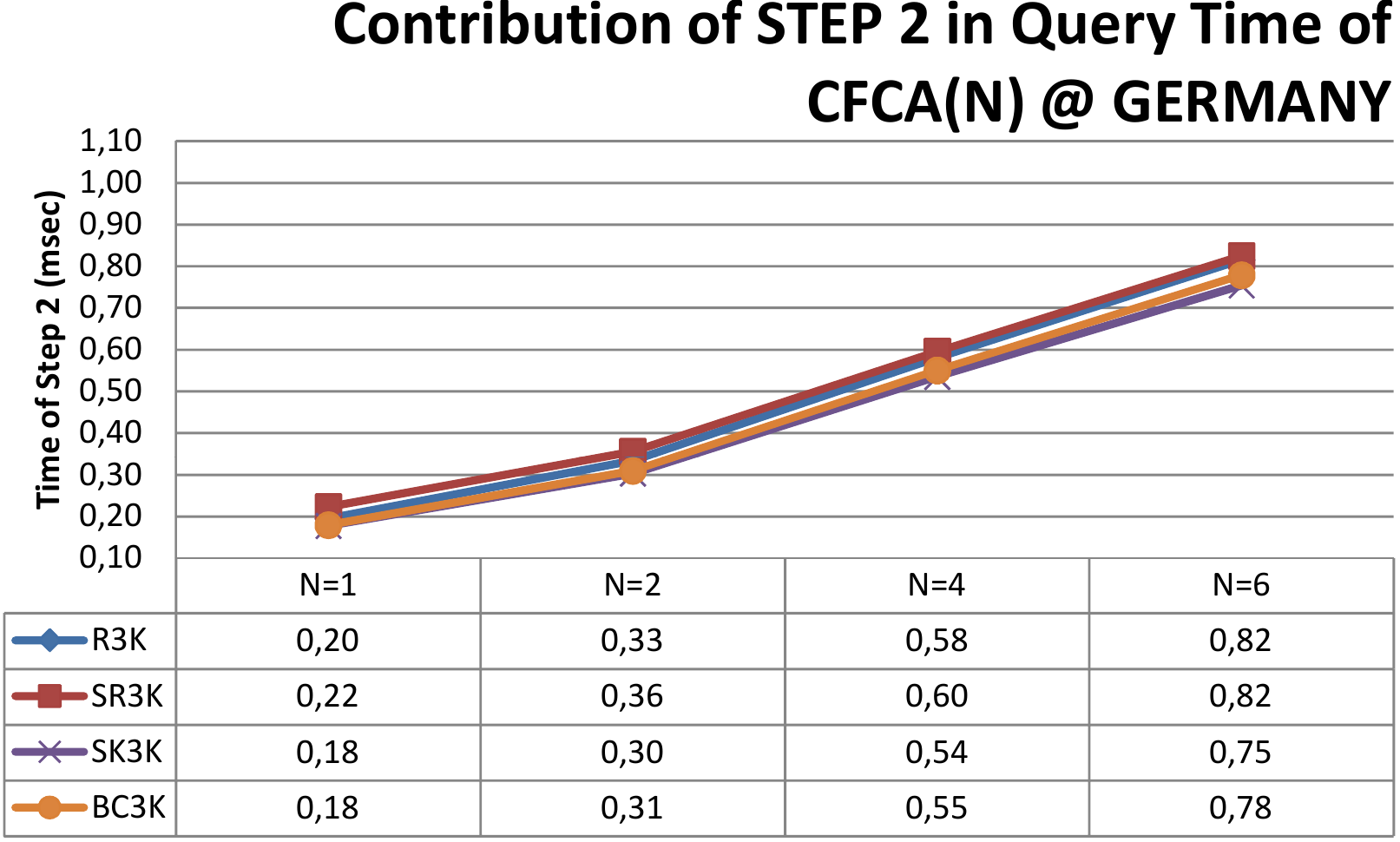}
	\\
	\includegraphics[height=4.25cm]{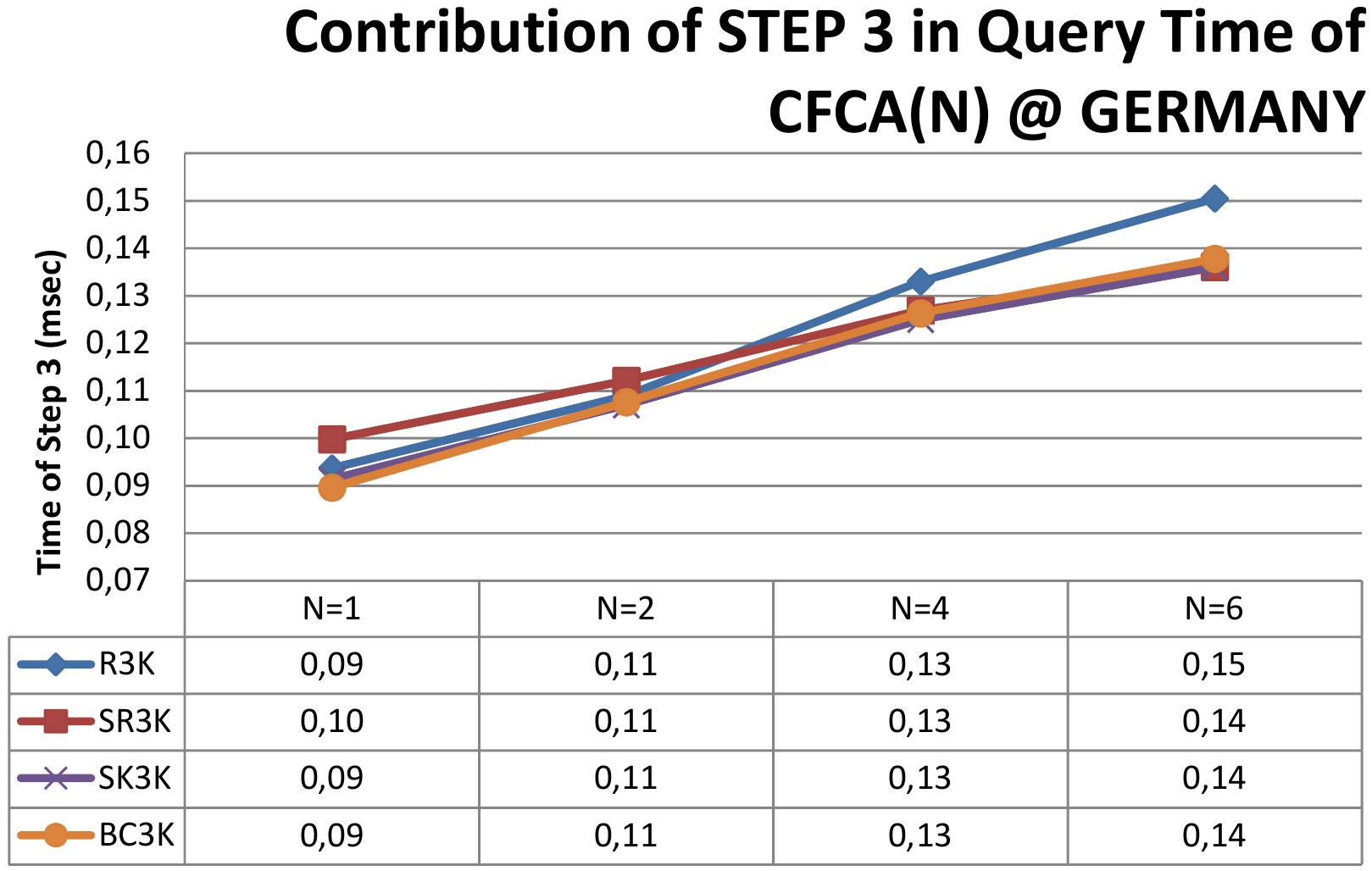}
	\end{tabular}
}%
\caption{\label{figure:GERMANY_CFLAT-PER-STEP-TIME-PERFORMANCE}
	Comparison of running times per step of $\alg{CFCA(N)}$, at $1.32$sec resolution, for a query set of $50,000$ \emph{random queries} in Germany.
}
\end{figure}

Observe finally that for Germany the speedups within the two measures (absolute running times, and ``touched'' arcs) are analogous. This is due to the fact that, since we have a quite small landmark set size this time, step 1 actually dominates the computational effort in this case.

\section{Exploring Outliers in Relative Errors}

\begin{figure}[htb!]
	\centerline{%
	\includegraphics[width=6.75cm]{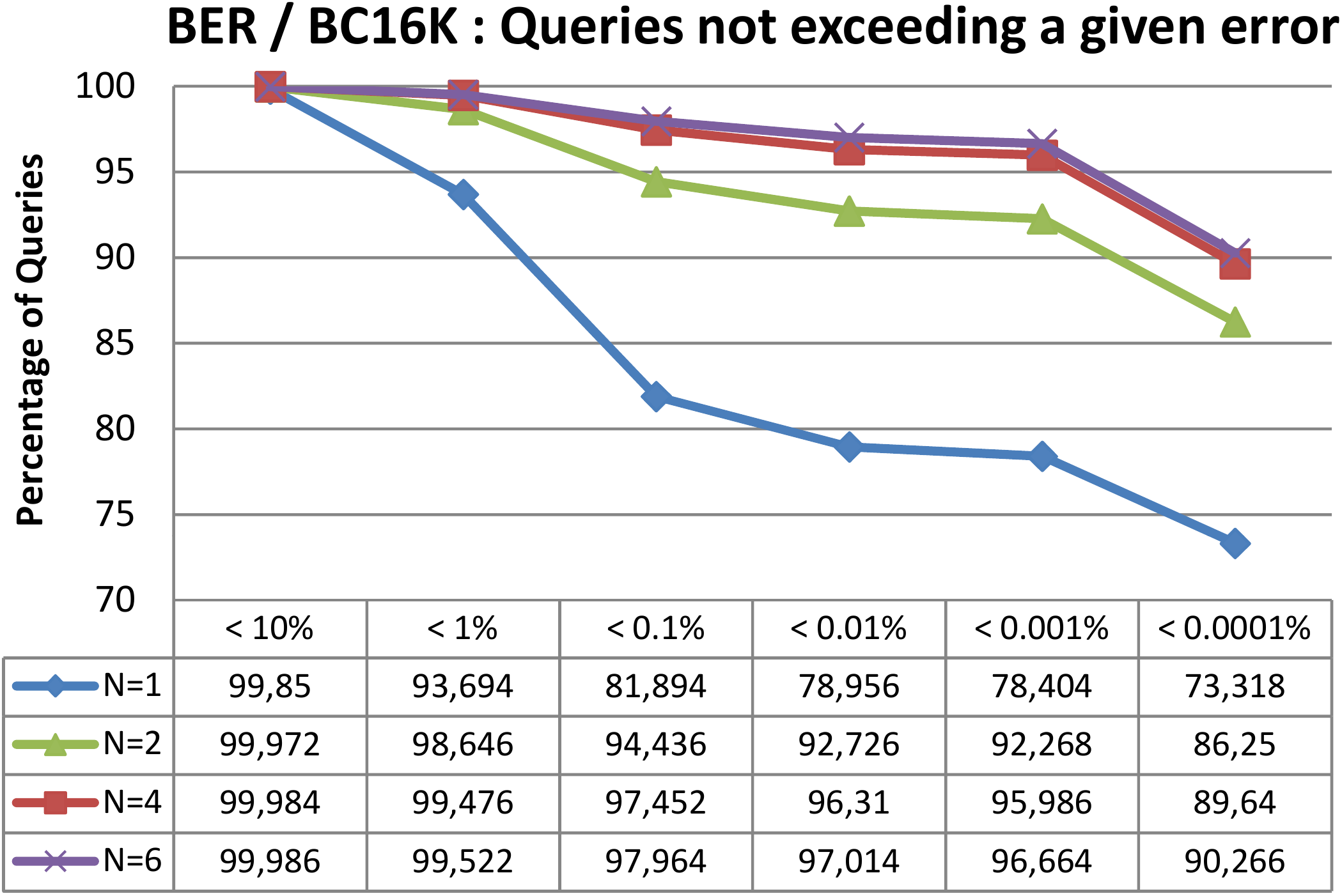}
	~
	\includegraphics[width=6.75cm]{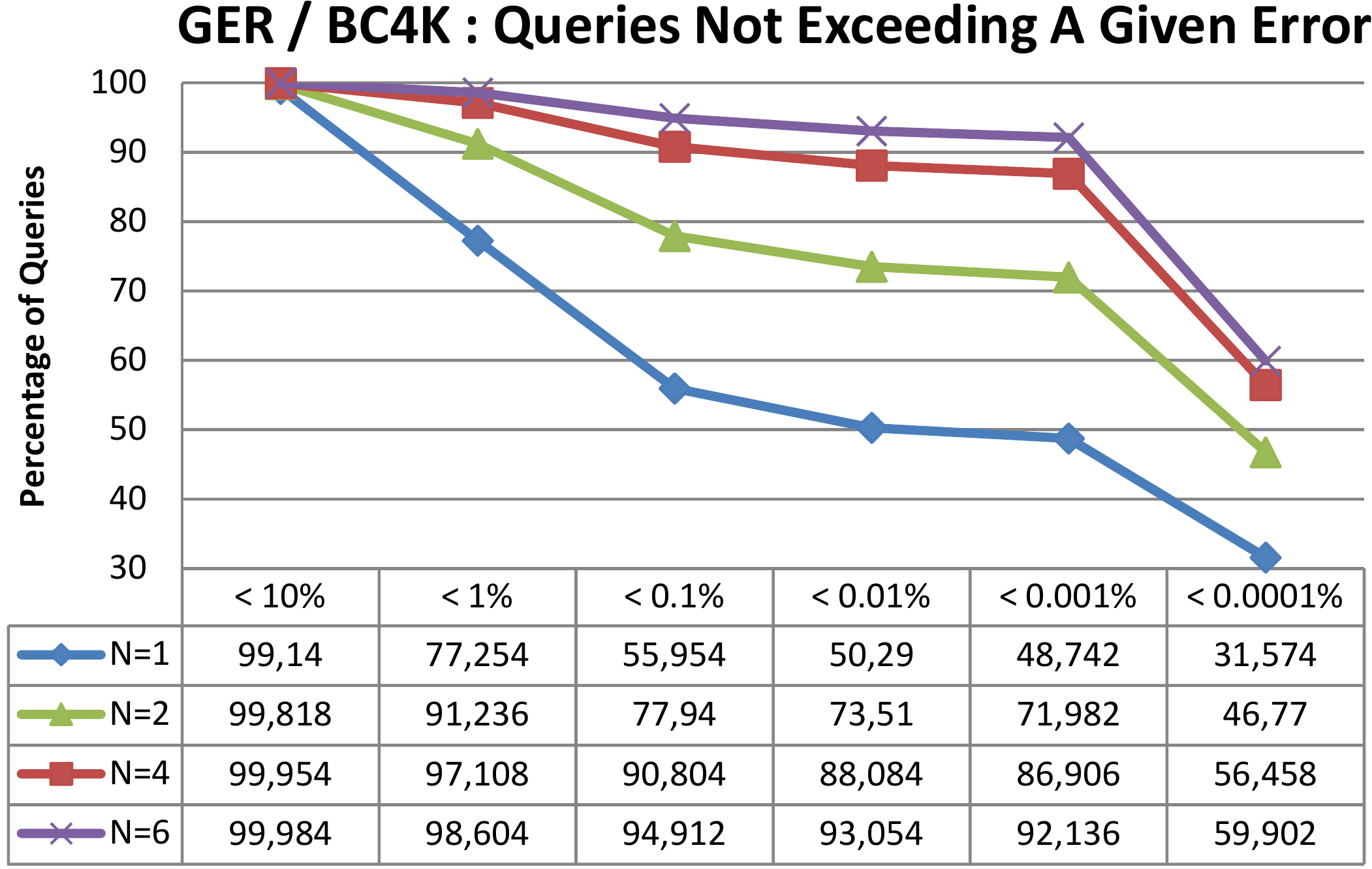}}
	\caption{\label{figure:error-quantiles-overview}
	Tails of the error percentages of $\alg{CFCA(N)}$, for $50,000$ randomly chosen queries in the instance of Berlin with the BC16K landmark set, and for the instance of Germany with the BC4K landmark set.}
	\end{figure}

The purpose of our next experiment was to delve into the details of the relative error of $\alg{CFCA(N)}$. We study the quantiles of the relative error for serving $50,000$ random queries, for  BC16K at Berlin, and for BC4K at Germany.
Figure~\ref{figure:error-quantiles-overview} presents the results of this experimentation.
	
It is worth mentioning that in Berlin, with BC16K-landmarks we can have almost $99.52$\% of queries with error less than $1$\%, and $97.96$\% with error less than $0.1$\%. 
As for Germany, for BC4K-landmarks we can have $98.6$\% of the queries answered with an error less than $1$\% and $94.9$\% of them with error less than $0.1$\%.

\section{Discussion on State-Of-Art Oracles and Speedup Techniques.}
\label{section:discussion-on-sota}

$\alg{CFLAT}$ achieves a significant improvement compared to $\alg{FLAT}$
	\cite{2016-Kontogiannis-Michalopoulos-Papastavrou-Paraskevopoulos-Wagner-Zaroliagis}.
Concerning preprocessing requirements for Berlin (resp.~Germany), $\alg{FLAT}$ consumed \emph{compressed} space $14$MiB ($25.7$MiB) and time $59.4$sec ($90$sec), whereas $\alg{CFLAT}$ requires uncompressed space $2.58$MB ($27.44$MiB)
[or compressed space $0.702$MiB ($8.26$MiB)] and $3.306$sec ($29.32$sec), per landmark.
As for the query performance, $\alg{FCA}(1)$ achieved $0.081$msec ($1.269$msec) and $0.771$\% ($1.534$\%), whereas $\alg{CFCA(1)}$ achieves $0.077$msec ($0.683$msec) and $0.18$\% ($0.831$\%), despite the fact that it also pays for the path construction.

We now proceed with the comparison of $\alg{CFLAT}$ with state-of-art speedup heuristics. In particular, we consider the speedup heuristics $\alg{inex.TCH}$~\cite{2013-Batz-Geisberger-Sanders-Vetter} (only for Germany), and $\alg{TDCRP}$~\cite{2016-Baum-Dibbelt-Pajor-Wagner}, $\alg{KaTCH}$\footnote{\url{https://github.com/GVeitBatz/KaTCH} with checksum 70b18ad0791a687c554fbfe9039edf79bc3a8ff3.}, $\alg{FreeFlow}$, $\alg{TD\mbox{-}S}$ and $\alg{TD\mbox{-}S\mbox{+}A}$ from \cite{2016-Strasser}, $\alg{FLAT}$~\cite{2016-Kontogiannis-Michalopoulos-Papastavrou-Paraskevopoulos-Wagner-Zaroliagis}, and $\alg{DijFreeFlow}$, $\alg{CFLAT}$ in this work. 

It should be once more noticed that $\alg{KaTCH}$, $\alg{DijFreeFlow}$, $\alg{FLAT}$ and $\alg{CFLAT}$ were experimented on our own machine, with exactly the same sets of uniformly and randomly selected queries. For the other algorithms we could only report (unscaled) the measurements of their experimentation by their authors, since we do not have the source codes at our disposal. For the sake of comparison and a posteriori verification, we provide the two random query sets that we have used in \url{http://150.140.143.218:8000/public/}. 

For Berlin, the only experimentally evaluated speedup techniques we are aware of are $\alg{TDCRP}$ \cite{2016-Baum-Dibbelt-Pajor-Wagner} and $\alg{TD\mbox{-}S}$, $\alg{TD\mbox{-}S\mbox{+}A}$~\cite{2016-Strasser}. We have also experimented with $\alg{KaTCH}$, but the observed performance is dominated by most of the other algorithms, except for $\alg{TD\mbox{-}S\mbox{+}A}$. 
$\alg{TDCRP}$ requires
$21$min of preprocessing time on a $16$-core machine,
$31$MiB of preprocessing space, and
achieves query performance (average query-time and relative error) $0.28$msec and $1.47$\%.
For an analogous amount of preprocessing \emph{work},
$\alg{CFLAT}$ preprocesses R500 in $27$min, exploiting $12$ threads on a $6$-core machine, consuming $1.3$GiB ($0.34$GiB compressed) space. It achieves query performance varying from $0.356$msec and $1.915$\% (for $N=1)$, to $1.848$msec and $0.102$\% (for $N=6$).
If query-time is the main goal, then with BC8K+R8K $\alg{CFCA}$ achieves query performance varying from $0.076$msec and $0.192$\% (for $N=1$), to $0.226$msec and $0.022$\% (cf. Figure~\ref{figure:BERLIN_CFLAT-16K-R+BC-MIX}).

The sweet spot of $\alg{CFLAT}$ w.r.t. the trade-off between query performance and preprocessing, seems to be  for $4$K landmarks: BC4K is preprocessed in $3$h$44$min consuming $10.4$GiB ($2.8$GiB compressed) space (cf. Figure~\ref{figure:BERLIN+GERMANY_PREPROCESSING-RANDOM}) and $\alg{CFCA}$ then achieves query-performance varying from $0.088$msec and $0.521$\%, to $0.367$msec and $0.021$\% (cf. Figure~\ref{figure:BERLIN_CFLAT-16K-R+BC-MIX}).
Moreover, for BC16K $\alg{CFCA}(6)$ provides stretch less than $1$\% for $99.522$\% of the $50,000$ queries (cf. Figure~\ref{figure:error-quantiles-overview}).
As for $\alg{FreeFlow}$, $\alg{TD\mbox{-}S}$ and $\alg{TD\mbox{-}S\mbox{+}A}$ \cite{2016-Strasser}, it is certainly the case that these are quite simple algorithms which achieve remarkable performances. Their rationale is analogous to that of $\alg{CFLAT}$: Certain paths for carefully selected time-windows (rather time-stamped shortest-path trees of $\alg{CFLAT}$) are chosen, whose arcs induce a quite small subgraph in which $\alg{TDD}$ is executed. The difference with our oracle is that, instead of having the combinatorial structures automatically positioned in time, based on the time-dependent metric (as $\alg{CFLAT}$ does in order to achieve a required approximation guarantee), manual selection of time-windows (by trial-and-error) is used during the preprocessing of $\alg{TD\mbox{-}S}$ and $\alg{TD\mbox{-}S\mbox{+}A}$.
For running times, $\alg{CFLAT}$ can be faster than all these algorithms, e.g., for BC8K+R8K and $\alg{CFCA(1)}$. 
Concerning their remarkable error performances, it should be noted that for $\alg{FreeFlow}$ we tried to verify the reported errors by running our own version ($\alg{DijFreeFlow}$). $\alg{DijFreeFlow}$  is not based on $\alg{CH}$, but on running (static) $\alg{Dijkstra}$ on the free-flow metric and then computing the time-dependent travel-time along the chosen path. 
At least for the common query-set that we use in all our experiments, the error guarantees for $\alg{FreeFlow}$ are much worse than the ones reported in \cite{2016-Strasser}. 

For Germany, we compare $\alg{CFLAT}$ with all the considered oracles and speedup heuristics.
$\alg{TDCRP}$ requires total preprocessing time $4$h$41$min on a $16$-core machine, using $0.361$GiB preprocessing space, and achieves query performance $1.17$msec and  $0.68$\%.
$\alg{inex.TCH}(0.1)$, on the other hand, preprocesses the instance in $6$h$18$min, consuming $1.34$GiB space, and achieves query performance $0.7$msec and $0.02$\%, and worst-case error $0.1$\%.
For an analogous amount of preprocessing \emph{work}, $\alg{CFLAT}$ preprocesses R1K in $8$h$9$min using $12$ threads of our $6$-core machine consuming $26.8$GiB ($8.1$GiB compressed) space, cf. Figure~\ref{figure:BERLIN+GERMANY_PREPROCESSING-RANDOM}. $\alg{CFCA}$ achieves query performance varying from $2.175$msec and $1.582$\% (for $N=1$), to $11.974$msec and $0.071$\% (for $N=6$).
If query-time is the main goal, then $\alg{CFLAT}$ preprocesses the hybrid landmark set BC3K+R1K in $32$h$35$min consuming $107.2$GiB ($32.3$GiB compressed) space, see Figure~\ref{figure:BERLIN+GERMANY_PREPROCESSING-RANDOM}. $\alg{CFCA}$ achieves then query performance varying from $0.683$msec and $0.831$\% (for $N=1)$, to $4.104$msec and $0.031$\% (for $N=6$), see Figure~\ref{figure:GERMANY_CFLAT-4K-R+BC-MIX}.
Moreover, for BC4K $\alg{CFCA}(6)$ provides an error at most $1$\% for $98.604$\% of the $50,000$ queries (cf. Figure~\ref{figure:error-quantiles-overview}).
$\alg{KaTCH}$ is clearly worse than $\alg{CFLAT}$. Indeed, the performance of $\alg{KaTCH}$ significantly deviates from the reported performances of all variants of $\alg{inex.TCH}$, and is dominated by all oracles and speedup heuristics. One possible explanation might be that our own query set triggered some sort of bug in $\alg{KaTCH}$, but it is impossible for us to verify this.  

Finally, the query performances of $\alg{FreeFlow}$, $\alg{TD\mbox{-}S}$ and $\alg{TD\mbox{-}S\mbox{+}A}$ for Germany are comparable to those of $\alg{CFLAT}$, but the reported errors are much better. Again, we tried to verify the reported errors by running our own version ($\alg{DijFreeFlow}$). 
At least for the common query-set that we use in all our experiments, the error guarantees for $\alg{FreeFlow}$ are much worse than the ones reported in \cite{2016-Strasser}.

Concerning temporal changes in the time-dependent data, the live-traffic updating procedure of $\alg{CFLAT}$'s preprocessed data, among $1,000$ $15$-min randomly chosen disruptions, takes (per disruption)
$0.275$sec in Berlin for updating on average $48$ affected BC4K-landmarks,
and $37.676$sec in Germany for updating on average $4$ affected BC3K-landmarks (cf. Section~\ref{section:live-traffic-updating}).

\section{Live Traffic Updating}
\label{section:live-traffic-updating}

As was done in \cite{2016-Kontogiannis-Michalopoulos-Papastavrou-Paraskevopoulos-Wagner-Zaroliagis}, we conducted an experiment to assess the responsiveness of $\alg{CFLAT}$ to live-traffic updates.
In particular, the goal is, when a disruption occurs ``on the fly'' (e.g., the abrupt and unforeseen congestion, or even blockage of a road segment for half an hour due to a car accident), how fast the oracle can take into account, for the affected route plans that have already been suggested or will be suggested in the near future, the temporal traffic-related information.
We thus consider dynamic scenarios where there is a stream of live-traffic reports about abnormal delays on certain road segments (arcs), along with a time-window $[r_s, r_e]$, of typically small duration, in which the disruption occurs.

Our update step involves the recomputation of min-travel-time-path summaries for a subset of landmarks in the vicinity of the disruption. In particular, for a disrupted arc $a=uv$ of disruption duration $[r_s,r_e]$, we run a (static) $\alg{Backward\mbox{-}Dijkstra}$ from $u$ under the free-flow metric, with travel time radius of at most $r_e-r_s$. The limited travel time radius is used to trace only the nearest landmarks that may actually be affected by the disruption, leaving unaffected all the ``faraway'' landmarks. The goal is to update as soon as possible the recommendations for the drivers who are close to the area of disruption.
For each affected landmark $\ell$, we consider a \term{disruption-times window} $[t_s, t_e]$, containing the latest departure-times from $\ell$ for arriving at the tail $u$ at any time in the interval $[r_s, r_e]$ in which the disruption occurs.
We then compute \emph{temporal} travel-time summaries for each affected landmark and disruption-times window. This computation is conducted as in the preprocessing phase.
Using a $15\mbox{-min}$ radius for the disruptions, we executed $1,000$ live-traffic updates for the instances of Berlin and Germany, for the landmark set BC4K and BC3K, respectively.
For Berlin, the average number of affected landmarks was $48$ for Berlin, and the updating procedure of the affected landmarks' summaries requires average time $0.275$sec, using $12$ threads on our $6$-core machine.
As for Germany, the average number of affected landmarks was only $4$, and the updating procedure of the affected landmarks' summaries requires average time $37.68$sec, again using $12$ threads on our $6$-core machine. 

%% file: 2017-ESA-B-SUBMISSION_CFLAT-EVALUATION.bbl
\begin{thebibliography}{10}
\begin{footnotesize}

\bibitem{2014-KAHIP}
{KaHIP -- Karlsruhe High Quality Partitioning}, May 2014.

\bibitem{2007-Bader-Kintali-Madduri-Mihail}
D.~Bader, S.~Kintali, K.~Madduri, M.~Mihail:
\newblock Approximating betweenness centrality.
\newblock {\em Algorithms and Models for the Web-Graph (WAW)}, pp. 124--137, Springer (2007)

\bibitem{2013-Batz-Geisberger-Sanders-Vetter}
G.~V.~Batz, R.~Geisberger, P.~Sanders, C.~Vetter:
\newblock Minimum time-dependent travel times with contraction hierarchies.
\newblock {\em J. of Experimental Algorithmics}, \textbf{18}(1.4):1--43, ACM (2013).

\bibitem{2016-Baum-Dibbelt-Pajor-Wagner}
M.~Baum, J.~Dibbelt, T.~Pajor, D.~Wagner:
\newblock Dynamic time-dependent route planning in road networks with user preferences.
\newblock {\em Experimental Algorithms (SEA)}, \textbf{LNCS~9685}:33--49, Springer (2016)

\bibitem{2012-Dehne-Omran-Sack}
F.~Dehne, M.~T. Omran, and J.-R.~Sack.
\newblock Shortest paths in time-dependent FIFO networks.
\newblock {\em Algorithmica}, \textbf{62}(1-2):416--435 (2012)

\bibitem{2011-Delling_TDSHARC}
D.~Delling:
\newblock {Time-Dependent SHARC-Routing}.
\newblock {\em Algorithmica}, \textbf{60}(1):60--94 (2011)

\bibitem{2012-Delling-Nannicini}
D.~Delling, G.~Nannicini:
\newblock Core routing on dynamic time-dependent road networks.
\newblock {\em Informs J. on Computing}, \textbf{24}(2):187--201 (2012)

\bibitem{2007-Delling-Wagner}
D.~Delling, D.~Wagner:
\newblock Landmark-based routing in dynamic graphs.
\newblock {\em Experimental Algorithms
  (WEA'07)}, \textbf{LNCS~4525}:52--65, Springer (2007)

\bibitem{2009-Delling-Wagner}
D.~Delling, D.~Wagner:
\newblock Time-dependent route planning.
\newblock {\em Robust and Online Large-Scale Optimization}, \textbf{LNCS~5868}:207--230, Springer (2009)

\bibitem{2010-Demiryurek-Kashani-Shahabi}
U.~Demiryurek, F.~Banaei-Kashani, C.~Shahabi:
\newblock A case for time-dependent shortest path computation in spatial networks.
\newblock {\em SIGSPATIAL Advances in Geographic Information Systems (GIS)}, pp. 474--477, ACM (2010).

\bibitem{1969-Dreyfus}
S.~E.~Dreyfus:
\newblock An appraisal of some shortest-path algorithms.
\newblock {\em Operations Research}, \textbf{17}(3):395--412, 1969.

\bibitem{2014-Foschini-Hershberger-Suri}
L.~Foschini, J.~Hershberger, S.~Suri:
\newblock On the complexity of time-dependent shortest paths.
\newblock {\em Algorithmica}, \textbf{68}(4):1075--1097 (2014).

\bibitem{2009-Hilger-Koehler-Moehring-Schilling}
M.~Hilger, E.~K{\"o}hler, R.~H.~M{\"o}hring, H.~Schilling:
\newblock Fast point-to-point shortest path computations with arc-flags.
\newblock {\em The Shortest Path Problem: Ninth DIMACS Implementation
  Challenge}, \textbf{74}: 41--72, AMS (2009)

\bibitem{2015-Kontogiannis-Michalopoulos-Papastavrou-Paraskevopoulos-Wagner-Zaroliagis}
S.~Kontogiannis, G.~Michalopoulos, G.~Papastavrou, A.~Paraskevopoulos, D.~Wagner, C.~Zaroliagis:
\newblock Analysis and experimental evaluation of time-dependent distance oracles.
\newblock {\em Algorithm Engineering and Experiments (ALENEX)}, pp. 147--158, SIAM (2015)

\bibitem{2016-Kontogiannis-Michalopoulos-Papastavrou-Paraskevopoulos-Wagner-Zaroliagis}
S.~Kontogiannis, G.~Michalopoulos, G.~Papastavrou, A.~Paraskevopoulos, D.~Wagner, C.~Zaroliagis:
\newblock Engineering oracles for time-dependent road networks.
\newblock {\em Algorithm Engineering and Experiments (ALENEX)}, pp. 1--14, SIAM (2016)

\bibitem{2016-Kontogiannis-Wagner-Zaroliagis}
S.~Kontogiannis, D.~Wagner, C.~Zaroliagis:
\newblock Hierarchical oracles for time-dependent networks. 
\newblock {\em Algorithms and Computation (ISAAC)}, \textbf{64}(47):1-13, LIPICS (2016) 

\bibitem{2014-Kontogiannis-Zaroliagis}
S.~Kontogiannis, C.~Zaroliagis:
\newblock Distance oracles for time-dependent networks.
\newblock {\em Algorithmica}, \textbf{74}(4):1404--1434 (2016)

\bibitem{2013-PGL}
G.~Mali, P.~Michail, A.~Paraskevopoulos, C.~Zaroliagis:
\newblock A new dynamic graph structure for large-scale transportation networks.
\newblock {\em Algorithms and Complexity (CIAC)}, \textbf{LNCS~7878}:312--323, Springer (2013)

\bibitem{2012-Nannicini-Delling-Liberti-Schultes}
G.~Nannicini, D.~Delling, L.~Liberti, D.~Schultes:
\newblock Bidirectional {A*} search on time-dependent road networks.
\newblock {\em Networks}, \textbf{59}:240--251 (2012)

\bibitem{2014-Omran-Sack}
M.~Omran and J.-R.~Sack:
\newblock Improved approximation for time-dependent shortest paths.
\newblock {\em Computing and Combinatorics (COCOON)}, \textbf{LNCS~8591}:453--464, Springer (2014)

\bibitem{1990-Orda-Rom}
A.~Orda, R.~Rom:
\newblock Shortest-path and minimum delay algorithms in networks with time-dependent edge-length.
\newblock {\em J. of the ACM}, \textbf{37}(3):607--625, ACM (1990)


\bibitem{2016-Strasser}
B.~Strasser:
\newblock Intriguingly simple and efficient time-dependent routing in road networks.
\newblock ArXiv technical report (arxiv:1606.06636v1), Karlsruhe Institute of
  Technology (2016)

\end{footnotesize}
\end{thebibliography}
